\documentstyle[epsfig,natbib2,natbibmnfix]{mn}

\newcommand{\be}{\begin{equation}}
\newcommand{\ee}{\end{equation}}
\newcommand{\bea}{\begin{eqnarray}}
\newcommand{\eea}{\end{eqnarray}}
\newcommand{\bc}{\begin{center}}
\newcommand{\ec}{\end{center}}

\renewcommand{\vec}[1]{ {\bmath #1} } 

\newcommand{\dd}{{\rm d}}

\renewcommand{\thefootnote}{\fnsymbol{footnote}}

\title{Cosmological SPH simulations: The entropy equation}

\author[V.~Springel and L.~Hernquist] {\parbox{18cm}{Volker
Springel$^{1}$\footnotemark[1] and Lars
Hernquist$^2$\footnotemark[2]}\vspace{0.3cm}\\ $^1$Max-Planck-Institut
f\"{u}r Astrophysik, Karl-Schwarzschild-Stra\ss{}e 1, 85740 Garching
bei M\"{u}nchen, Germany\\ $^2$Harvard-Smithsonian Center for
Astrophysics, 60 Garden Street, Cambridge, MA 02138, USA}

\begin{document}

\maketitle
\begin{abstract}
We discuss differences in simulation results that arise between the
use of either the thermal energy or the entropy as an independent
variable in smoothed particle hydrodynamics (SPH).  In this context,
we derive a new version of SPH that, when appropriate, manifestly
conserves both energy and entropy if smoothing lengths are allowed to
adapt freely to the local mass resolution.  To test various
formulations of SPH, we consider point-like energy injection, as in
certain models of supernova feedback, and find that powerful
explosions are well represented by SPH even when the energy is
deposited into a single particle, provided that the entropy equation
is integrated.  If the thermal energy is instead used as an
independent variable, unphysical solutions can be obtained for this
problem.

We also examine the radiative cooling of gas spheres that collapse and
virialize in isolation and of halos that form in cosmological
simulations of structure formation.  When applied to these problems,
the thermal energy version of SPH leads to substantial overcooling in
halos that are resolved with up to a few thousand particles, while the
entropy formulation is biased only moderately low for these halos
under the same circumstances.  For objects resolved with much larger
particle numbers, the two approaches yield consistent results.  We
trace the origin of the differences to systematic resolution effects
in the outer parts of cooling flows.  When the thermal energy equation
is integrated and the resolution is low, the compressional heating of
the gas in the inflow region is underestimated, violating entropy
conservation and improperly accelerating cooling.

The cumulative effect of this overcooling can be significant.  In
cosmological simulations of moderate size, we find that the fraction
of baryons which cool and condense can be reduced by up to a factor
$\sim 2$ if the entropy equation is employed rather than the thermal
energy equation, partly explaining discrepancies with semi-analytic
treatments of galaxy formation.  We also demonstrate that the entropy
method leads to a greatly reduced scatter in the density-temperature
relation of the low-density Ly-$\alpha$ forest relative to the thermal
energy approach, in accord with theoretical expectations.

\end{abstract}

\section{Introduction}

\renewcommand{\thefootnote}{\fnsymbol{footnote}}
\footnotetext[1]{E-mail: volker@mpa-garching.mpg.de}
\footnotetext[2]{\hspace{0.03cm}E-mail: lars@cfa.harvard.edu} Smoothed
particle hydrodynamics (SPH) was introduced by \citet{Lu77} and
\citet{Gi77} as an alternative to grid-based fluid solvers and has
developed into a mature and popular simulation technique.  The
Lagrangian nature of SPH enables it to adjust to the large dynamic
range posed by problems such as the formation of galaxies and
large-scale structure in a manner that is difficult to match with
non-adaptive Eulerian methods.  It has been shown that SPH can produce
accurate results in many situations \citep[e.g.][]{St93}, albeit at
the expense of an artificial viscosity which broadens shocks over
several smoothing lengths.

The use of SPH to study the formation and evolution of galaxies was
pioneered more than a decade ago by a number of researchers
\citep*[][among
others]{Ev88,Ev90,He89b,He89,Nav91,Ba91,Ka91,Hi91,Th92,KaHeWe92}.
Together with subsequent applications, these works demonstrated the
reliability of the approach and ultimately influenced our
understanding of e.g. quasar absorption line systems, the intracluster
medium, and galaxy interactions.  While cosmological SPH simulations
have also provided a successful instrument for exploring galaxy
formation, it has recently become clear that subtle numerical
properties of existing algorithms severely influence characteristics
of model galaxies, such as their luminosities and colours.

These numerical complications follow once radiative cooling of the gas
is included.  Cooling is critically important in the conventional
paradigm of galaxy formation, because it enables baryons to accumulate
in the centres of halos, where they act as reservoir of cold, dense
gas for forming stars.  Thus, cosmological hydrodynamic simulations
must incorporate these dissipative effects to determine the formation
sites and stellar masses of galaxies.  Ultimately, an appropriate
treatment of star formation and associated feedback processes is also
required.  However, it is clear that reliable cooling rates,
uncompromised by numerical effects, are essential if the simulations
are to predict accurate galaxy properties.

Unfortunately, calculating the proper amount of gas that can
radiatively cool is a difficult problem, especially in hierarchical
cosmologies where the most abundant objects are of low mass and hence
will invariably be relatively poorly resolved.  Naively, it is
expected that coarse resolution will tend to inhibit cooling because the
radiative cooling rate per unit volume is a sensitive function of
local density and high density peaks will be washed out by smoothing.
Near the resolution limit of SPH, this effect will suppress cooling
entirely, since halos will not harbour sufficient quantities of high
density gas if they contain $\sim N_{\rm sph}$ gas particles, where
$N_{\rm sph}$ is the number of SPH smoothing neighbours.  When the
mass resolution is poor, gas particles can also be heated by two-body
interactions with heavy dark matter particles \citep{Stein97}, which
further reduces the effective cooling rates.

However, inadequate resolution can also act in the opposite sense and
{\em enhance} the cooling rate in halos.  This can arise from,
e.g. `in-shock cooling' \citep{Hutch00}.  In SPH, shock fronts are
broadened and gas passing through them is heated only `slowly' to
post-shock temperatures.  During this transition, gas can radiate
energy, artificially depressing post-shock temperatures and reducing
the cooling time of post-shock gas.  This effect is particularly
severe when the functional dependence of the cooling rate on
temperature exhibits a peak between pre- and post-shock temperatures
\citep{Mar2001}.  The gas can then cool severely in a broadened shock,
in extreme cases converting it incorrectly into an isothermal shock.
Additional problems can develop in density discontinuities between the
hot phase of a halo and the condensed cold gas that has already
cooled.  Owing to limited resolution, SPH estimates of the density for
hot particles passing close to cold particles will be biased high if
they sample the dense cold phase, leading to accelerated cooling
\citep{Pea99,Th98,Cr00}.

Given these complications, and the highly nonlinear nature of
radiative cooling, it is unclear whether or not cooling rates obtained
by existing cosmological SPH codes are reliable, in particular with
respect to the majority of halos which are resolved with relatively
few particles.  However, despite the fundamental importance of this
issue, only limited attempts have been made to examine the systematic
dependence of the results on numerical resolution or technique
\citep[among others][]{Th98}.  \citet{Abadi00} found good agreement
between SPH predictions and a variant of the self-similar cooling wave
solution derived by \citet{Ber89}.  While self-similar tests are
useful, it is difficult to validate codes in typical cosmological
settings from them, because in reality the cooling function is not
self-similar, and the objects forming early on contain few particles.
The latter is a particularly important restriction, because for
sufficiently large particle numbers SPH yields the proper continuum
limit.

In this study, we are especially interested in deviations from the
correct behaviour when the resolution is coarse.  To examine this
issue, we compare entropy formulations of SPH \citep{He93} with
variants of the standard algorithm in which the thermal energy
equation is integrated.  By appealing to a variational principle, we
also derive a novel version of SPH which manifestly conserves both the
energy and the entropy (under appropriate conditions) even when the
smoothing lengths are fully adaptive. In order to investigate the
numerical properties of these formulations of SPH in situations where
cooling is important, we study collapsing and virializing gas spheres
in isolation, and we compute a set of full cosmological
simulations of structure formation.

We also explore the behaviour of SPH when feedback scenarios are
invoked that deposit large quantities of energy into single particles,
as in the method proposed recently by \citet{Spr01hybrid} for
modelling starbursts and associated galactic winds and outflows.
\citet{Benz90} have argued that in explosive outbursts, energy must be
deposited in a smooth manner for SPH to produce reasonable results,
and a similar concern has been raised by Klypin (2001, private
communication).  We show that singular point explosions do indeed
present severe difficulties for some of the standard formulations of
SPH, but also demonstrate that reasonable solutions are obtained when
the entropy equation is integrated, thereby justifying the numerical
validity of feedback schemes that make use of point-like energy
injection.

The outline of our paper is as follows.  In Section 2, we briefly
review several common treatments of SPH. We then derive a new
conservative formulation in terms of the entropy equation in
Section~3.  In Section 4, we start our comparisons by considering
strong explosions in a uniform background.  In Section 5, we analyse
systematic differences between various formulations of SPH in a series
of collapse simulations of gas spheres that undergo radiative cooling.
We then extend this analysis to small cosmological simulations of
structure formation in Section 6.  Finally, we conclude and summarise
in Section 7.

\section{Entropy formulation of SPH}

In the conventional implementation of SPH \citep[e.g.][]{Mo92}, the
density of each particle is computed according
to \be \rho_i = \sum_{j=1}^N
m_j W(|\vec{r}_{ij}|,h_i), \label{eqndens}\ee where
$\vec{r}_{ij}\equiv \vec{r}_i - \vec{r}_j$, and $W(r,h)$ is the SPH
kernel\footnote{We assume that the kernel drops to zero at
$r=h$.}. Commonly, variable smoothing lengths are employed so that the
number of neighbours for each particle with $|\vec{r}_{ij}| \le h_i$
is maintained at a nearly fixed value $N_{\rm sph}$.  This adaptive
kernel estimation implicitly obeys the continuity equation.  The
thermal energy and momentum equations can then be integrated in time
according to \be \frac{\dd u_i}{\dd t} = -\frac{{\cal L}_i}{\rho_i} +
\frac{1}{2}\sum_{j=1}^N m_j \left( \frac{P_i}{\rho_i^2} +
\frac{P_j}{\rho_j^2} +\Pi_{ij}\right) \vec{v}_{ij} \cdot \nabla_i
\overline{W}_{ij}\, , \label{eqnu} \ee and \be \frac{\dd
\vec{v}_i}{\dd t} = -\sum_{j=1}^N m_j \left( \frac{P_i}{\rho_i^2} +
\frac{P_j}{\rho_j^2} +\Pi_{ij}\right) \nabla_i \overline{W}_{ij} \, ,
\label{eqnacc} \ee 
respectively \citep{Mo83,Mo92,He93,SprGadget2000}. Here,
\be \overline{W}_{ij}= \frac{1}{2}\left[ W(|\vec{r}_{ij}|, h_i) +
W(|\vec{r}_{ij}|, h_j)\right] \ee is a symmetrised kernel
\citep{He89}, and $\Pi_{\rm ij}$ denotes the artificial viscosity,
for which we adopt a standard form as in \citet{St96}.  Note that we
have also included an emissivity per unit volume ${\cal L}_i={\cal
L}(\rho_i, u_i)$ to describe external sinks or sources of 
energy due to radiative cooling or heating.

Some authors prefer to employ $\overline{W}_{ij}= W(|\vec{r}_{ij}|,
[h_i+h_j]/2)$ in the above equations, so that smoothing lengths are
symmetrised rather than the kernels.  It is also possible to apply a
similar procedure to the estimation of the local density \citep{He89}.
However, of larger importance are the choices made for the
symmetrisation of the pressure terms in the equations of motion and in
the energy equation.

As a variant of pair-wise symmetrisation with arithmetic means,
\citet{He89} suggested the use of geometric means instead, which
appeared to provide better integration stability for the thermal
energy in their numerical tests.  In this case, the SPH equations read
\be \frac{\dd u_i}{\dd t} = -\frac{{\cal L}_i}{\rho_i}
+\frac{1}{2}\sum_{j=1}^N m_j \left(2 \frac{\sqrt{P_i
P_j}}{\rho_i\rho_j} +\Pi_{ij}\right) \vec{v}_{ij} \cdot \nabla_i
\overline{W}_{ij}\, ,
\label{eqnu-geom}
\ee and \be \frac{\dd \vec{v}_i}{\dd t} = -\sum_{j=1}^N m_j \left( 2
\frac{\sqrt{P_i P_j}}{\rho_i\rho_j} +\Pi_{ij}\right) \nabla_i
\overline{W}_{ij} \, .
\label{eqnacc-geom}
\ee A number of SPH codes have employed this formulation \citep[among
others][]{He89,Ka96,Da97,Ca98,SprGadget2000}.

While both of the above formulations are pleasingly symmetric between
particles $i$ and $j$, there is actually no compelling need to
distribute the pressure work done by a particle pair equally between
them. In fact, if one simply uses the SPH estimate for the local
velocity divergence to derive the energy equation, one obtains \be
\frac{\dd u_i}{\dd t} = -\frac{{\cal L}_i}{\rho_i} + \sum_{j=1}^N m_j
\left( \frac{P_i}{\rho_i^2} + \frac{1}{2}\Pi_{ij}\right) \vec{v}_{ij}
\cdot \nabla_i \overline{W}_{ij}\, .
\label{eqnu-as}
\ee This form of the thermal energy equation has frequently been used
in SPH computations instead of a symmetrised form
\citep{Ev88,Rasio91,Na93,St93,Ne93,Cou95,Hu97,Th98}.  It conserves
energy just as well, but as \citet{Cou95} pointed out, it
produces less scatter in the entropy.  Note that
this formulation cannot give rise to negative temperatures,
as can occur for other formulations of the energy equation
in certain situations.

In principle, there are many other forms that the dynamical equations
can take in SPH.  It is clear, however, that differences between
various implementations will in general be small in the limit of
relatively smooth flows.  However, this need not be true for extreme
cases, such as point-explosions and cooling flows, as we discuss in
what follows.  In our tests, we are especially interested in how well
different versions of SPH perform when physical conditions are unusual
and numerical resolution is only poor or moderate, as is often the
case in cosmological simulations of galaxy formation.

A somewhat more fundamental change in the numerical scheme is to
formulate SPH in terms of dynamical equations for the entropy
\citep{Lu77,Benz87,He93}, rather than the internal energy. We can
characterise the specific entropy $s$ of a fluid element in terms of
an entropic function $A(s)$, defined by \be P= A(s)\rho^\gamma ,  \ee
where $\gamma$ is the adiabatic index.
Rather than following the evolution of the internal energy, we can
integrate \be \frac{\dd A}{\dd t}= -\frac{\gamma-1}{\rho^\gamma} {\cal
L} \label{Aevolv}, \ee for an inviscid fluid.  Note that a convective
time derivative is used here so that $A(s)$ is conserved for each
fluid element in an adiabatic flow (${\cal L} = 0$).  In this
approach, the temperature is inferred from \be u=
\frac{A(s)}{\gamma-1}\rho^{\gamma -1} . \ee If shocks occur, the
entropic function $A(s)$ can vary with time even in the absence of
other sources or sinks of entropy. In order to allow for this
possibility, an artificial viscosity needs to be introduced into SPH.
For example, a suitable SPH discretisation of equation (\ref{Aevolv})
is given by \be \frac{\dd A_i}{\dd t} =
-\frac{\gamma-1}{\rho_i^\gamma} {\cal L}(\rho_i,u_i) \; +\;
\frac{1}{2}\frac{\gamma-1}{\rho_i^{\gamma-1}}\sum_{j=1}^N m_j \Pi_{ij}
\vec{v}_{ij}\cdot\nabla_i \overline{W}_{ij} \,,
\label{eqnentropy}
\ee which shows that entropy is {\em only} generated by the artificial
viscosity in shocks, and by external sources of heat, if they are
present.  In the entropy formulation of SPH, this expression is
integrated in place of equation (\ref{eqnu}). Note that this approach
provides tight control on sources of entropy and, in particular, it is
possible to manifestly guarantee that the specific entropy of a
particle can {\em only grow} in time (assuming for the moment that
external sources of entropy are unimportant).  This property can be
exploited to obtain sharper shock profiles, or to devise shock
detection algorithms.

In the continuum limit, employing equation (\ref{Aevolv}) is
equivalent to solving the gas dynamics by means of the equation for
the internal energy.  However, when the number of neighbours defining
smoothed estimates in SPH is finite, as is {\it always} the case in
practice, the two formulations exhibit differences, with one or the
other yielding a better approximation to the continuum solution.  In
particular, \citet{He93} has shown that while integrating the internal
energy in SPH results in good energy conservation, entropy is not
conserved even for purely adiabatic flows.  On the other hand, if the
entropy is integrated, the total energy is not necessarily
conserved. \citet{He93} also showed that these errors are primarily
caused by the use of variable smoothing and the neglect of relevant
terms in the dynamical equations.  However, even for fixed smoothing
lengths, simultaneous conservation of energy and entropy is not
manifest in the above formulations of SPH, but is only guaranteed in
the continuum limit; i.e.~for a very fine sampling of the fluid.

There is perhaps a tendency to take violations of entropy conservation
less seriously than those of total energy, which may explain why most
SPH implementations have been made by integrating the thermal energy
equation together with the equations of motion, resulting in good
energy conservation.  However, as we demonstrate below, this choice
can lead to significant violations of entropy conservation in certain
situations which are important for cosmological simulations of galaxy
formation.

\section{A fully conservative formulation of SPH}

In principle, it is possible to construct Lagrangian treatments of
hydrodynamics that behave better in their conservation properties, for
example based on spatial Voronoi tesselations \citep{Ser00}.
Alternatively, for SPH, one can explicitly account for terms arising
from the variation of the smoothing lengths; the so-called $\nabla h$
terms.  \citet{Ne93,Ne94} have shown that conservation of energy and
entropy can be improved substantially if such $\nabla h$ terms are
included \citep[see also][]{Serna96}.  However, this approach leads to
somewhat cumbersome forms for the dynamical equations and introduces
noise into smoothed estimates owing to the dependence of the
correction terms on the single most distant neighbour.  Consequently,
this version of SPH has not found widespread usage in astrophysical
applications.

With this in mind, we now derive a new formulation of SPH which
employs variable smoothing lengths and which conserves energy {\em
and} entropy (when appropriate) by construction. To this end, consider
the Lagrangian \be L(\vec{q},\dot\vec{q})= \frac{1}{2}\sum_{i=1}^N m_i
\dot\vec{r}_i^2 - \frac{1}{\gamma -1 }\sum_{i=1}^N m_i A_i
\rho_i^{\gamma -1} \ee in the independent variables
$\vec{q}=(\vec{r}_1,\ldots,\vec{r}_N, h_1,\ldots,h_N)$, where the
thermal energy acts as the potential generating the motion of SPH
particles.  The densities $\rho_i$ are functions of $\vec{q}$, as
defined by equation (\ref{eqndens}).  The quantities $A_i$ are treated
as constants; i.e.~the flow is assumed to be strictly adiabatic for
now.

We select smoothing lengths by requiring that a fixed mass is
contained within a smoothing volume, viz.  $({4\pi}/{3}) h_i^3 \rho_i
= M_{\rm sph}$, where $M_{\rm sph}= \overline{m} N_{\rm sph}$ relates
the mass $M_{\rm sph}$ to the typical number $N_{\rm sph}$ of
smoothing neighbours for an average particle mass $\overline{m}$.
These equations provide $N$ constraints \be \phi_i(\vec{q})\equiv
\frac{4\pi}{3} h_i^3 \rho_i - M_{\rm sph} = 0
\label{eqnconstr}
\ee on the coordinates of the Lagrangian. 

We now obtain the equations of motion from \be \frac{\dd}{\dd
t}\frac{\partial L}{\partial \dot q_i} -\frac{\partial L}{\partial
q_i} = \sum_{j=1}^N \lambda_j \frac{\partial \phi_j}{\partial q_i}\, ,
\ee where $N$ Lagrange multipliers $\lambda_i$ have been introduced.
The second half of these $2N$ equations gives the Lagrange multipliers
as \be \lambda_i= \frac{3}{4\pi}
\frac{m_i}{h_i^3}\frac{P_i}{\rho_i^2}\,
\left[{1+\frac{3 \rho_i}{h_i}\left(\frac{\partial \rho_i}{\partial
h_i} \right)^{-1}}\right]^{-1} .  \ee With this result, the first half
of the equations then yields \be m_i\frac{\dd \vec{v}_i}{\dd t} = -
\sum_{j=1}^N m_j \frac{P_j}{\rho_j^2} \left[ 1+ \frac{h_j}{3\rho_j}
\frac{\partial \rho_j}{\partial h_j} \right]^{-1} \nabla_i{ \rho_j} \, .
\ee Using \be \nabla_i{ \rho_j} = m_i \nabla_i W_{ij}(h_j)
+\delta_{ij}\sum_{k=1}^N m_k \nabla_i W_{ki}(h_i) \, , \ee we finally
obtain the equations of motion \be \frac{\dd \vec{v}_i}{\dd t} = -
\sum_{j=1}^N m_j \left[ f_i \frac{P_i}{\rho_i^2} \nabla_i W_{ij}(h_i)
+ f_j \frac{P_j}{\rho_j^2} \nabla_i W_{ij}(h_j) \right],
\label{eqnmot} 
\ee where the $f_i$ are defined by \be f_i = \left[ 1 +
\frac{h_i}{3\rho_i}\frac{\partial \rho_i}{\partial h_i} \right]^{-1}
\, , \ee and the abbreviation $W_{ij}(h)=
W(|\vec{r}_{i}-\vec{r}_{j}|, h)$ has been used.

Note that because the potential (thermal) energy of the Lagrangian
depends only on coordinate differences, the pairwise force in equation
(\ref{eqnmot}) is automatically anti-symmetric.  Total energy,
entropy, momentum, and angular momentum are therefore all manifestly
conserved, provided that the smoothing lengths are adjusted locally to
ensure constant mass resolution as defined by equation
(\ref{eqnconstr}).  Although not explicitly present, the $\nabla h$
terms are thus consistently included to all orders.

Moreover, the equations of motion (\ref{eqnmot}) are remarkably
similar to some prior implementations of SPH.  In fact, our equations
of motion reduce to those employed by \citet{Th92} if one sets
$f_i=f_j=1$.  Hence, the modifications needed to realize the benefits
noted above in existing SPH codes are minor, necessitating only a few
additional computations.  In particular, the quantities
$\frac{\partial\rho_i}{\partial h_i}$ can be easily computed along
with the densities themselves.  A slightly more involved alteration is
required for the algorithm which updates smoothing lengths, since it
is now necessary to ensure that a `constant mass' resides within a
smoothing volume rather than a constant number of neighbours.  The
requisite changes are minimal, however, if an iterative bisection
algorithm is used to determine the $h_i$ and $\rho_i$, as for example
in the parallel version of {\small GADGET} \citep{SprGadget2000}.

In order to complete our derivation, we now incorporate an artificial
viscosity to allow for the handling of shocks.  We invoke a viscous
force \be \left. \frac{\dd \vec{v}_i}{\dd t}\right|_{\rm visc.}  =
-\sum_{j=1}^N m_j \Pi_{ij} \nabla_i\overline{W}_{ij} \, ,
\label{eqnvisc}
\ee just as in the standard formulation of SPH, and add it to the
acceleration given by equation (\ref{eqnmot}).  The resulting
dissipation of kinetic energy is exactly balanced by a corresponding
increase in thermal energy if the entropy is evolved according to
equation (\ref{eqnentropy}).

In the following sections, we describe detailed tests of various
implementations of SPH.  In this context, we distinguish between
`energy' formulations of SPH, where the thermal energy equation is
integrated, and `entropy' formulations, where the entropic function is
integrated. We will consider three variants of the energy approach: a
`standard' one described by equations (\ref{eqnu}) and (\ref{eqnacc});
one where the symmetrisation is done using a geometric mean; and one
where the standard form for the equations of motion is combined with
the asymmetric form of the energy equation.  We contrast these schemes
with a `standard' entropy formulation, where equation (\ref{eqnu}) is
replaced by the entropy equation (\ref{eqnentropy}), and we finally
investigate the performance of our new SPH formulation given by
equations (\ref{eqnentropy}), and (\ref{eqnmot}) together with
(\ref{eqnvisc}), which we refer to as the `conservative entropy
approach'. The different combinations of SPH equations used are
summarised in Table~\ref{tab1}.

\begin{table}
\begin{center}
\begin{tabular}{lc}
\hline
Description of SPH scheme & Equations used \\
\hline 
energy, `standard'  &  (\ref{eqnu}), (\ref{eqnacc}) \\ 
energy, geometric-mean &  (\ref{eqnu-geom}), (\ref{eqnacc-geom}) \\ 
energy, asymmetric &  (\ref{eqnu-as}), (\ref{eqnacc}) \\
entropy, `standard'  &  (\ref{eqnentropy}), (\ref{eqnacc}) \\ 
entropy, `conservative'  &  (\ref{eqnentropy}), 
(\ref{eqnmot})+(\ref{eqnvisc}) \\  
\hline \\
\end{tabular}
\end{center}
\begin{caption}
{\label{tab1} The different formulations of SPH considered in this paper.}
\end{caption}
\end{table}

\section{Point-like energy injection}

In the feedback model of \citet{Spr01hybrid}, starbursts are
accompanied by the release of substantial amounts of thermal energy,
injected into individual SPH particles which represent the hot ISM
left behind by starbursts.  These particles can be sufficiently
energetic to entirely escape from galactic halos.

As we now discuss, the standard formulation of SPH is not well-suited
for dealing with `delta-function' energy distributions such as this.
For definiteness, we consider the propagation of spherical
Taylor-Sedov blast waves, which describe the gas dynamics resulting
from the point-like injection of energy into a homogeneous medium of
negligible pressure.  This problem thus serves as a test case for the
numerical behaviour of a code when starburst-`explosions' are
produced.  The analytic similarity solution for a strong explosion is
well known \citep[e.g.][]{Lan66}.  After a time $t$, the blast wave
propagates a distance \be R(t)= \beta \left(\frac{E
t^2}{\rho}\right)^{1/5}\, , \ee where the constant $\beta$ depends on
the adiabatic index $\gamma$ ($\beta=1.15$ for $\gamma=5/3$), $E$ is
the explosion energy, and $\rho$ describes the initial density of the
ambient gas.  Directly at the spherical shock front, the gas density
jumps to a maximum compression of $\rho'/ \rho =
(\gamma+1)/(\gamma-1)$, with most of the mass inside the sphere being
swept up into a thin radial shell.  Behind the shock, the density
rapidly declines and ultimately vanishes towards the explosion centre.
Modelling this problem accurately in 3D is a challenge for any
hydrodynamical code.

\begin{figure}
\bc
\resizebox{8cm}{!}{\includegraphics{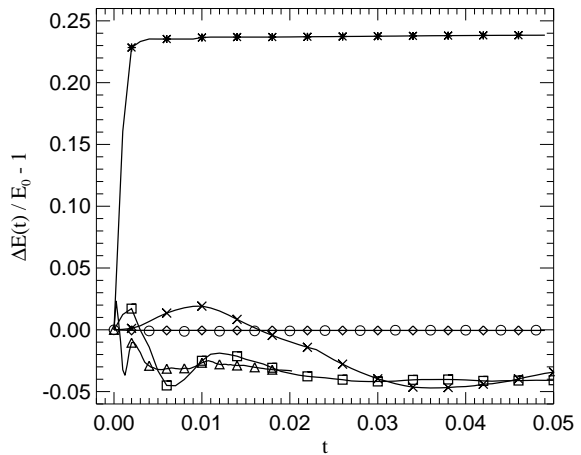}}%
\caption{Deviation of the total energy from the initial explosion
energy as a function of time for a number of different simulations.
The large positive deviation that reaches a maximum error of $\sim
24\%$ is for a $32^3$ run where the initial energy is added to a
single particle and the thermal energy equation is integrated in the
standard form.  In this case, energy conservation is violated because
the code prevents the occurrence of unphysical negative temperatures
in the early phase of the evolution. When the initial energy is
deposited smoothly instead, this is prevented, and energy is well
conserved (diamonds).  Crosses, boxes, and triangles indicate results
for $16^3$, $32^3$ and $64^3$ single point explosions where the code
instead integrates the entropy equation and the equations of motion in
a standard form.  Initially, a fluctuation with a characteristic
pattern is observed. The maximum error is about $\sim 4\%$, but at
later times, energy conservation is reasonable. However, when our new
conservative entropy formulation is employed, energy is again well
conserved (circles).
\label{figExplEnergy}}
\ec
\end{figure}

\begin{figure*}
\bc
\resizebox{5.2cm}{!}{\includegraphics{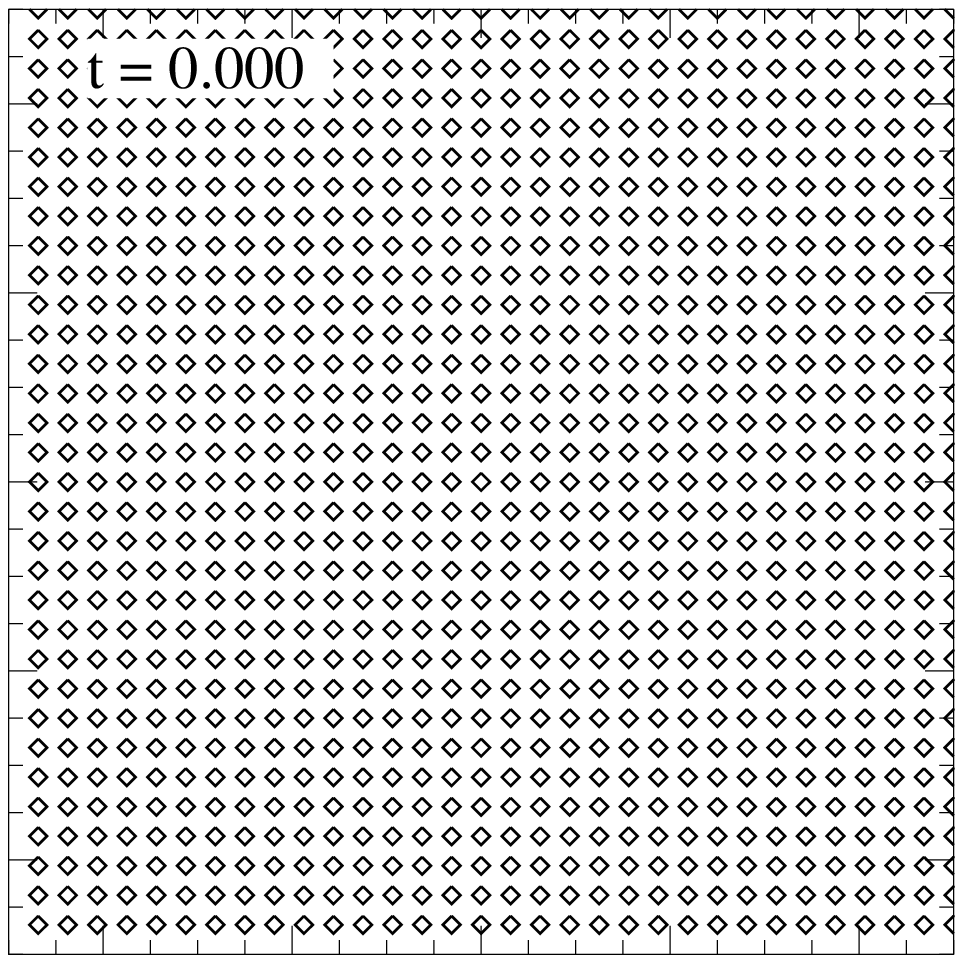}}%
\resizebox{5.2cm}{!}{\includegraphics{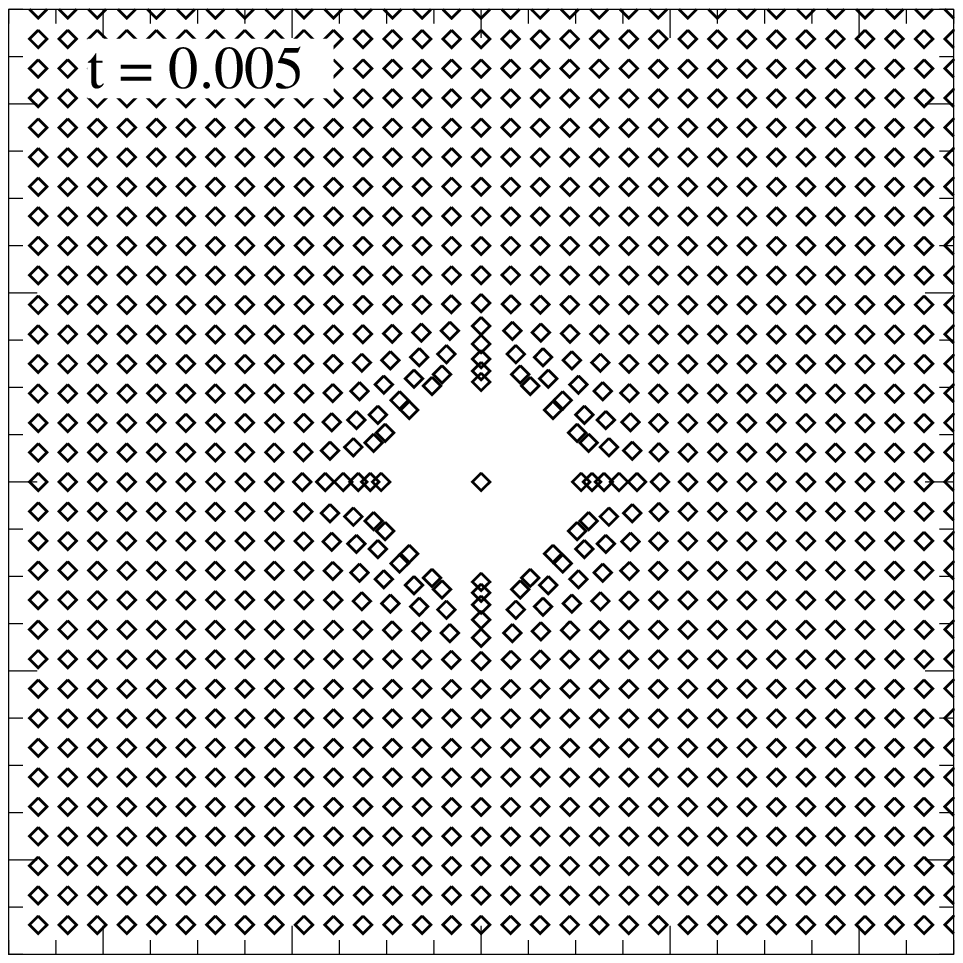}}%
\resizebox{5.2cm}{!}{\includegraphics{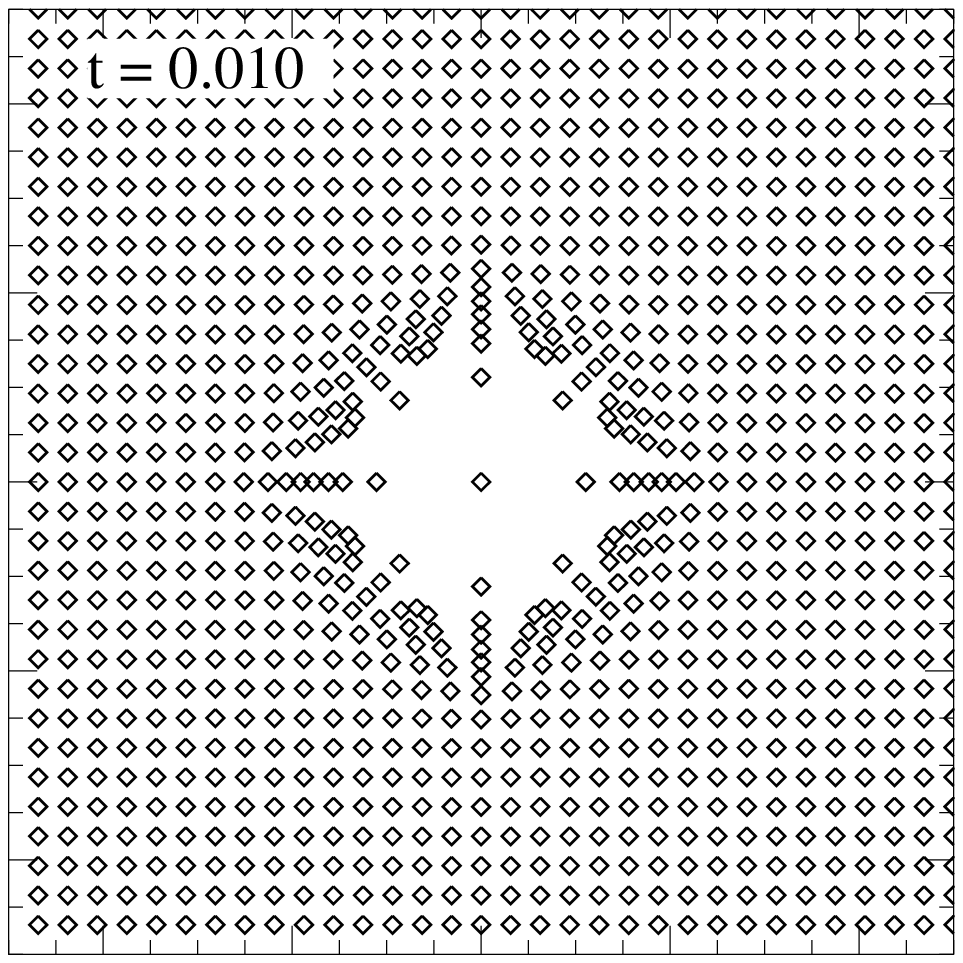}}\\
\resizebox{5.2cm}{!}{\includegraphics{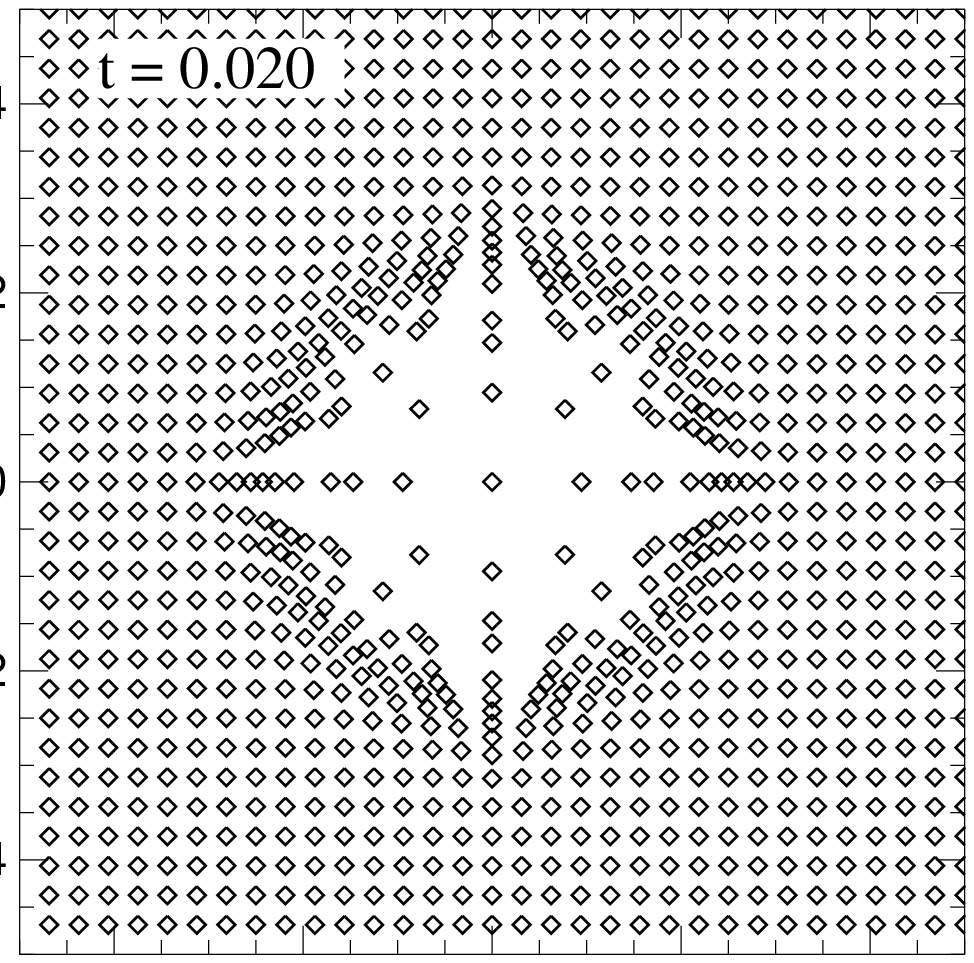}}%
\resizebox{5.2cm}{!}{\includegraphics{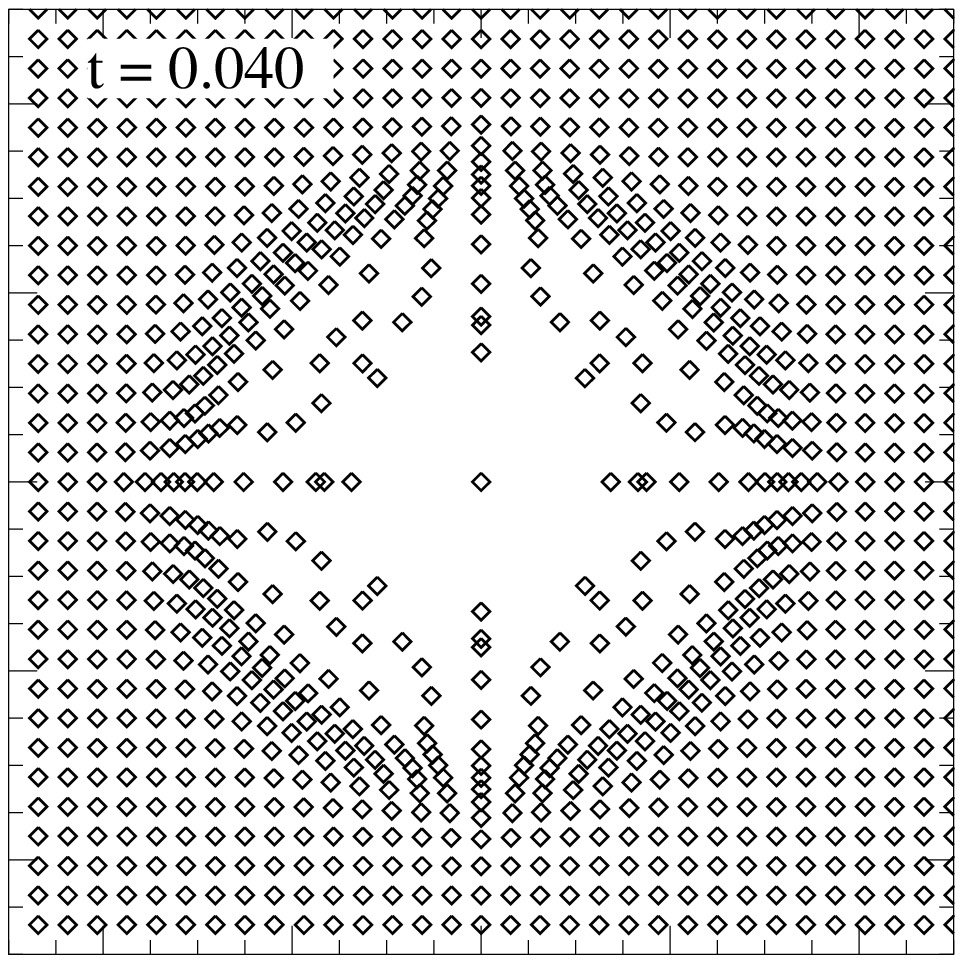}}%
\resizebox{5.2cm}{!}{\includegraphics{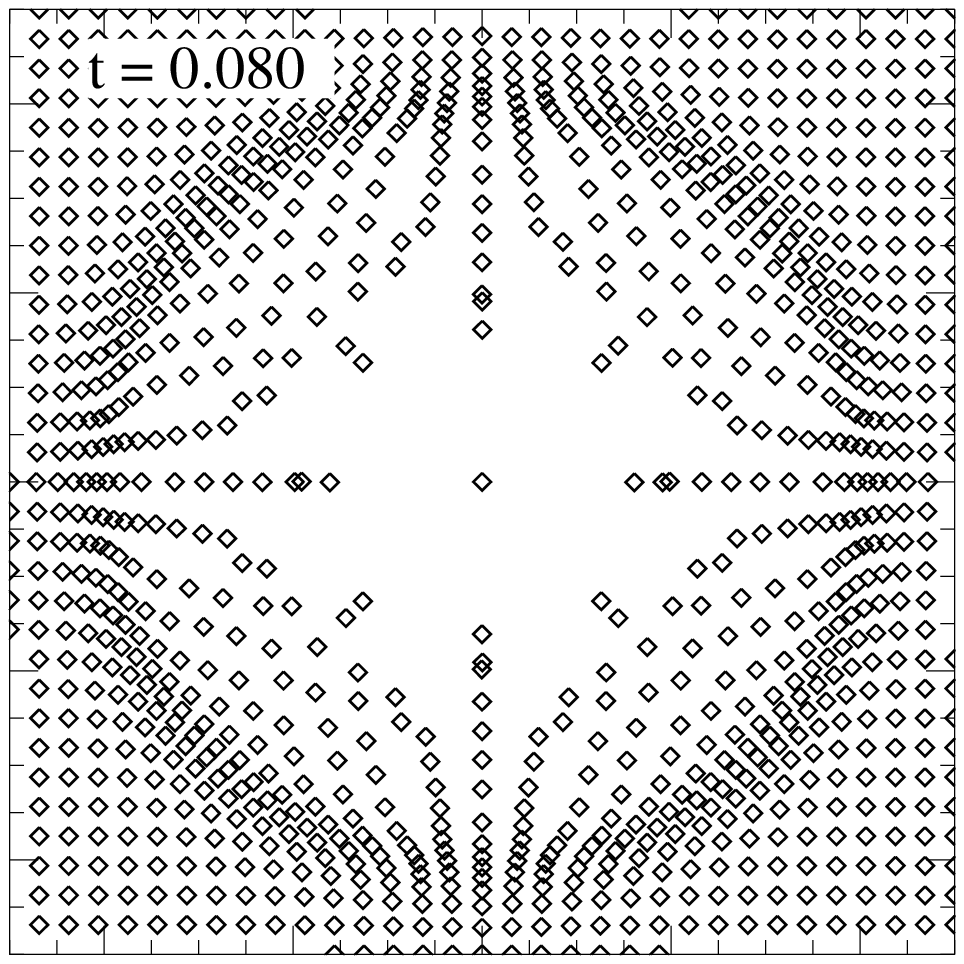}}\\
\ \\
\caption{Point-explosion triggered by the injection of an energy $E=1$
into a single particle in a pressure-less medium of unit density.
Initially, the particles were distributed on a $32^3$ Cartesian grid;
the midplane is shown in all panels. The evolution was computed in
terms of the entropy equation, using our new conservative formulation.
\label{figExpl1}}
\ec
\end{figure*}

In the following, we describe explosions set up in periodic boxes of
unit length per side and unit density, using a Cartesian grid to
initialise particle distributions. At time $t=0$, we inject an
explosion energy $E=1$ into a single particle, and compute the
subsequent 3D-evolution using the parallel SPH-code {\small GADGET}
\citep{SprGadget2000}.

\begin{figure*}
\bc
\resizebox{5.33cm}{!}{\includegraphics{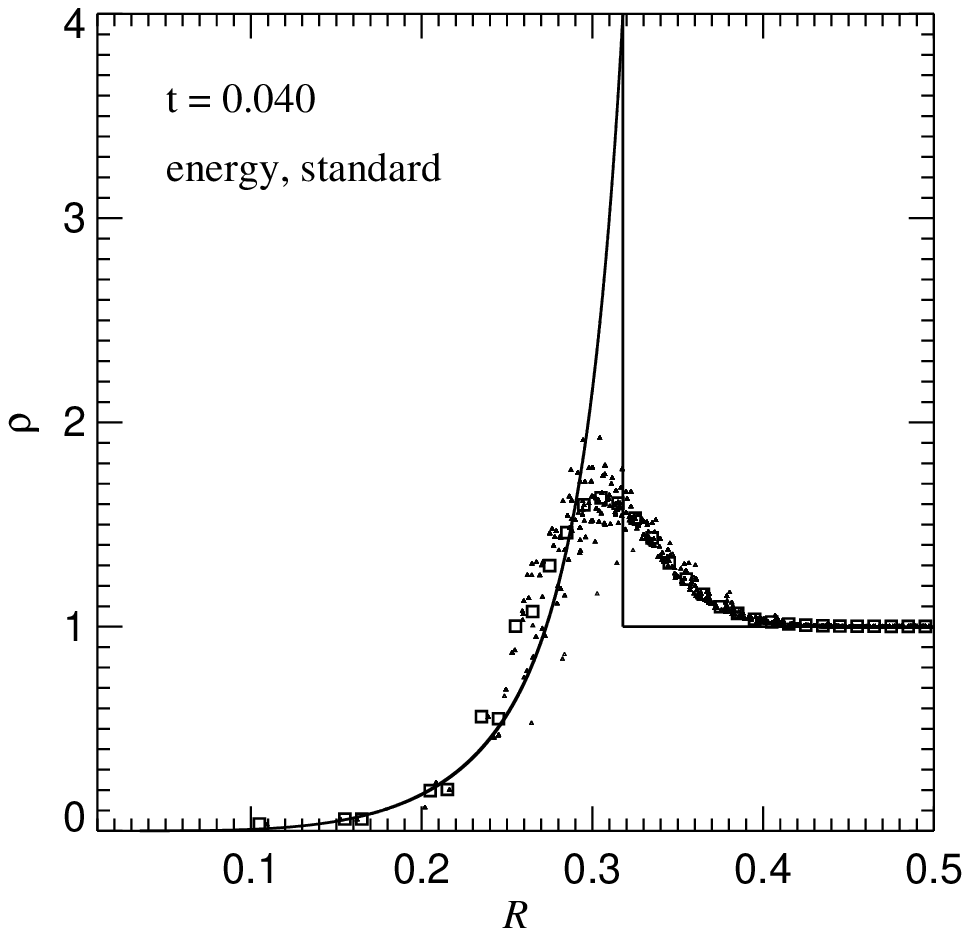}}%
\resizebox{5.33cm}{!}{\includegraphics{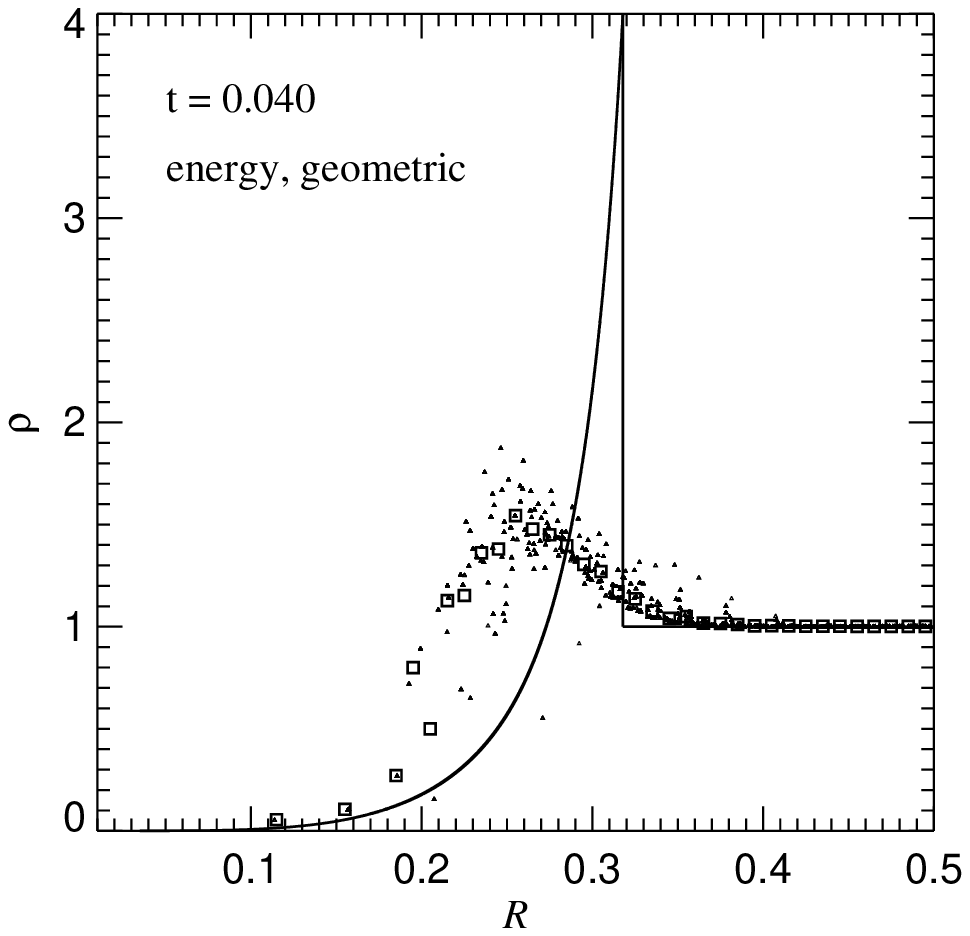}}%
\resizebox{5.33cm}{!}{\includegraphics{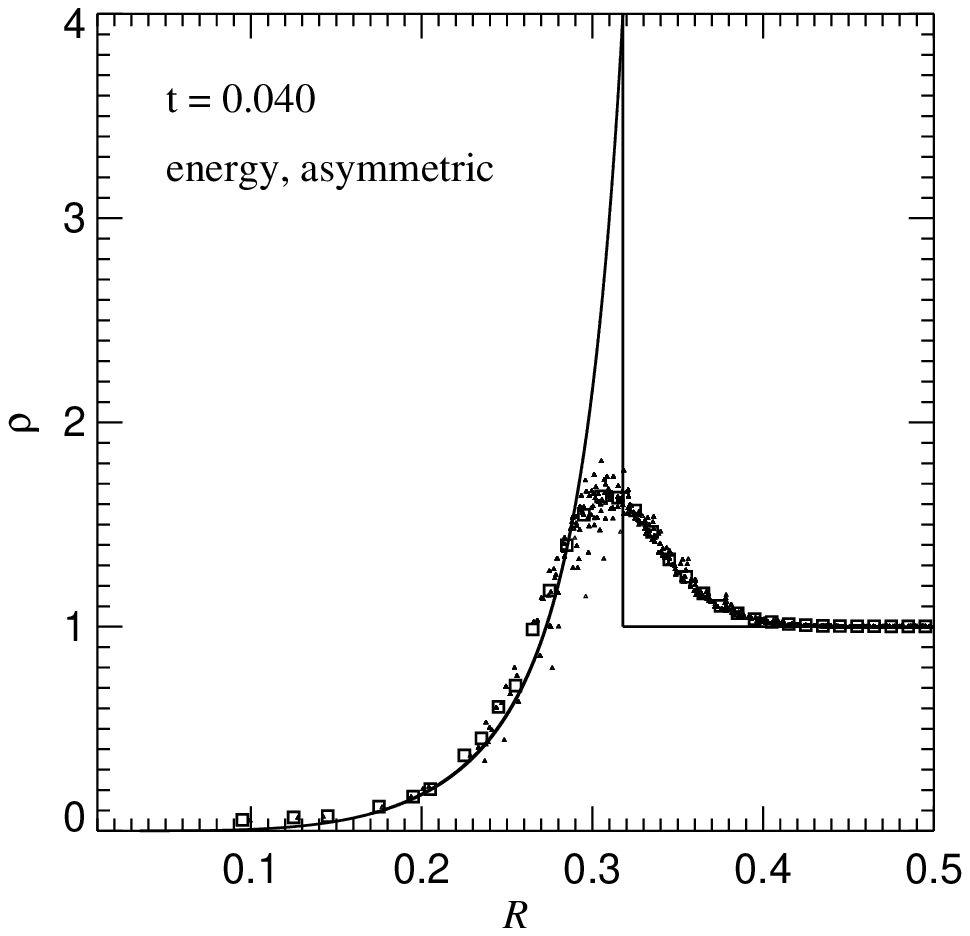}}\\%
\resizebox{5.33cm}{!}{\includegraphics{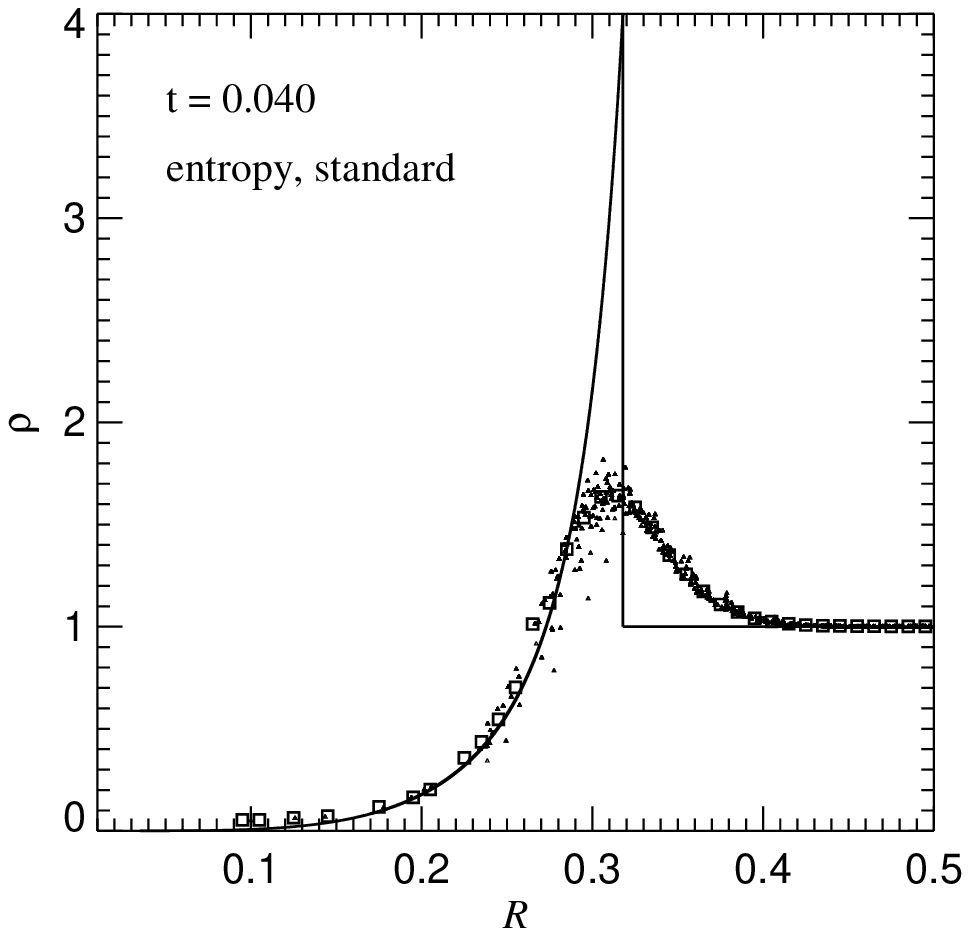}}%
\resizebox{5.33cm}{!}{\includegraphics{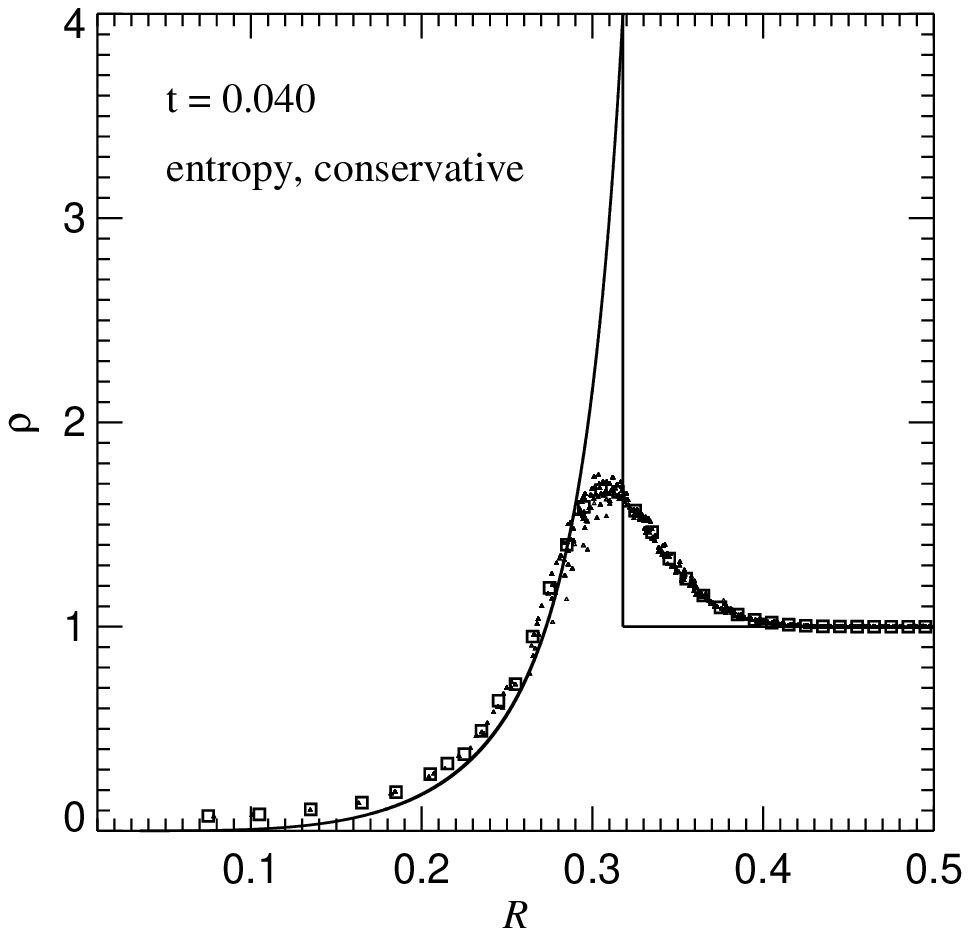}}\\%
\caption{Radial density distribution at a time $t=0.04$ after the
triggering of an explosion in a $32^3$ particle distribution, with the
initial explosion energy smoothed by the SPH kernel.  Results for
different formulations of SPH are shown.  Top: Integration of the
thermal energy, from left to right: in its standard form, with
geometric mean symmetrisation, and with the asymmetric form of the
energy equation.  Bottom: Integration of the entropy equation in the
standard form (left) and with the new conservative formulation
(right).  Small points indicate distances and densities measured from
individual particles, while boxes denote spherically averaged values.
Solid lines show the analytical Sedov solution (adiabatic index
$\gamma= 5/3$).
\label{figExpl2}}
\ec
\end{figure*}

In the `standard' formulation of SPH (integration of the thermal
energy equation and arithmetic mean for the pressure terms), a perhaps
surprising phenomenon is observed: The direct neighbours of the
explosion seed are quickly driven to unphysical negative temperatures.
This is a result of the pair-wise symmetrisation of the pressure terms
in the energy equation.  First, the neighbouring particles of the
explosion seed are pushed away.  Once they gain velocity, the pressure
work done by the particle pairs needs to be supplied by a
corresponding decline of thermal energy.  Due to the symmetrisation,
the (large) rate of decline of thermal energy is {\em equally}
distributed between the particles of each pair.  For a strong
explosion, this quickly consumes any thermal energy the neighbouring
particles of the explosion seed might have had previously, and drives
them towards negative temperatures.  Formally, total energy would still
be conserved in this case, but our code suppresses unphysical negative
temperatures and enforces a minimum value of zero for the thermal
energy, thereby leading to an unavoidable, strong violation of energy
conservation in this case. Note that this behaviour is not a
consequence of time integration errors, but is a result of the initial
conditions and the particular set of equations used to evolve them.
In our tests, we observe energy violations of up to $\sim 25\%$,
generated entirely in the initial phase.  After a short time, however,
the energy distribution of the central particles becomes sufficiently
`smooth' that a successful explosion still develops.  Of course, the
initial `cooling' of the neighbouring particles and the error in the
explosion energy are substantial artifacts of the solution.  These
problems can be avoided completely if the asymmetric form of the
thermal energy equation is used, in which case energy is conserved
accurately for these point explosions.

If a geometric mean is instead used to symmetrise the pressure terms,
it is clear that the maximum rate of decline of internal energy in
particle pairs that involve the explosion seed becomes smaller.
However, for the problem of a single point explosion, this damping
leads to a completely unphysical solution: No explosion takes place.
This is because the geometric mean (nearly) vanishes if the background
is (nearly) pressureless.

However, integrating the entropy equation leads to a solution that is
as physically well behaved as the one for the asymmetric form of the
thermal energy equation.  As before, the explosion seed pushes
particles away, but because they are compressed, their temperatures
{\em increase}, as they should, because their entropy is required to
increase monotonically.  Unphysical negative temperatures cannot
result from the dynamics, regardless of how the energy is initially
distributed.  However, the total energy is not conserved as well with
this approach.  We observe a characteristic pattern of fluctuations in
the total energy, as shown in Figure~\ref{figExplEnergy}. There is a
maximum deviation of $\sim 4\%$ occurring in the early phase of the
expansion of the blast wave, but there is no long-term secular trend,
indicating that the entropy-method conserves energy quite well unless
the conditions are extreme.  The initial fluctuation is related to the
strong variations of the smoothing lengths near the beginning of the
simulation.  However, if our new conservative formulation of the
entropy approach is used, the variations of the smoothing lengths are
properly accounted for in the equations of motion and, so, energy is
conserved at all times.

In Figure~\ref{figExpl1}, we show an example of the time evolution of
an explosion set up in a $32^3$ grid by injecting the energy into a
single particle, and by using the conservative entropy-method for SPH.
The explosion is not exactly spherical, but is slightly modulated by
effects due to the geometric pattern of the Cartesian grid seen around
the exploding particle.  Note that the resolution available for
following the self-similar evolution of the blast wave effectively
grows with time. At a given time, the resolution can be assessed by
the ratio of the radius that the blast wave has reached to the initial
mean interparticle separation.  For example, at time $t=0.04$, the
effective resolution is 10.17 for the $32^3$ grid, with about 4405
particles contained inside the spherical shock front.

\begin{figure*}
\bc
\resizebox{5.33cm}{!}{\includegraphics{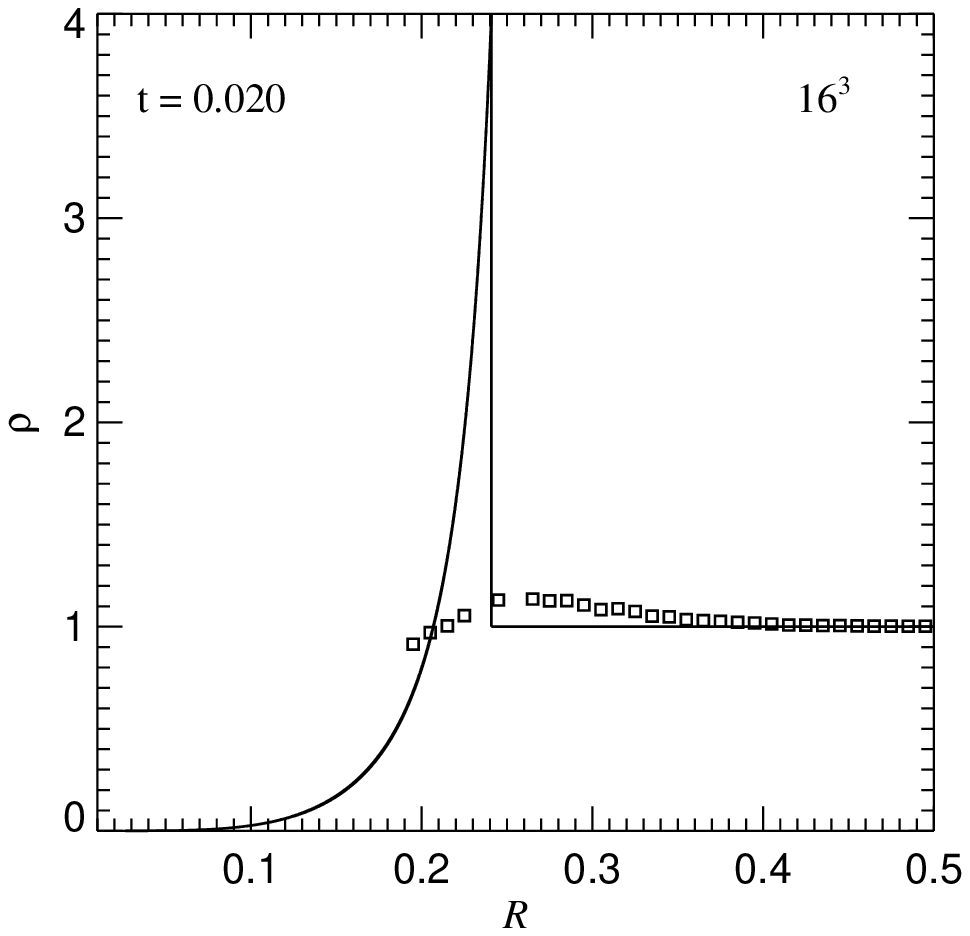}}%
\resizebox{5.33cm}{!}{\includegraphics{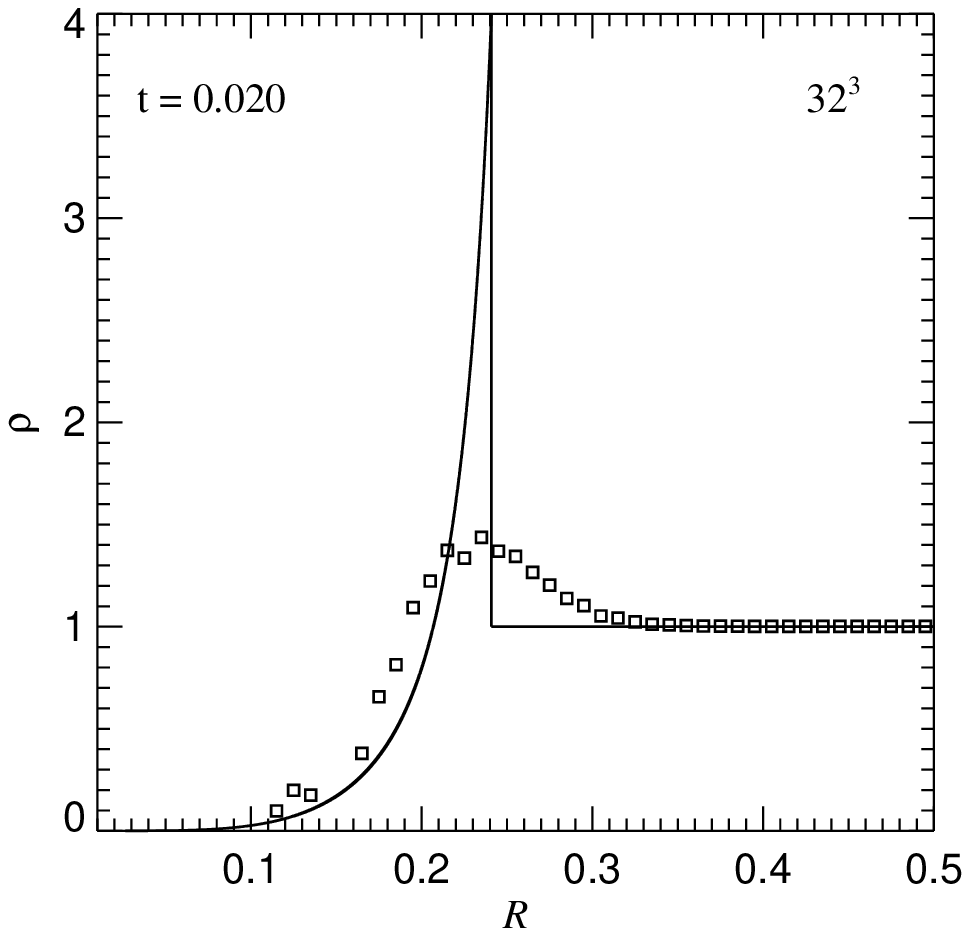}}%
\resizebox{5.33cm}{!}{\includegraphics{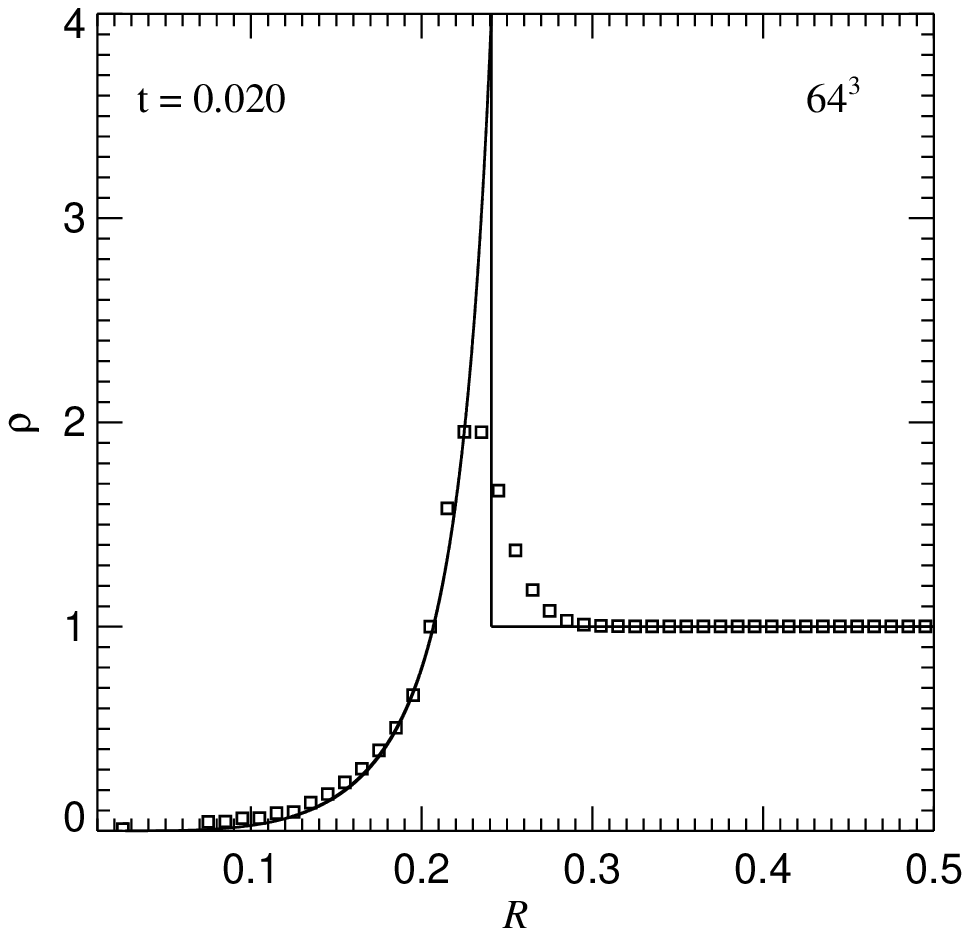}}\\
\caption{Radial density distribution at a time $t=0.02$ after the
triggering of an explosion in particle distributions of varying mean
interparticle separation. From left to right, $16^3$, $32^3$, and
$64^3$ particles are used in a box of unit side length. The effective
resolutions at time $t$ (taken to be the mean interparticle separation
in units of the radius of the blast wave) are 3.86, 7.71, and 15.42,
respectively, corresponding to roughly 240, 1921, and 15370 particles
inside the spherical shock wave.  Boxes mark spherically averaged
density values, and solid lines indicate the analytical Sedov
solution.
\label{figExpl3}}
\ec
\end{figure*}

In Figure~\ref{figExpl2}, we show radial density profiles for
different $32^3$ simulations at a time $t=0.04$ after the explosion,
computed with all five variants of SPH discussed above, but with the
initial explosion energy smoothed according to the SPH kernel.  Due to
the more regular initial conditions, the spherical symmetry of the
explosions is now nearly perfect.  The results for integrating the
entropy, or the thermal energy in its standard or asymmetric form, are
similar, but notice that our new conservative entropy-method produces
the smallest amount of scatter in the density estimates for particles
in the blast wave.  However, it is obvious that the variant of SPH
where pressure terms are symmetrised with a geometric mean gives a
considerably worse match to the analytic Sedov solution than the other
methods.  The shock wave appears to be delayed in time with this
technique.

\begin{figure*}
\bc
\resizebox{7.7cm}{!}{\includegraphics{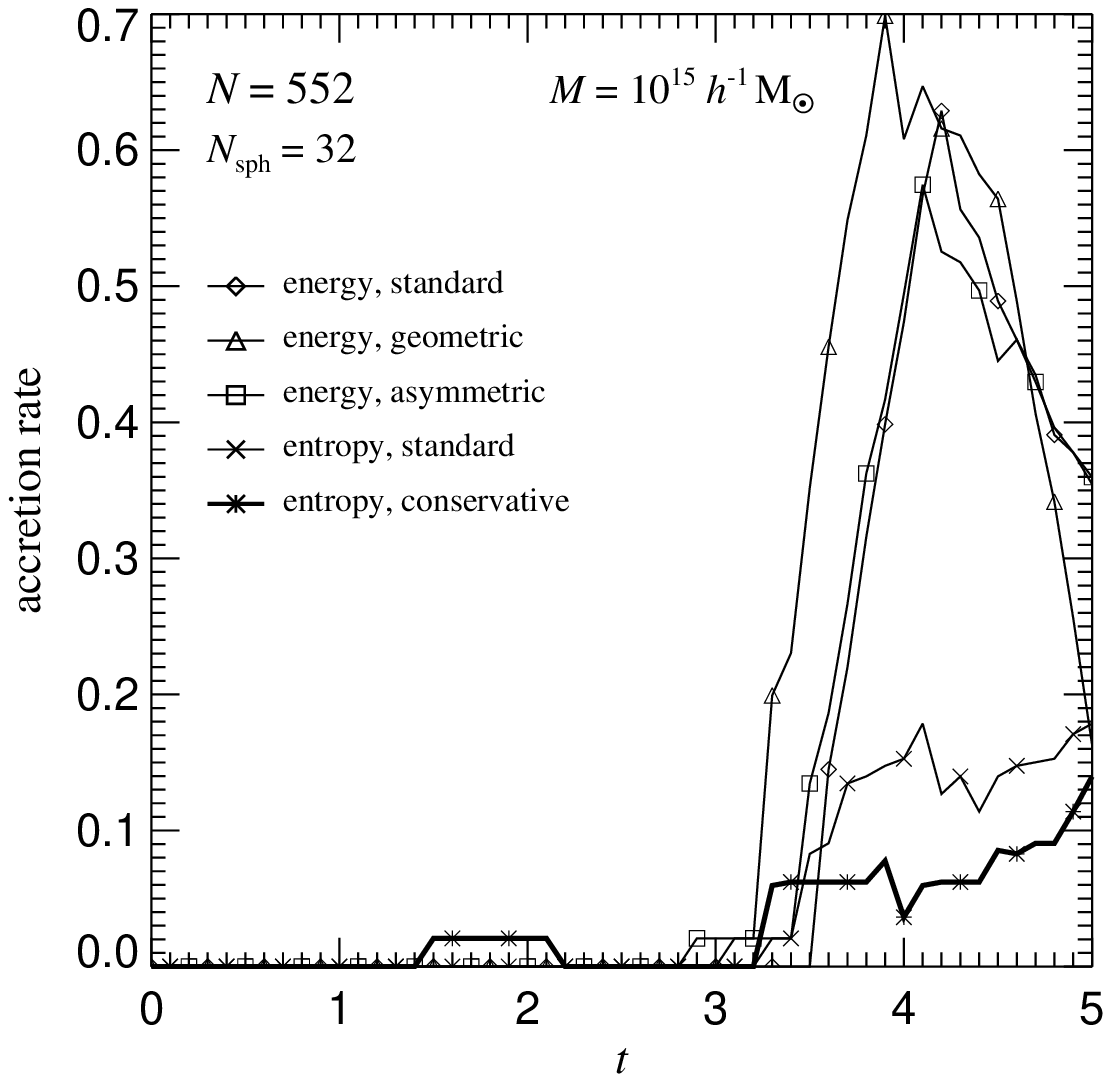}}%
\resizebox{7.7cm}{!}{\includegraphics{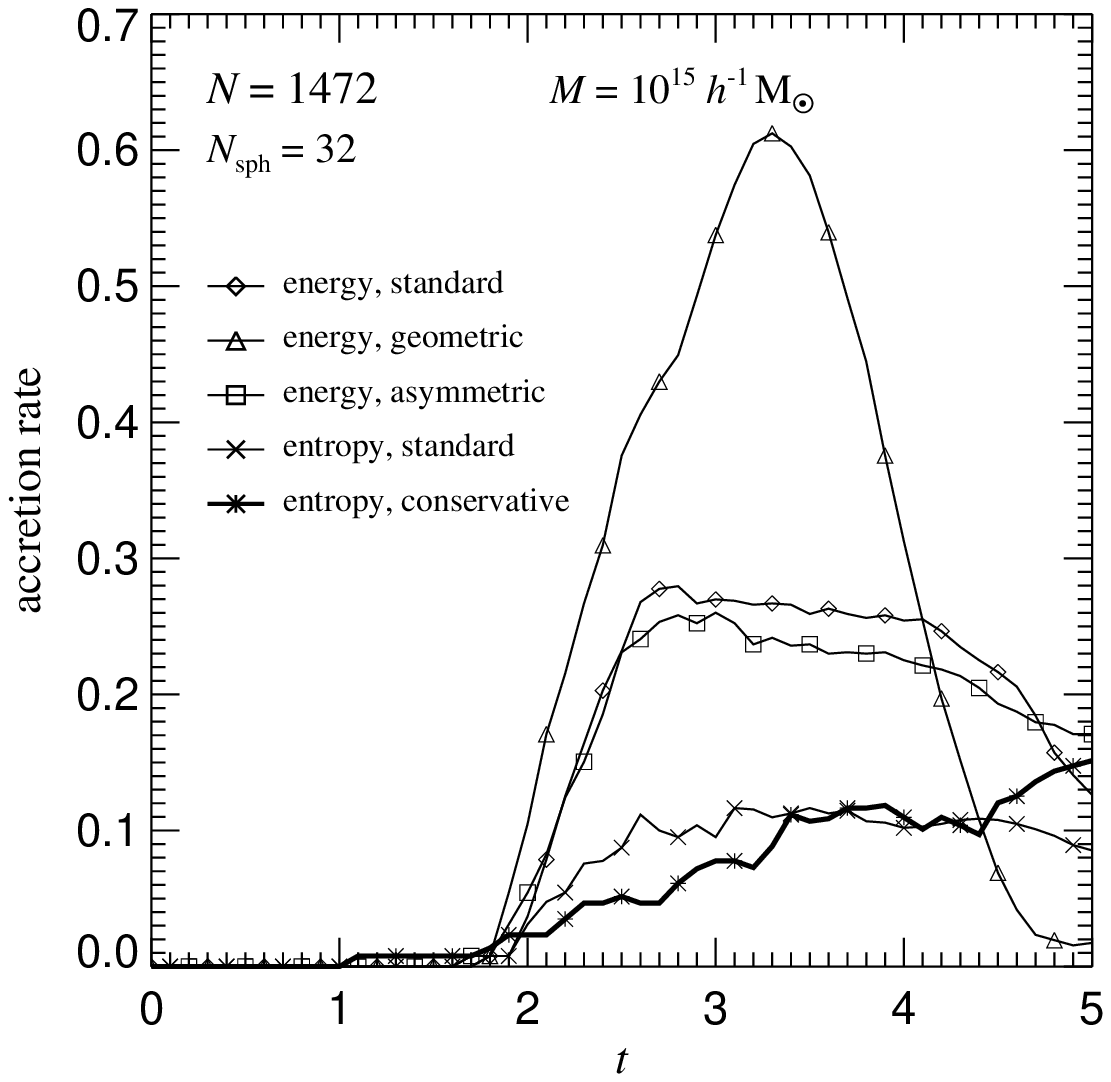}}\vspace{-0.4cm}\\
\resizebox{7.7cm}{!}{\includegraphics{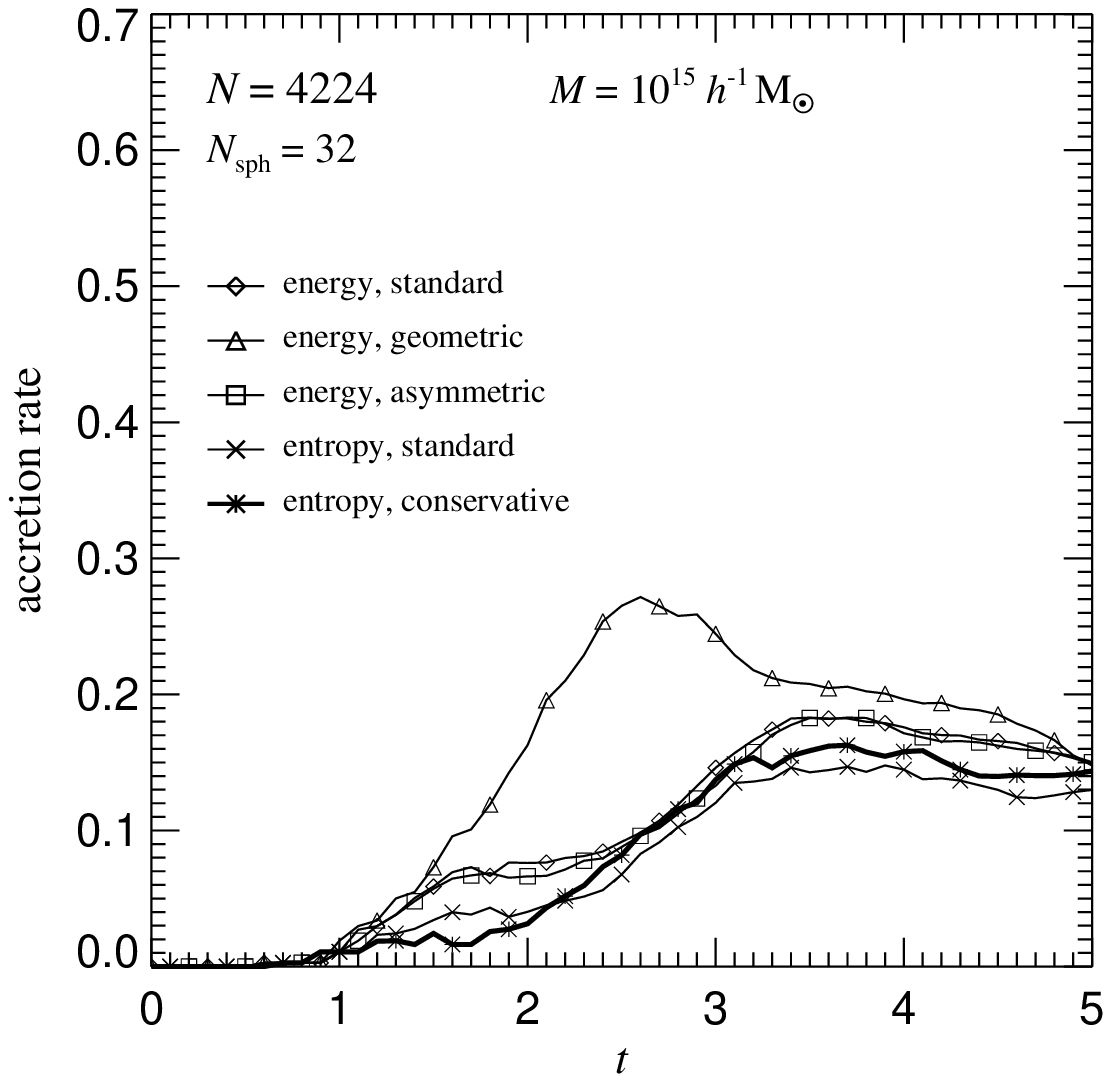}}%
\resizebox{7.7cm}{!}{\includegraphics{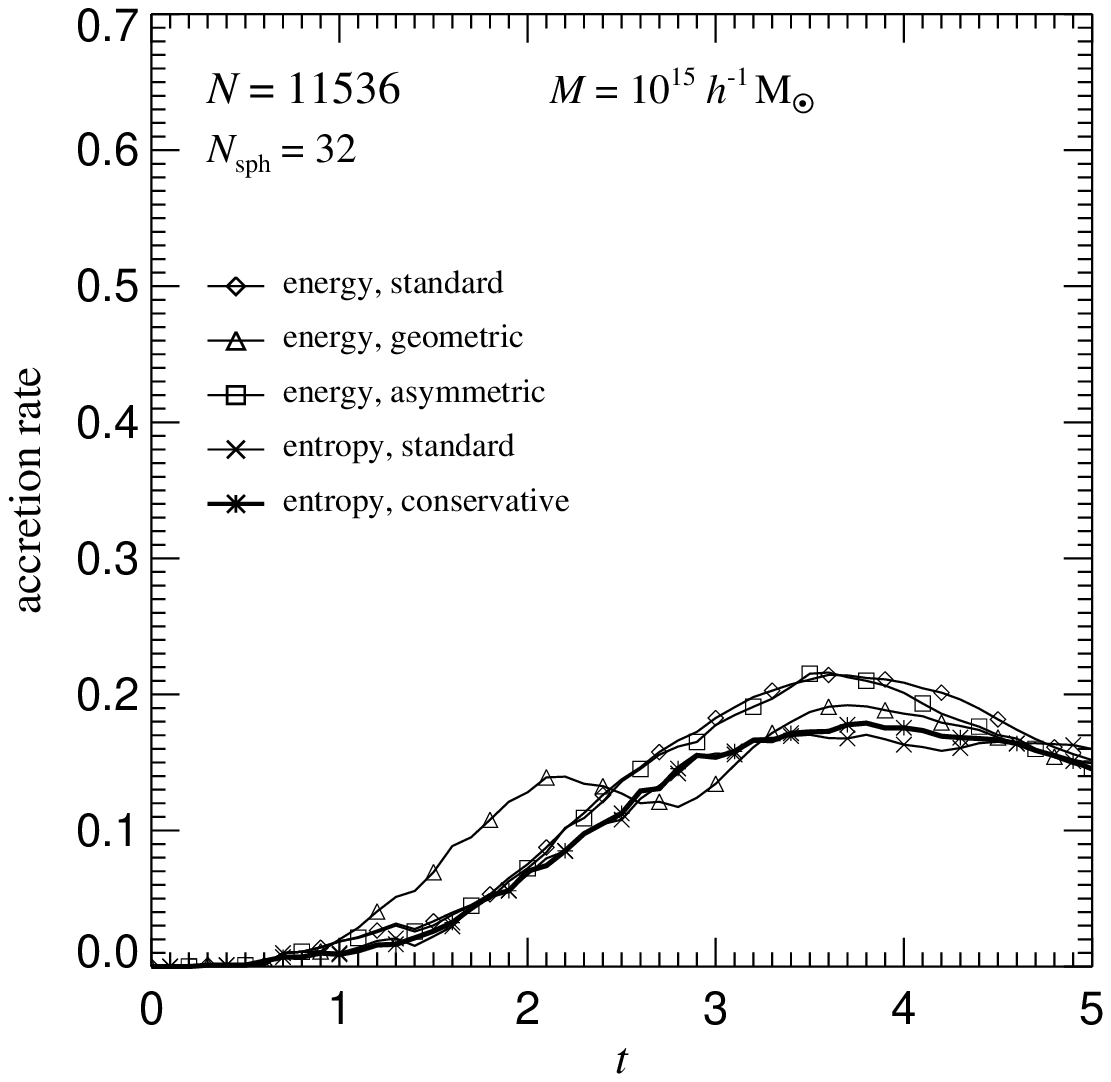}}\vspace{-0.4cm}\\
\resizebox{7.7cm}{!}{\includegraphics{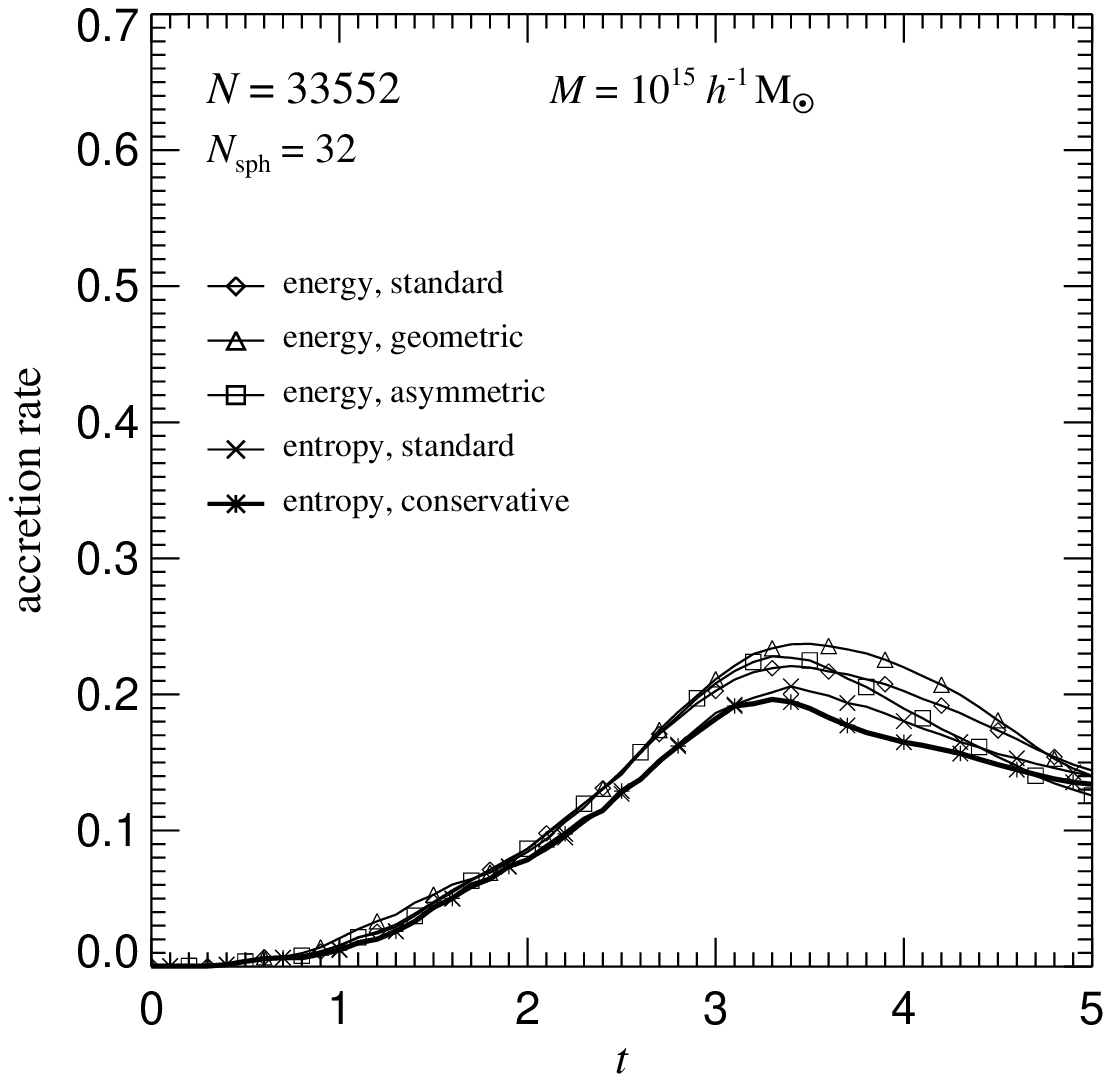}}%
\resizebox{7.7cm}{!}{\includegraphics{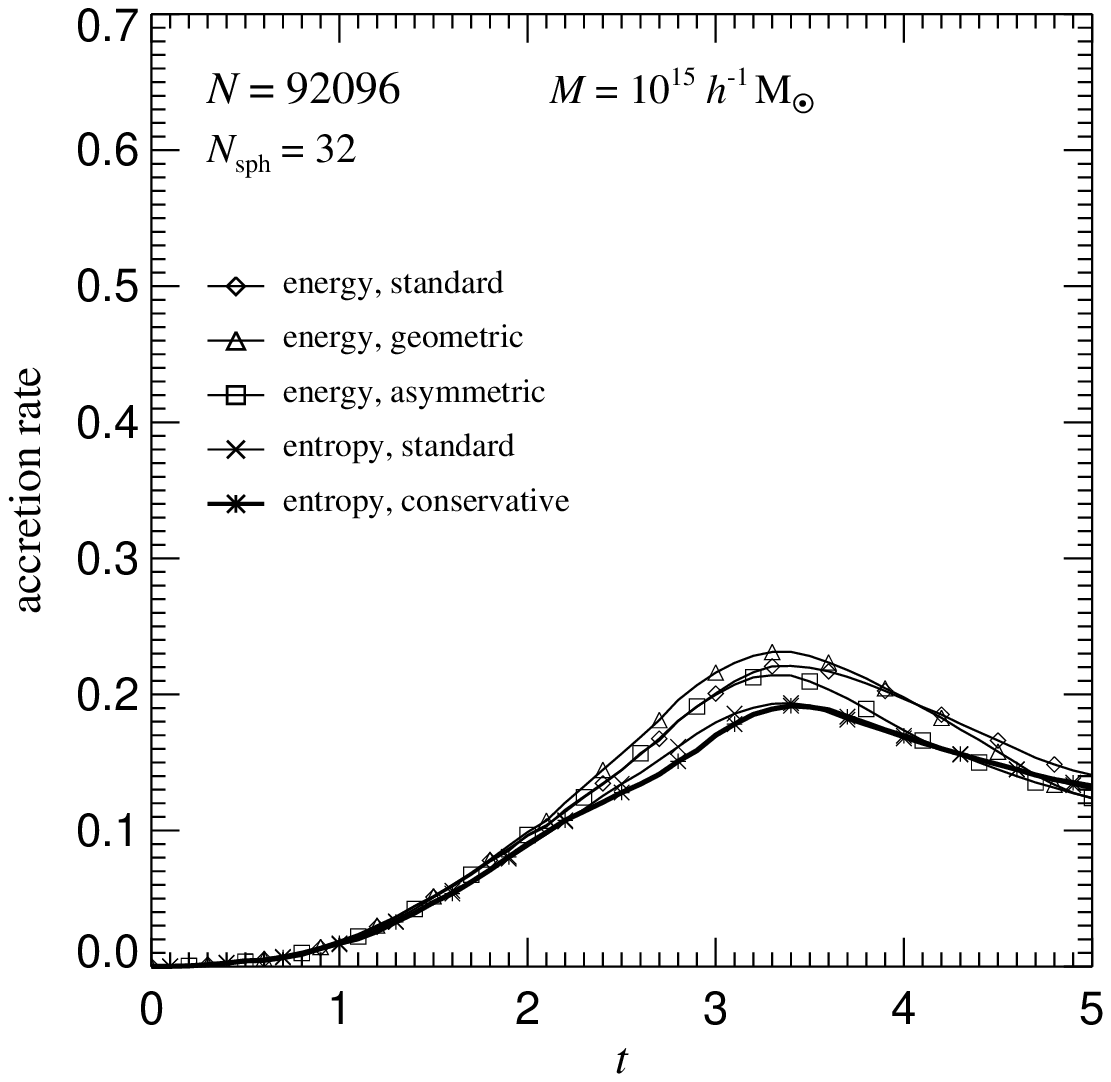}}\vspace{-0.4cm}\\
\caption{Rate at which cold gas accumulates in the centres of
collapsing and virializing gas spheres of mass
$M=10^{15}\,h^{-1}M_\odot$.  Each panel compares results obtained for
a given particle number with the five formulations of SPH listed in
Table~\ref{tab1}.  Symbols are used to mark the lines drawn for each
of the SPH variants, and the result for our new version of SPH is
highlighted by a bold line. The accretion rates are given in units of
the total mass per unit time, where unit time is $1\, h^{-1}{\rm
Gyr}$.
\label{figAccRatem15}}
\ec
\end{figure*}

Finally, in Figure~\ref{figExpl3} we show radial density profiles for
$16^3$, $32^3$, and $64^3$ simulations at a time $t=0.02$ after the
explosion. In these tests, the explosion energy was deposited into a
single particle, and the entropy equation was integrated.  This
resolution test makes it clear that it is difficult to resolve the
internal structure of a thin shock wave in 3D accurately, and that it
is challenging to obtain the expected maximum value of the density
compression, albeit increasing the number of particles drastically
improves the solution.  It is also possible to improve resolution by
modifying the smoothing algorithm of SPH so that it locally adjusts to
properties of the flow \citep{Shap96,Owen98}.  Note, however, that the
propagation of the shock front in time is described quite accurately
in all cases, as well as the density structure behind the blast wave.
This is true even for the very low resolution of $16^3$ and
illustrates the remarkable robustness of SPH in three dimensions.  The
$32^3$-result shown also compares very well to the one in
Figure~\ref{figExpl2}, where the energy had been deposited in a
smoothed fashion.

From these experiments we conclude that formulations of SPH in terms
of entropy or in terms of the asymmetric form of the energy equation
are able to produce quite reasonable results for the explosive release
of energy into a {\em single} particle.  It has been argued
\citep{Benz90,Owen98} that SPH can treat a strong explosion only if
the energy is initially deposited sufficiently smoothly; i.e.~if the
explosion energy is distributed according to the smoothing kernel into
a group of particles.  We think that this argument only really applies
when the thermal energy is integrated in its symmetrised form, where
neighbouring particles of a single explosion seed can suffer from an
unphysical decline in temperature and entropy.  While an initial
smoothing of the explosion energy can cure this problem, certain
starburst algorithms rely on the possibility of locally releasing a
large amount of energy in a single particle.  Note that using the
geometric mean to symmetrise the pressure terms does increase the
robustness of the thermal energy method, but yields noticeably worse
or even unphysical solutions for the explosion problem.

\section{Cooling within halos}
\label{sectioncool}

Radiative cooling within halos is a highly non-linear process that can
be severely influenced by numerical resolution.  In order to
straightforwardly examine such systematic effects, we consider a
series of idealised test problems consisting of gas spheres that
collapse and virialize under their own self-gravity.  We parameterise
the models by a `virial velocity' $V$.  Each gas sphere has a total
mass $M=V^3/(10GH_0)$ and is initially at rest with a density profile
$\rho(r) \propto 1/r$, a radius $R=V/(5H_0)$, and a thermal energy per
unit mass of $u=5.0\times 10^{-4} V^2$.  After being released, these
gas spheres collapse to their centres, bounce back, and virialize
through a strong outgoing shock.  As such, this problem is a version
of the `Evrard'-collapse \citep{Ev88}, a common test problem for SPH,
except that we allow the gas to cool radiatively like a plasma
composed of a primordial mix of helium and hydrogen.  The actual
cooling rates are computed as in \citet{Ka96}.  In order to arrive at
a clean inner boundary condition to allow for well defined resolution
studies, we set the gravitational softening to a fixed value of
$\epsilon = 1.5\times 10^{-3} V/H_0$ in all the tests, and we turn
cold gas within a distance of less than $0.25\epsilon$ from the origin
and with a density higher than $7.2\times 10^7 \rho_{\rm crit}$ into
stationary sink particles that only gravitationally interact with the
remainder of the gas.  This ensures that once it can cool
efficiently, the gas at the origin will essentially be fixed at a
density $7.2\times 10^7 \rho_{\rm crit}$, allowing a meaningful
attempt to achieve convergence for the entire density profile as the
mass resolution is increased.  For the particle numbers employed here,
the SPH smoothing lengths of particles in the sink region will be
smaller than $0.5\epsilon$; hence, they are no longer interacting with
particles in the actual cooling flow once they are converted into sink
particles.

\begin{figure*}
\bc
\resizebox{8cm}{!}{\includegraphics{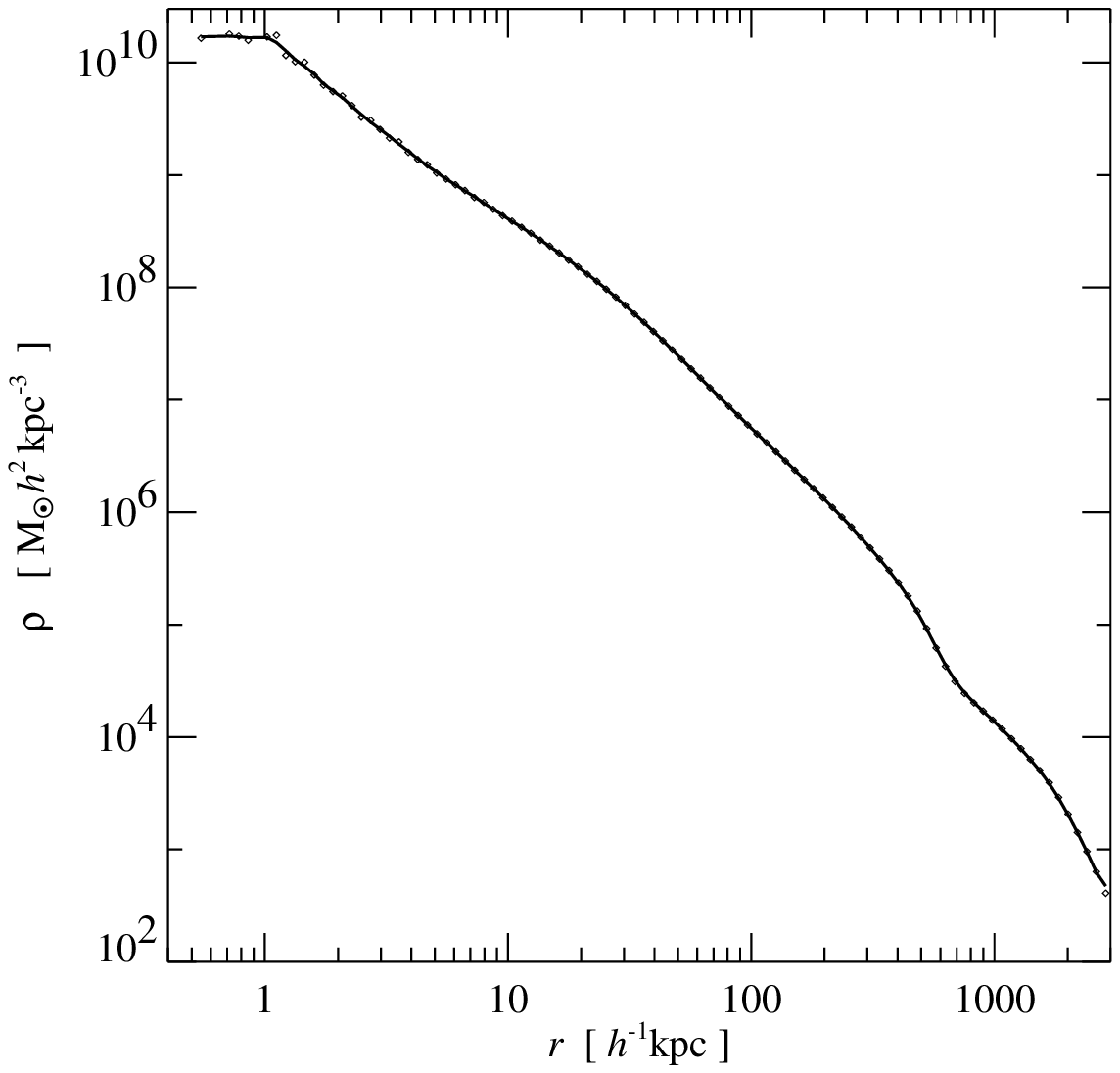}}%
\resizebox{8cm}{!}{\includegraphics{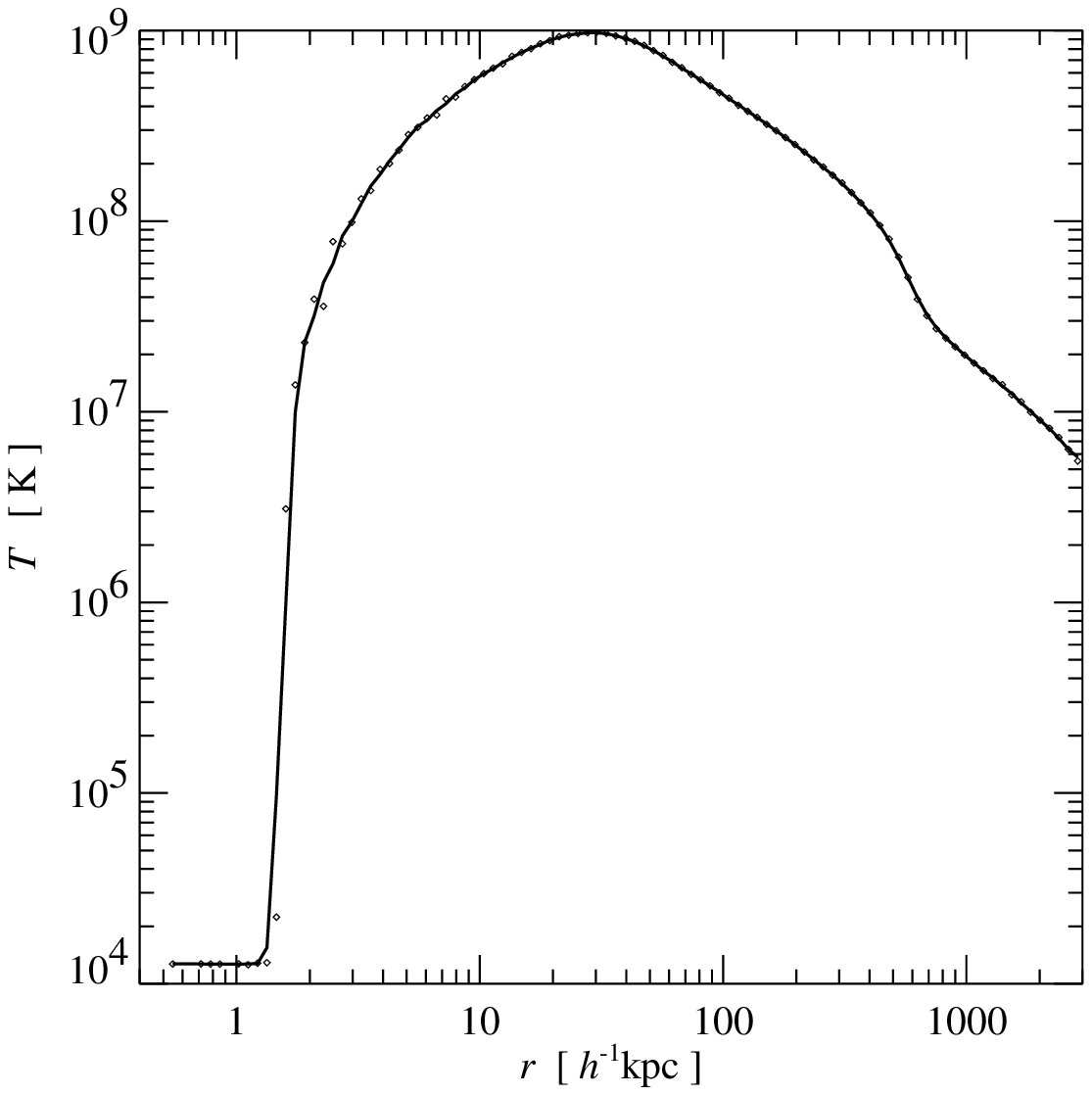}}\\
\resizebox{8cm}{!}{\includegraphics{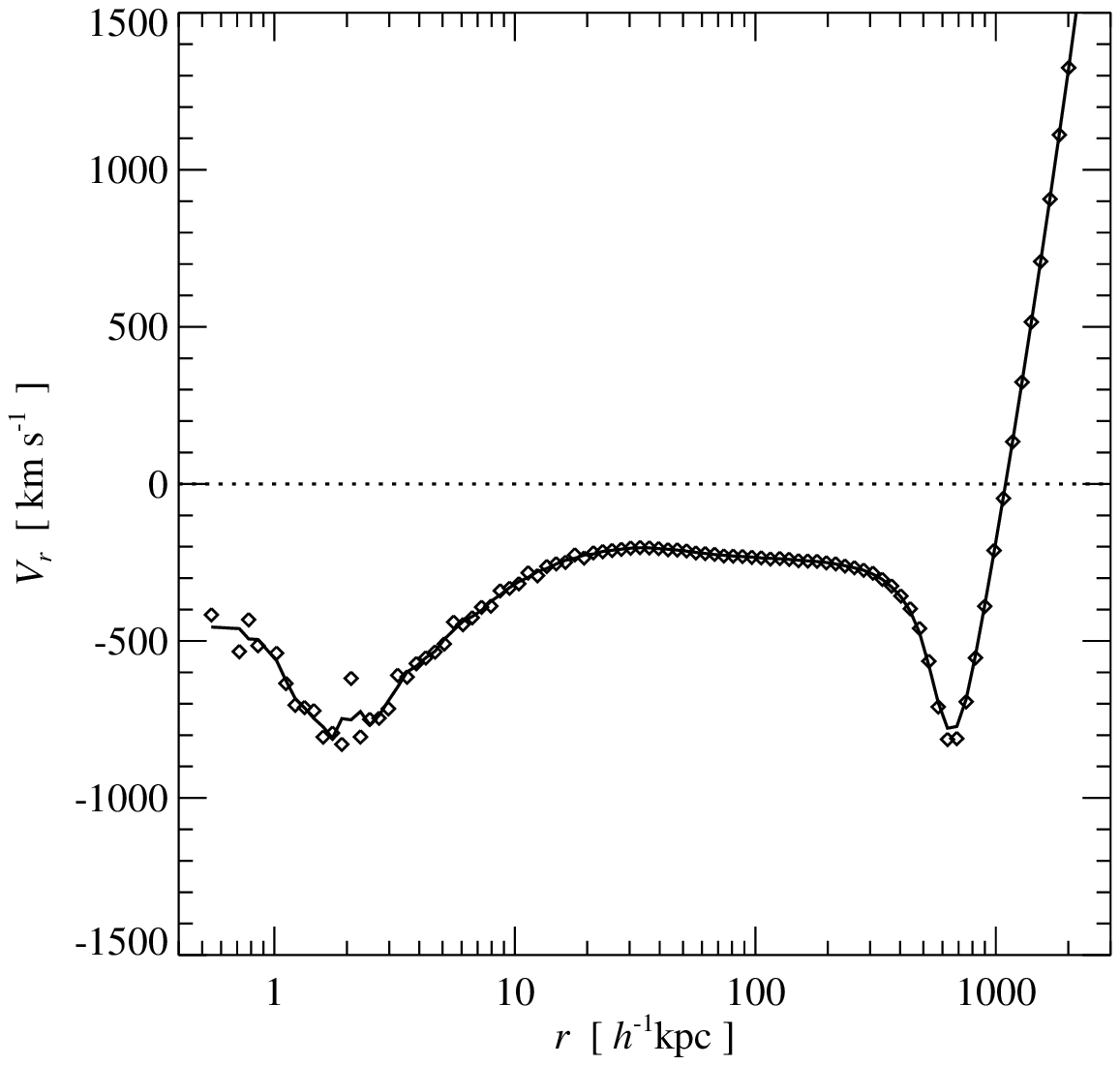}}%
\resizebox{8cm}{!}{\includegraphics{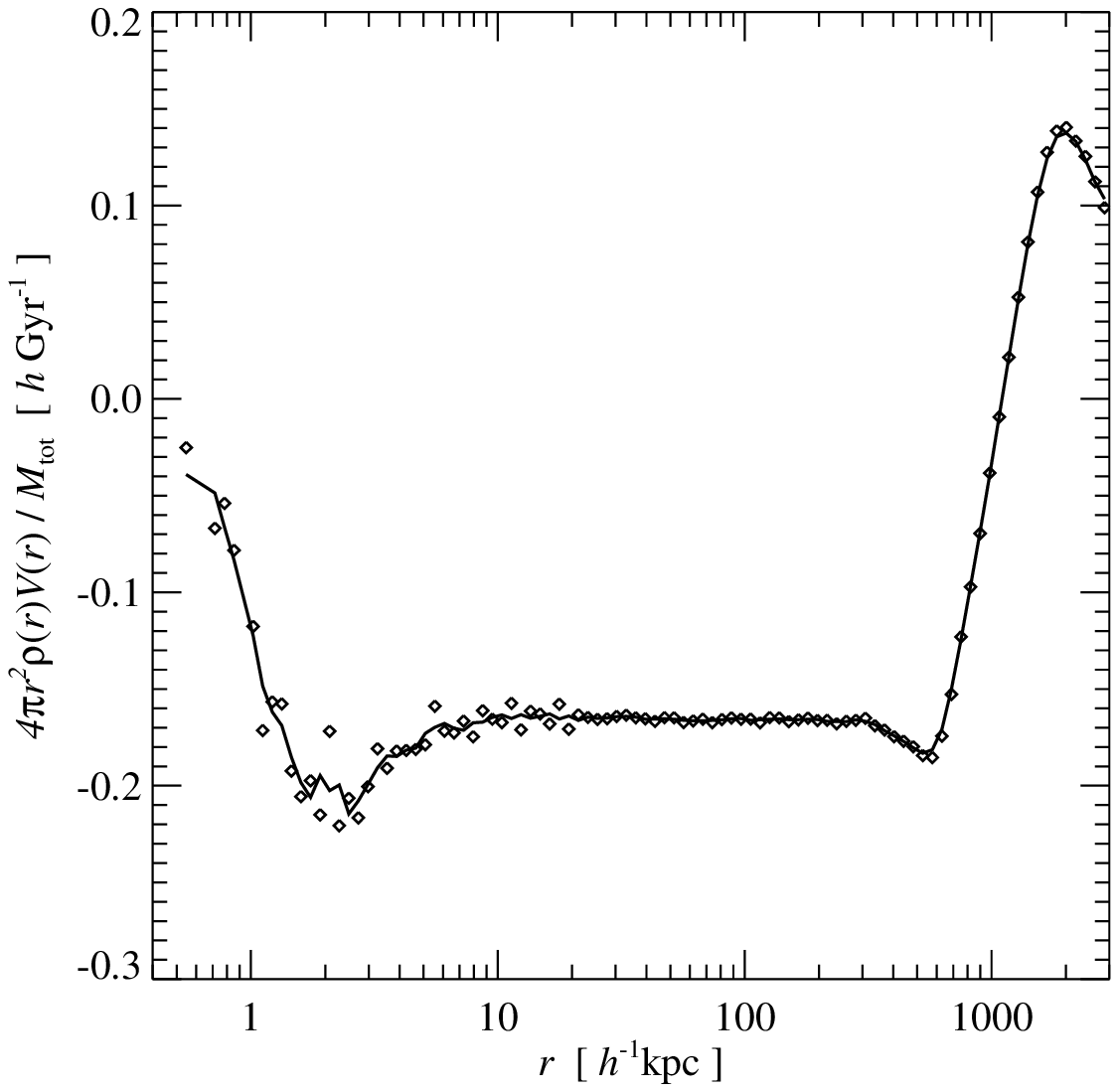}}\\
\caption{Radial profiles of density (top left), temperature (top
right), mean radial velocity (bottom left), and mean radial mass flux
(bottom right).  The profiles are shown for the collapse of a $10^{15}
h^{-1} {\rm M}_\odot$ gas sphere, at time $t=4.0$ after the start of
the simulation.  The particular model shown here was evolved with the
new conservative entropy formulation, using 92096 particles for the
initial conditions.  In all panels, symbols represent averages
obtained for logarithmic bins in the radial coordinates, while the
lines are boxcar averages of these points.
\label{figProfiles}}
\ec
\end{figure*}

\begin{figure*}
\bc
\resizebox{8cm}{!}{\includegraphics{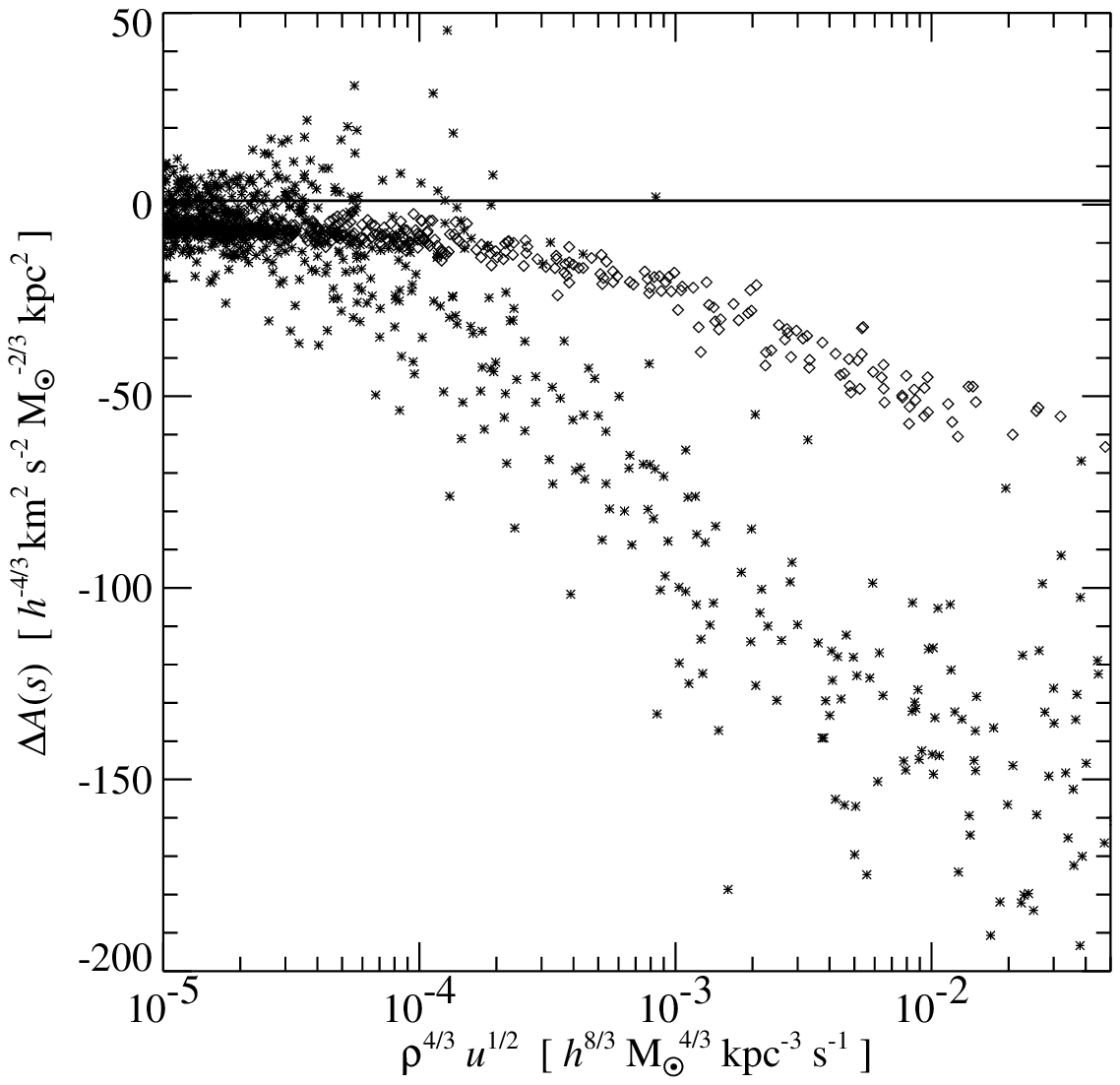}}%
\resizebox{8cm}{!}{\includegraphics{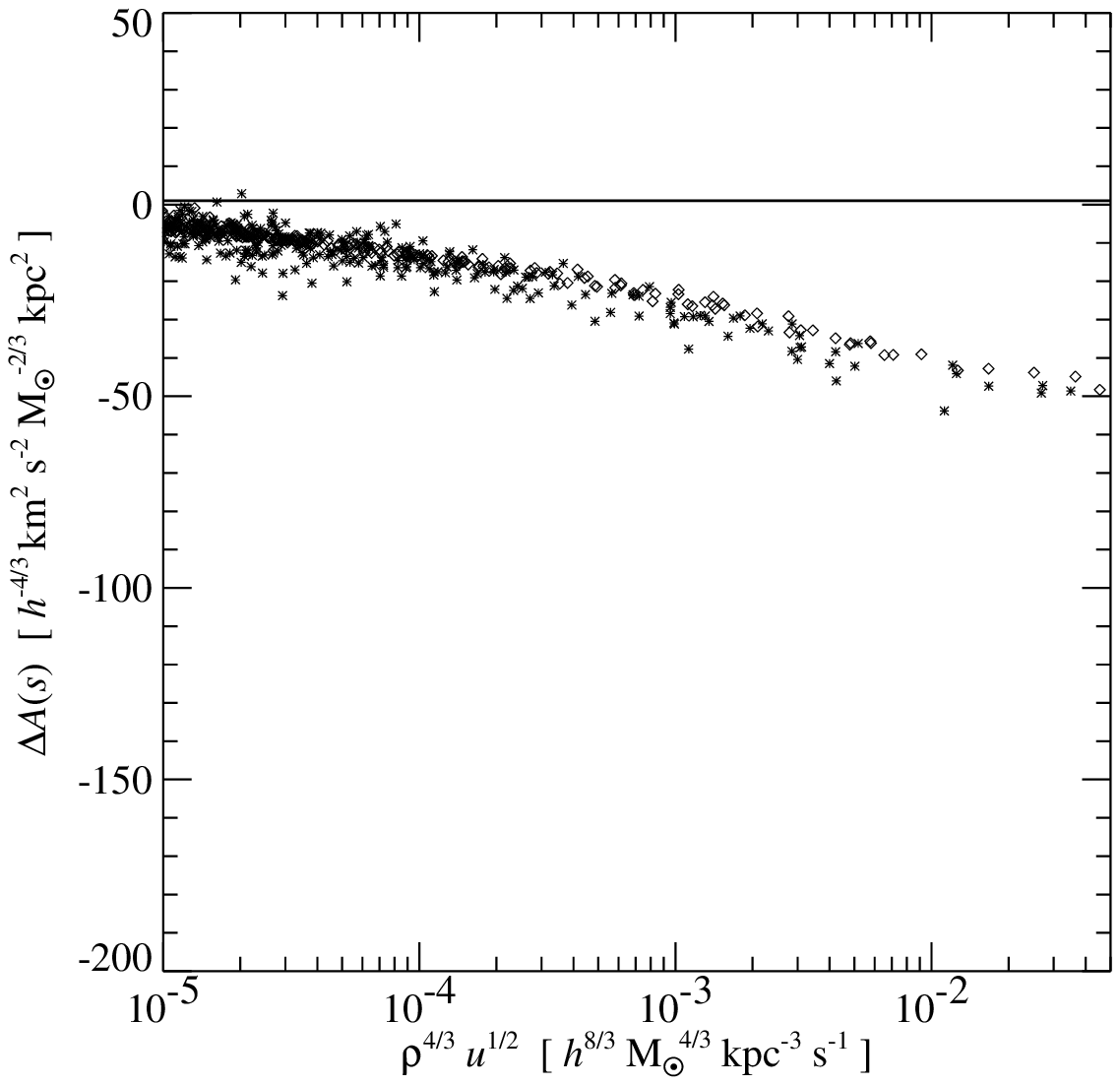}}\\
\caption{Decline of entropy of individual particles between two
subsequent output times from the gas sphere simulations, spaced
$\Delta t=0.1$ apart, and measured at $t=3.4$. The left panel compares
particles from the $N=1472$ simulation computed either using the
standard energy formulation (stars) or using the entropy formulation
(diamonds).  The particles are plotted versus $\rho^{4/3}u^{1/2}$
because the rate of decline of entropy due to cooling is expected to
be approximately proportional to this quantity.  The range of
$\rho^{4/3}u^{1/2}$ shown corresponds to the region of the cooling
flow where the gas begins to cool rapidly.  It is seen that the
particles in the thermal energy run loose entropy faster than can be
explained by cooling alone.  This error arises because the adiabatic
flow is not sufficiently well resolved, ultimately leading to
overcooling.  In the right panel, we show the same measurement for the
$N=11536$ runs.  Here the two formulations of SPH show a consistent
rate of entropy decline due to cooling.
\label{figEntrDecline}}
\ec
\end{figure*}

While this test problem is highly idealised, it is nevertheless
relevant for simulations of galaxy formation, because in that case the
virialization of halos and the subsequent cooling of their gas happens
in an analogous, yet dynamically more complicated manner.  Note that
we deliberately excluded dark matter from this test to highlight
gas-dynamical effects without introducing complexities arising from
collisionless dynamics.  An interesting alternative test would be to
examine self-similar cooling wave solutions \citep{Ber89}.  While
having an analytic solution is an advantage, we prefer to use a set-up
that is closer to actual cosmological simulations with respect to
particle number, to collapse scenario, and to the cooling function.
This simplifies the interpretation of numerical effects.

Because the cooling function for primordial gas introduces
characteristic temperatures, the evolution depends on the mass scale
of the collapsing sphere.  We here consider spheres of mass
$M=10^{15}\,h^{-1}{\rm M}_\odot$, where $h=0.7$ parameterises the
Hubble constant $H_0=100\, h^{-1} {\rm km\,s^{-1}Mpc^{-1}}$.  These
gas spheres thus have the mass of a rich cluster of galaxies, and a
virial temperature of $\simeq 5\times 10^7\,{\rm K}$.  Note that
because these 'clusters' consist of {\em only} gas, their cooling
times will be shorter than those of real, dark-matter dominated
clusters of comparable mass.

In Figure~\ref{figAccRatem15}, we show the deposition rate of cold gas
in the halo centre as a function of time, computed for mass
resolutions ranging from 552 to 92096 particles, and for the five
variants of SPH listed in Table~\ref{tab1}.  The deposition rate is
given in units of total mass per unit time, where unit time is
$1/(10H_0)=0.98\,h^{-1}\,{\rm Gyr}$.  A value of 0.5 thus means an
accretion rate such that 50\% of the gas cools within $1.4\,{\rm Gyr}$
for $h=0.7$.

\begin{figure*}
\bc
\resizebox{6.5cm}{!}{\includegraphics{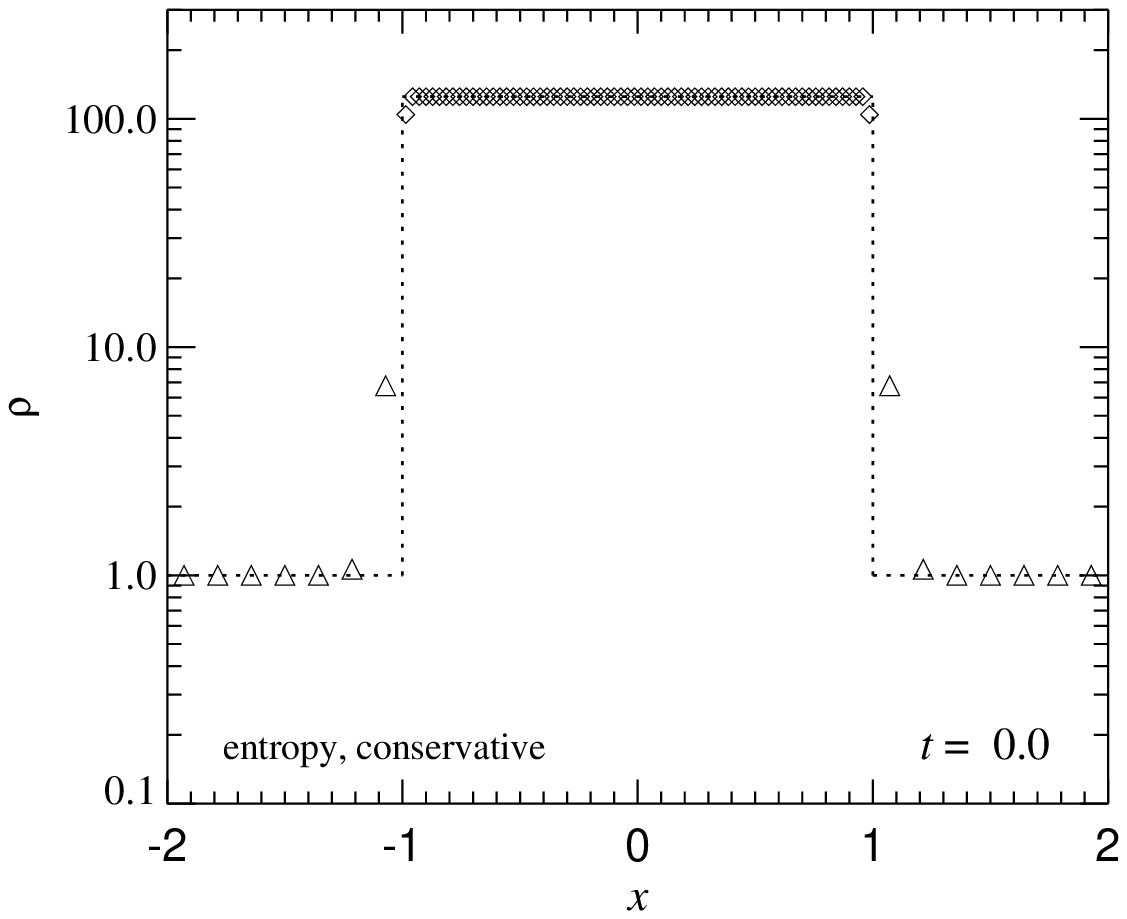}}%
\hspace*{-1.0cm}\resizebox{6.5cm}{!}{\includegraphics{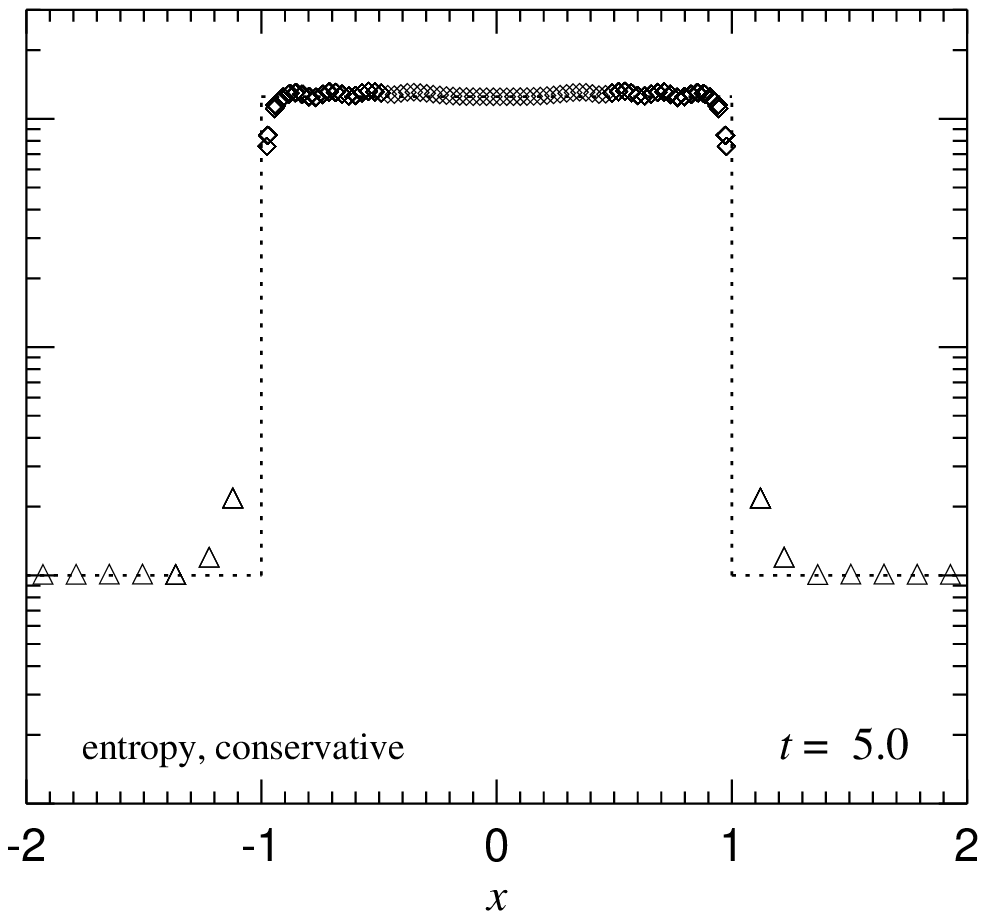}}%
\hspace*{-1.0cm}\resizebox{6.5cm}{!}{\includegraphics{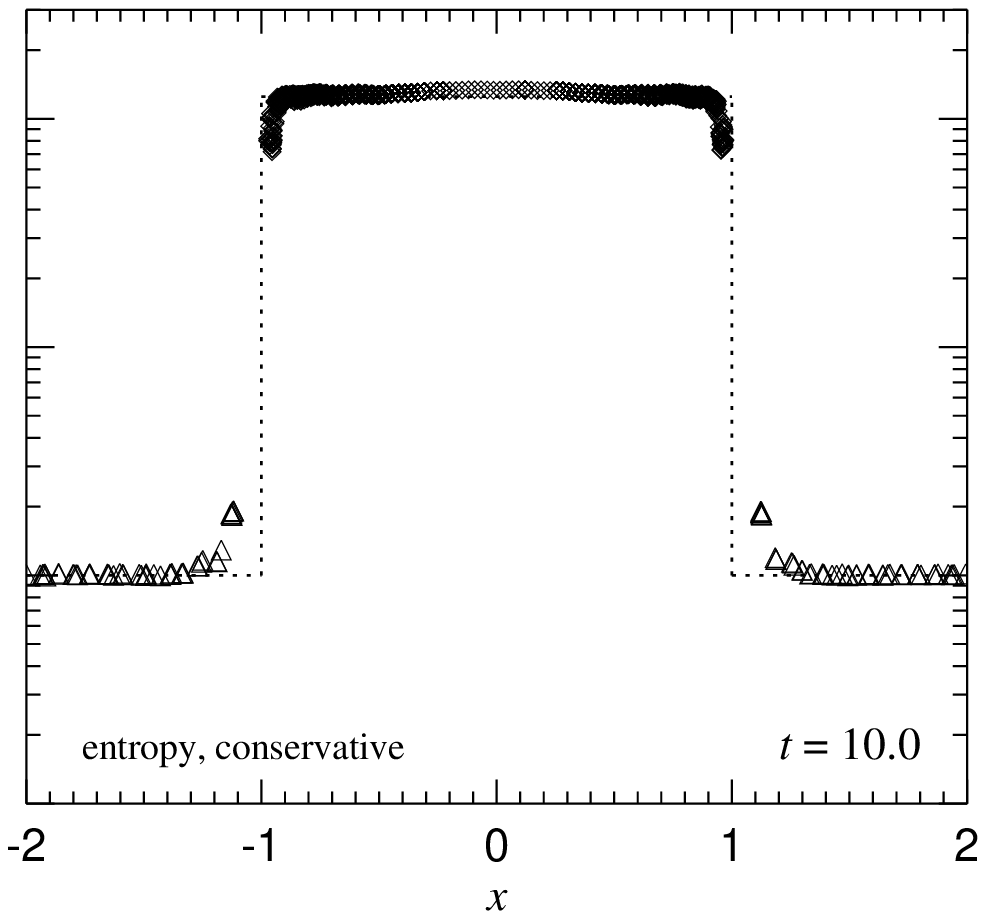}}\\%
\resizebox{6.5cm}{!}{\includegraphics{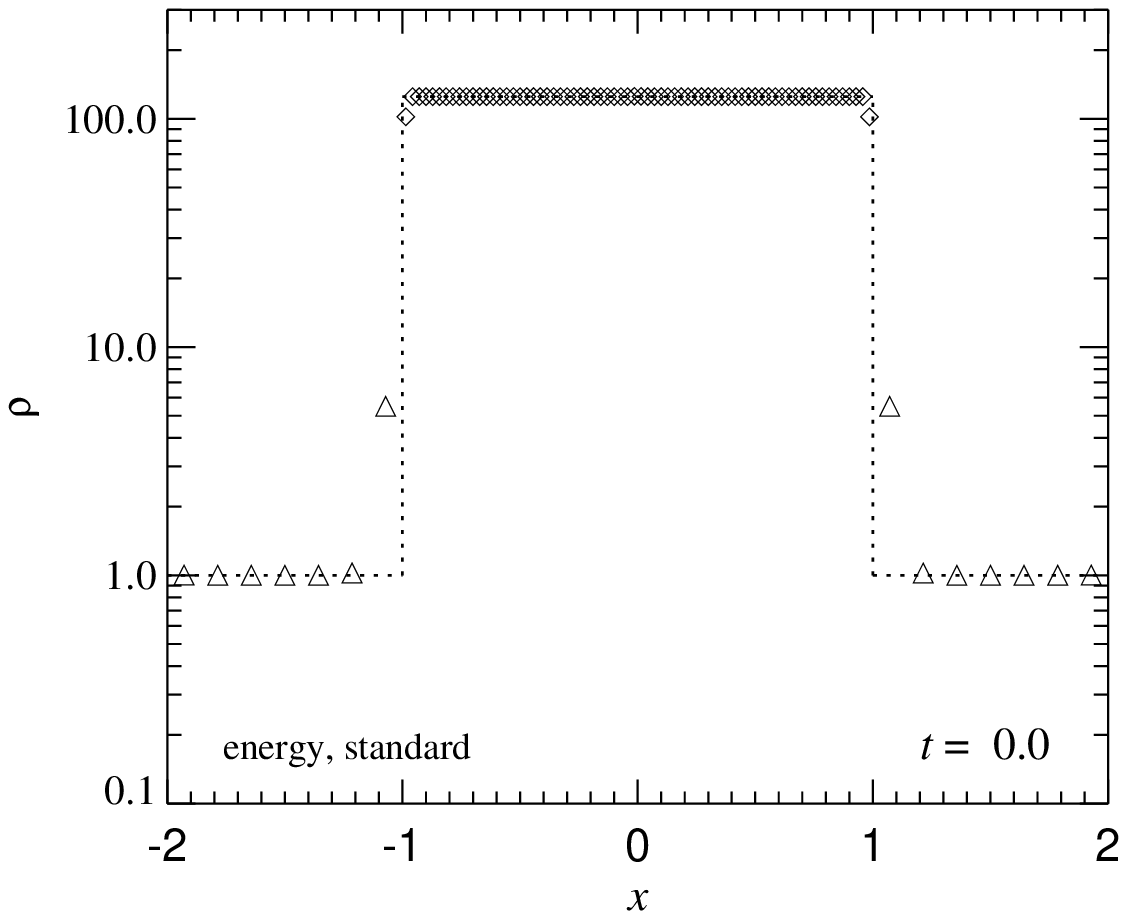}}%
\hspace*{-1.0cm}\resizebox{6.5cm}{!}{\includegraphics{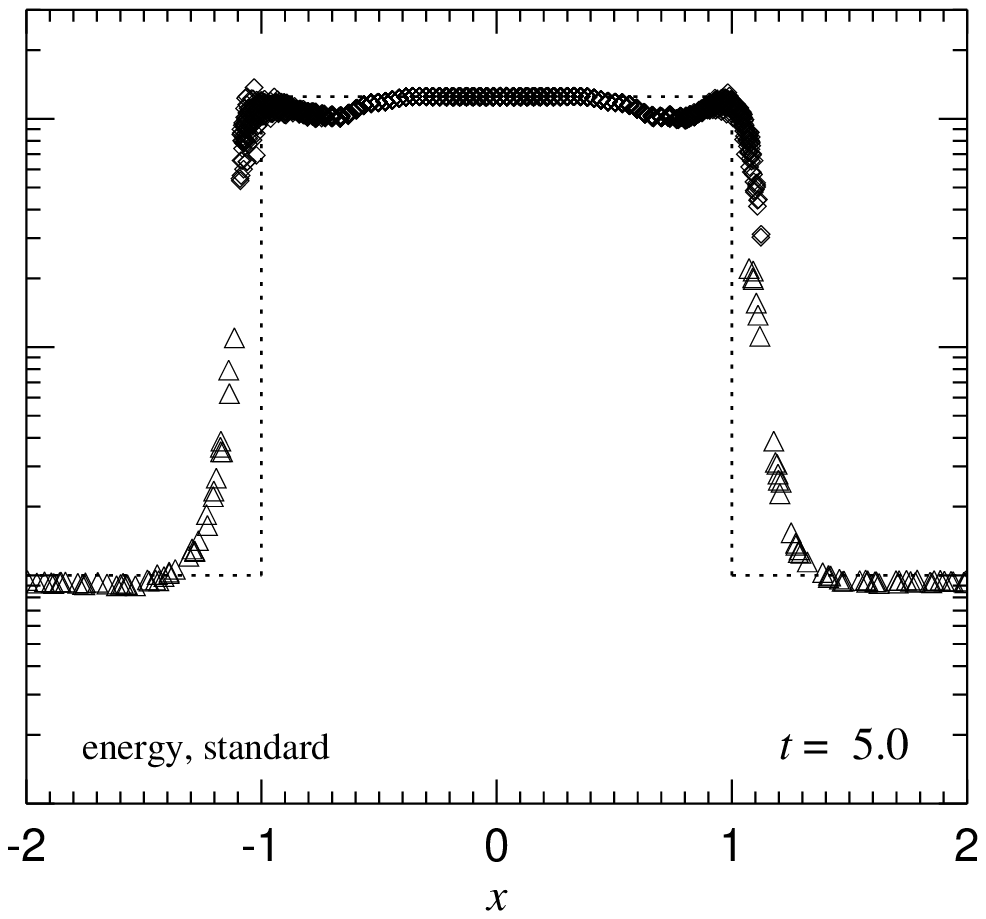}}%
\hspace*{-1.0cm}\resizebox{6.5cm}{!}{\includegraphics{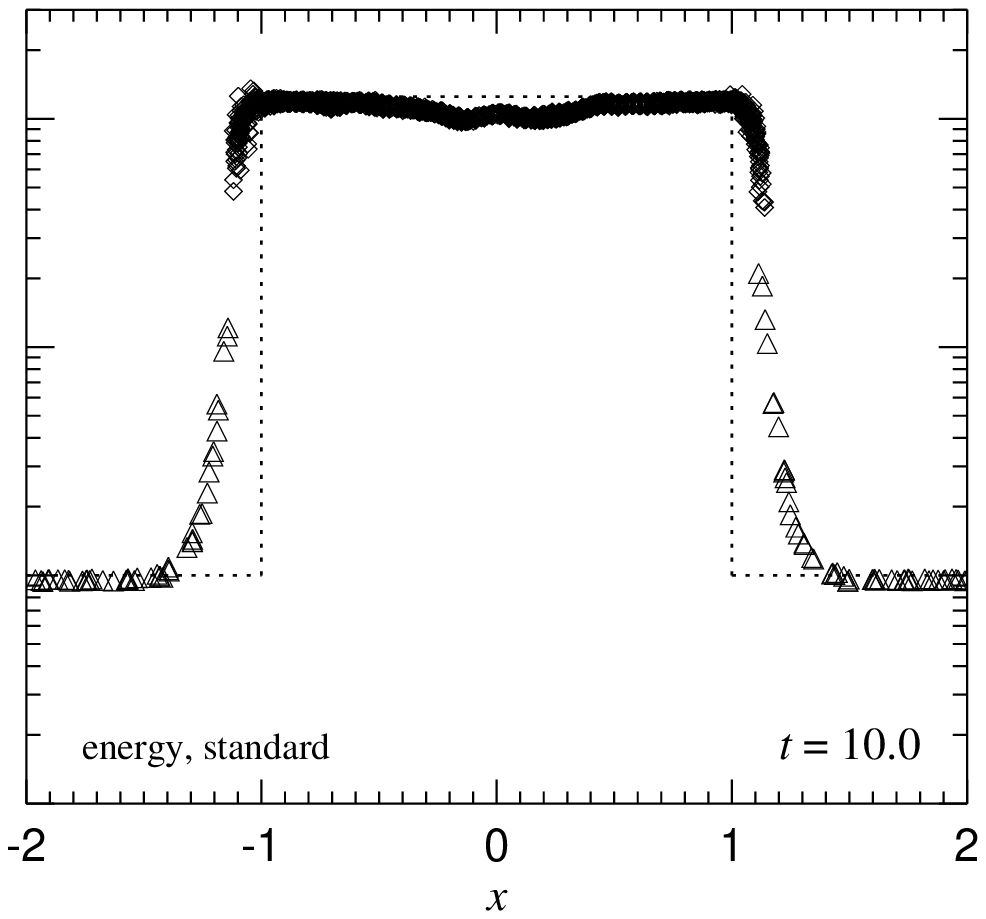}}\\%
\resizebox{6.5cm}{!}{\includegraphics{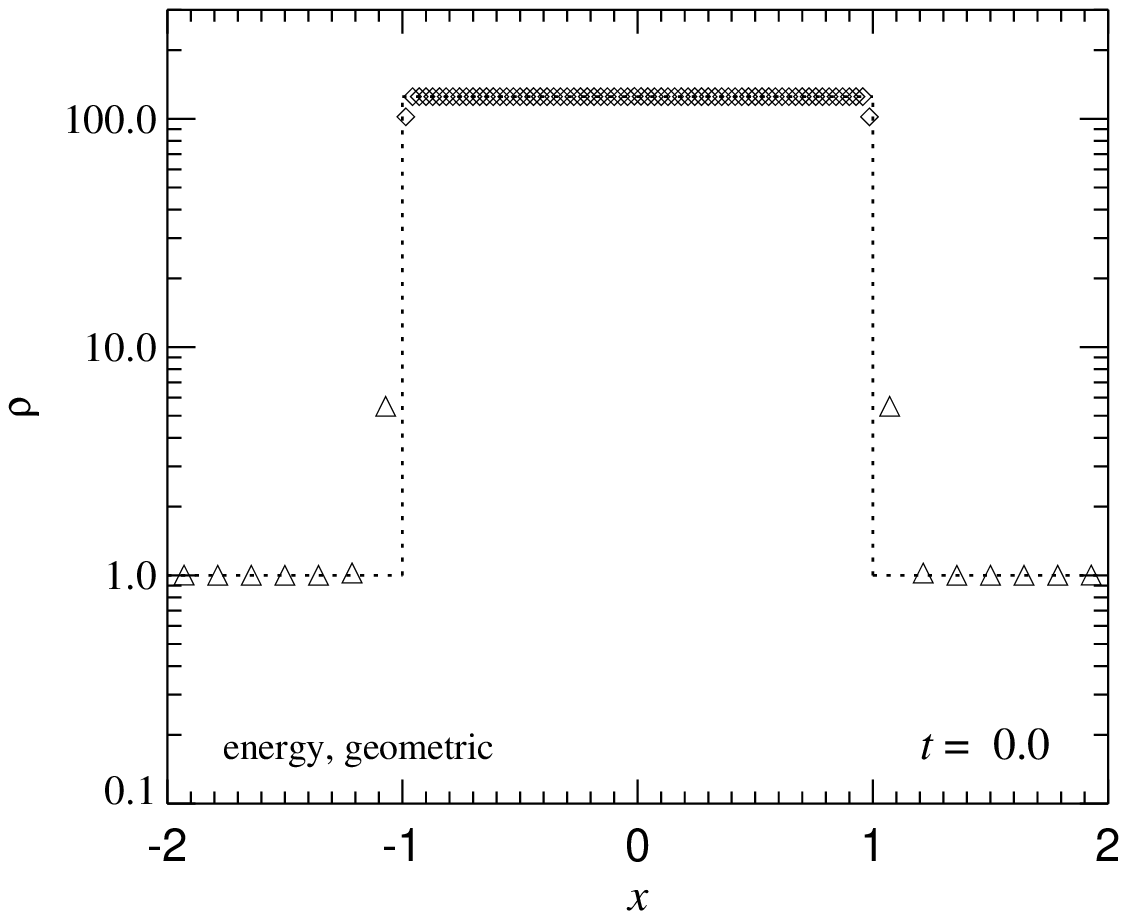}}%
\hspace*{-1.0cm}\resizebox{6.5cm}{!}{\includegraphics{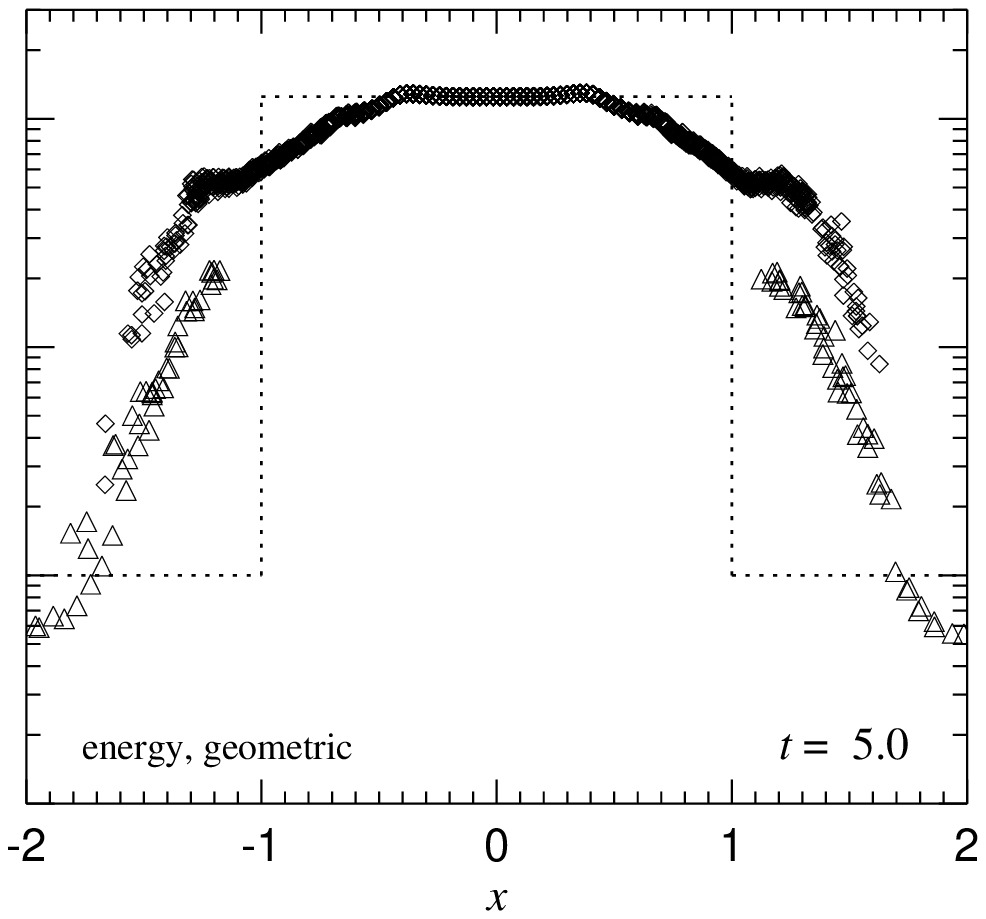}}%
\hspace*{-1.0cm}\resizebox{6.5cm}{!}{\includegraphics{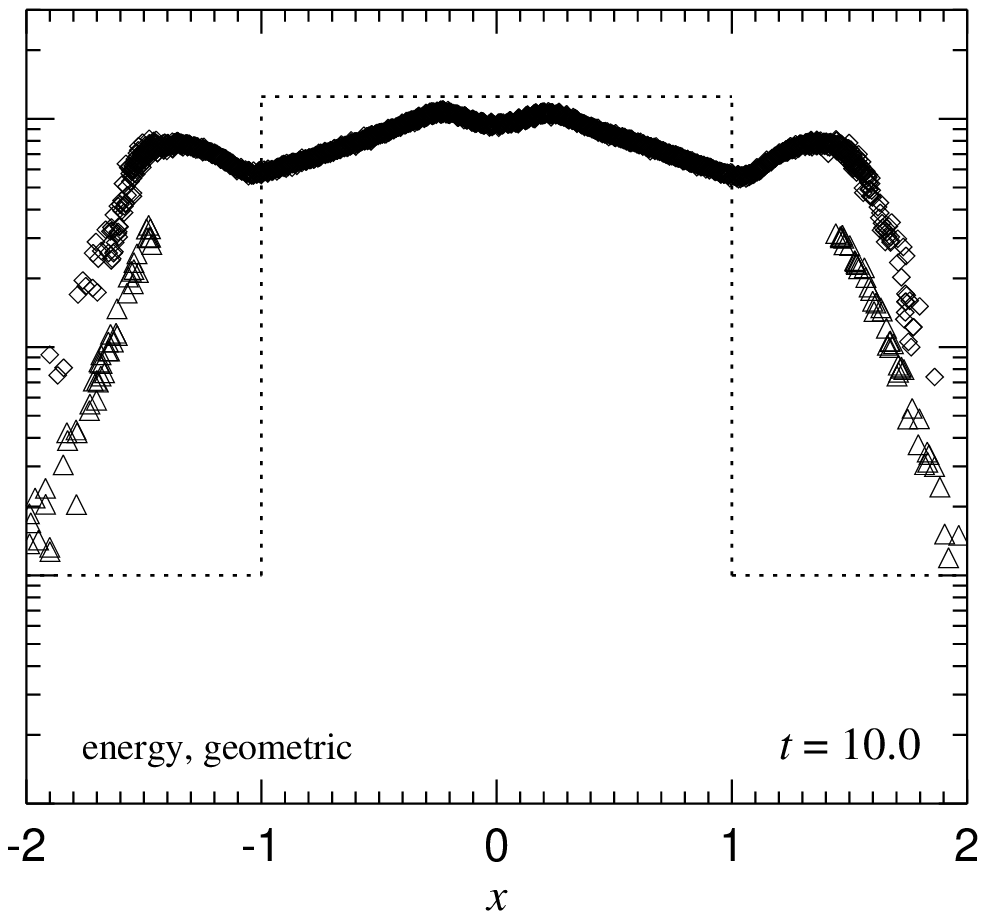}}\\%
\caption{Time evolution of a strong density discontinuity when evolved
with three different SPH variants, as indicated in the panels. The
simulations follow two phases of gas that are brought into contact at
time $t=0$ in a three-dimensional, periodic box of dimension $4\times
1 \times 1$. Initially, the gas is at rest and has equal pressure of
$P=1$, with the region $|x|>1$ being filled with gas at density
$\rho_1=1$, and the region $|x|<1$ being filled with gas at density
$\rho_2=125$.  The symbols show individual particle densities, with
triangles marking particles that were originally in the low-density
phase while those that were in the high-density phase are plotted as
diamonds.
\label{figphaseboundary}}
\ec
\end{figure*}

From Figure~\ref{figAccRatem15}, it is evident that at low resolution
there are strong systematic differences between the energy and entropy
formulations of SPH: When the thermal energy equation is integrated,
cooling rates are {\em substantially overestimated}, while with the
entropy equation, they are {\em slightly underestimated} compared to
the `converged' values measured at very high resolution.  As the
particle number is increased, the various treatments yield results
that are in better and better agreement.  In itself, this is
reassuring, but it is unsettling that the cooling rate can be as
dramatically overestimated as is seen in the 552 and 1472 particle
runs which integrated the thermal energy equation.  After all, many
halos in cosmological simulations are resolved with far fewer
particles.  Therefore, all of these halos are likely subject to strong
overcooling.  Moreover, in hierarchical scenarios of structure
formation, larger systems will inherit this problem if their
progenitors suffered from overcooling.  Such behaviour may well
explain why cosmological SPH simulations predict more gas cooling than
is expected based on simple analytic models
\citep{Pea2000,Benson01,Yoshida2001}.

Note from Figure~\ref{figAccRatem15} that the low-resolution results
for the geometric symmetrisation of the pressure terms are
particularly discrepant compared to the high-resolution solutions.
Even the simulation with 1472 particles cools {\em all} the gas in the
initial collapse in this case, and for 4224 and 11536 particles there
are still strong `shoulders' in the accretion rate at early times
around $t\simeq 2$.  Such a shoulder is also present for the other
energy-based methods in the 4224 runs, but it is absent for the
conservative entropy formulation.

To determine the cause of the difference in the low-resolution
behaviour between the entropy and energy formulations of SPH, it is
instructive to examine the radial structure of the cooling flows in
the simulations.  In Figure~\ref{figProfiles}, we show profiles of the
density, the temperature, the radial velocity, and the radial mass
flux for one of our runs at time $t=4.0$, when the cooling flow has
been operative for some time and the accretion is almost stationary.
It is seen that there is an inflow region in the radial range $\sim
20-300 h^{-1} {\rm kpc}$ where the gas is moving inwards at nearly
constant radial velocity, and the density profile is steeply rising
with slope $\simeq -2$.  It is important to note that this inflow is
essentially adiabatic; the gas is not yet radiating efficiently, hence
the temperature is strongly rising as the gas is compressed on its
inbound path.  Close to the very centre, the gas begins to cool
efficiently, leading to an accelerated inflow, and eventually to a
rapid drop in the temperature to around $10^4\,K$.  It is clear that
an underestimate of the compressional heating of the gas in the
extended inflow region will artificially boost the overall accretion
rate.  In fact, we believe it is this effect which is responsible for
the overcooling in the thermal energy runs at low resolution, as we
now demonstrate.

In the left panel of Figure~\ref{figEntrDecline}, we show the change
$\Delta A(s)$ in the entropy of individual SPH particles between two
subsequent simulation outputs of the 1472 particle runs, for both the
standard energy and entropy methods.  Recall that $A(s)$ should change
only as a result of radiative losses, or owing to generation of
entropy by the artificial viscosity in shocks.  We plot $\Delta A(s)$
as a function of $\rho^{4/3}u^{1/2}$, which is the relevant dependence
of the entropy loss rate due to cooling in this temperature range.
When we compare particles from the thermal energy and the entropy runs
in this way, we see that $A(s)$ for the particles in the energy run
decreases {\em faster} than expected based on the cooling rate, where
this expectation is given by the locus of particles from the entropy
run.  This may be expressed differently by saying that the entropy of
these particles is violated -- it decreases faster than can be
accounted for by the external sinks of entropy.  It should be noted
that in spite of this, the total energy is well conserved for these
particles, i.e.~the total energy declines at exactly the rate at which
energy is radiated.

What we observe in Figure~\ref{figEntrDecline} is a discreteness
effect arising from insufficient resolution in the cooling flow, where
the adiabatic heating of the inflow is not adequately resolved. If the
particle number is increased, the resulting false loss of entropy is
avoided, as can be seen in the right panel of
Figure~\ref{figEntrDecline}, where we show the equivalent plot for the
11536 particle simulations.  We note that the geometric symmetrisation
of the thermal energy equation is particularly prone to this effect.
At low resolution, an infalling particle that interacts with particles
that are already much colder will not be heated sufficiently despite
its strong compression, because the geometric mean heavily suppresses
adiabatic work done between particle pairs at different temperatures.

In fact, we have found that this relative suppression of pair-wise
forces in the geometric symmetrisation technique can be strong enough
to render density discontinuities unstable. In order to highlight this
effect, we have considered a test problem where two `phases' of gas of
equal pressure are brought into contact along a planar intersection,
with the density of one of the phases being substantially higher than
that of the other. Since we take the gas to be initially at rest in
this problem, the resulting density discontinuity should remain stable
for all times.

However, as a result of its inherent smoothing, it is clear that SPH
will not be able to precisely maintain the sharpness of the initial
density discontinuity. Instead, we expect that the gas at the
intersection will undergo some small transient motion before it
settles into a relaxed state, such that the phase boundary ends up
being slightly washed out. We investigate this relaxation process with
a three-dimensional realisation of a density discontinuity of strength
$ \rho_2/\rho_1 =125$.  For
definiteness, we consider a periodic box of size $4\times 1\times 1$
where one half is filled with a $2\times 7^3$ particle grid at a
fiducial density $\rho_1=1$, and the other half is filled with a
$2\times 35^3$ grid. All particles are taken to have equal mass, such
that the density in the latter half is $\rho_2=125$. Temperatures are
assigned such that an equal pressure of $P=1$ in both phases (using
$\gamma=5/3$) is obtained.

In Figure~\ref{figphaseboundary}, we show a comparison of the time
evolution of the resulting particle system obtained for three
different SPH formulations, using $N_{\rm sph}=32$ smoothing
neighbours in each case. In the top three panels, we show the
evolution when the new conservative entropy formulation is used.  The
system shows the expected behaviour. Low-density particles adjacent to
the high-density phase `see' some of the dense particles, and are
hence estimated slightly high in density. Likewise, the density
estimates of particles close to the edge of the high-density phase
become biased low, such that the phase transition is sligthly washed out.
However, the phase boundary remains reasonably sharp, and its
stability is maintained for arbitrarily long times.

In comparison, the behaviour resulting for the standard energy
formulation of SPH is quite similar, but the sharpness of the initial
density discontinuity is not preserved quite as well.
In particular, a larger number of particles from
the original low-density phase is affected by higher density estimates
attained in the new equilibrium.  If the system was subject to
density-dependent cooling, this may cause an overestimate
of the cooling rates of particles close to the boundary.

This may however be seen as a small effect compared to the relative
performance of the geometric symmetrisation technique, which gives
a solution that is drastically worse than the ones obtained for the
other two methods.  In fact, as the lower row of panels in
Figure~\ref{figphaseboundary} shows, the phase boundary behaves in an
unstable way in this case, resulting in an {\em unphysical} evolution
of the system.  Particles from the low-density phase are literally
pressed into the high density phase. This happens because the
geometric symmetrisation renders the repulsive forces between particle
pairs of very different temperature largely ineffective. As a result,
each of the $7\times 7$ columns of low-density particles orthogonal to
the boundary penetrates as a finger into the high-density phase,
causing a partial overlap of the fluids, and the gradual decay of the
initial density discontinuity. Clearly, if the system was subject to
cooling, the hot low-density phase would suffer from a substantial
increase of its cooling rate because it would tend to overlap with the
dense cold phase, thereby becoming subject to overestimates of
density.

\begin{figure*}
\bc
\resizebox{8.2cm}{!}{\includegraphics{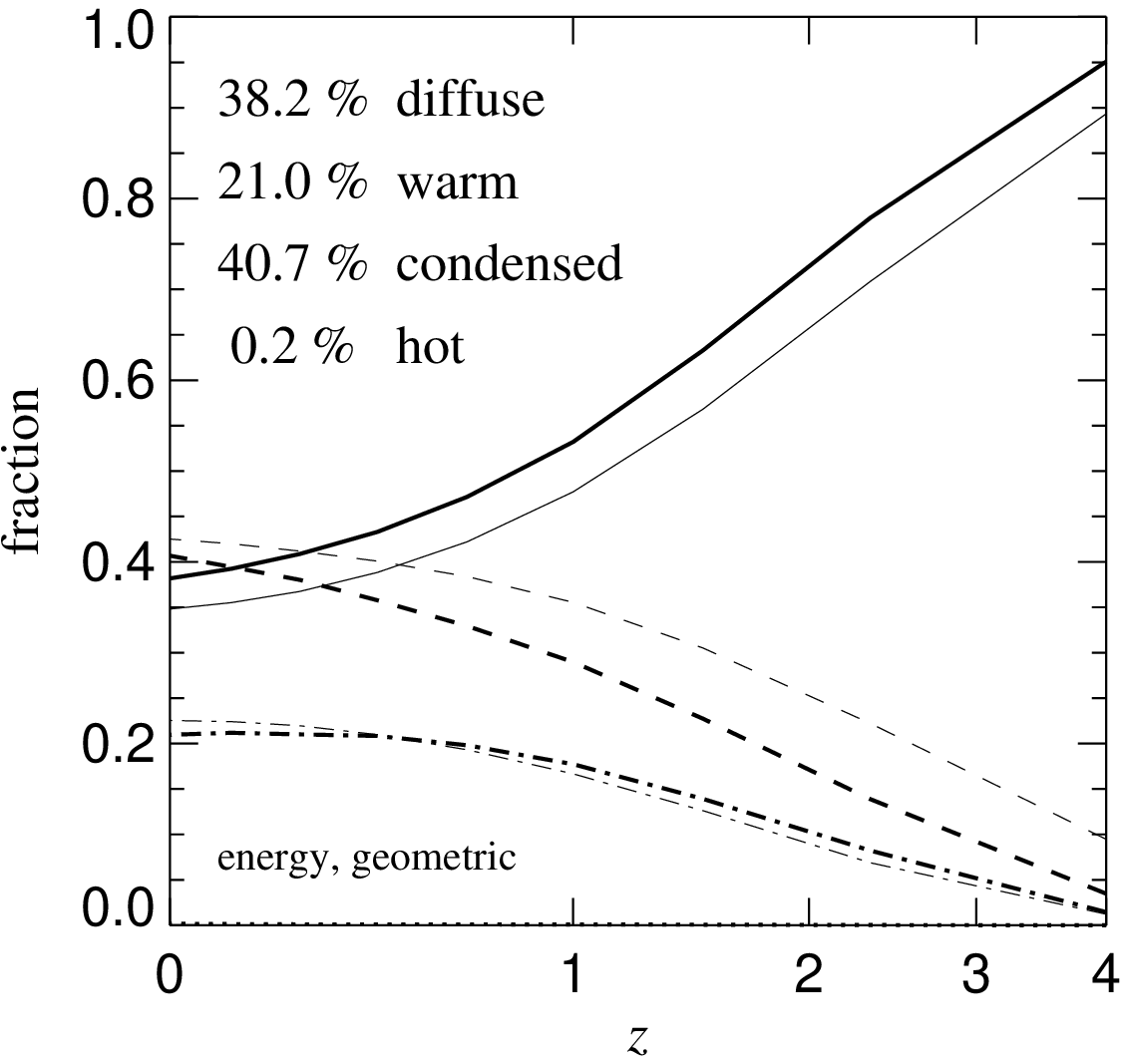}}%
\resizebox{8.2cm}{!}{\includegraphics{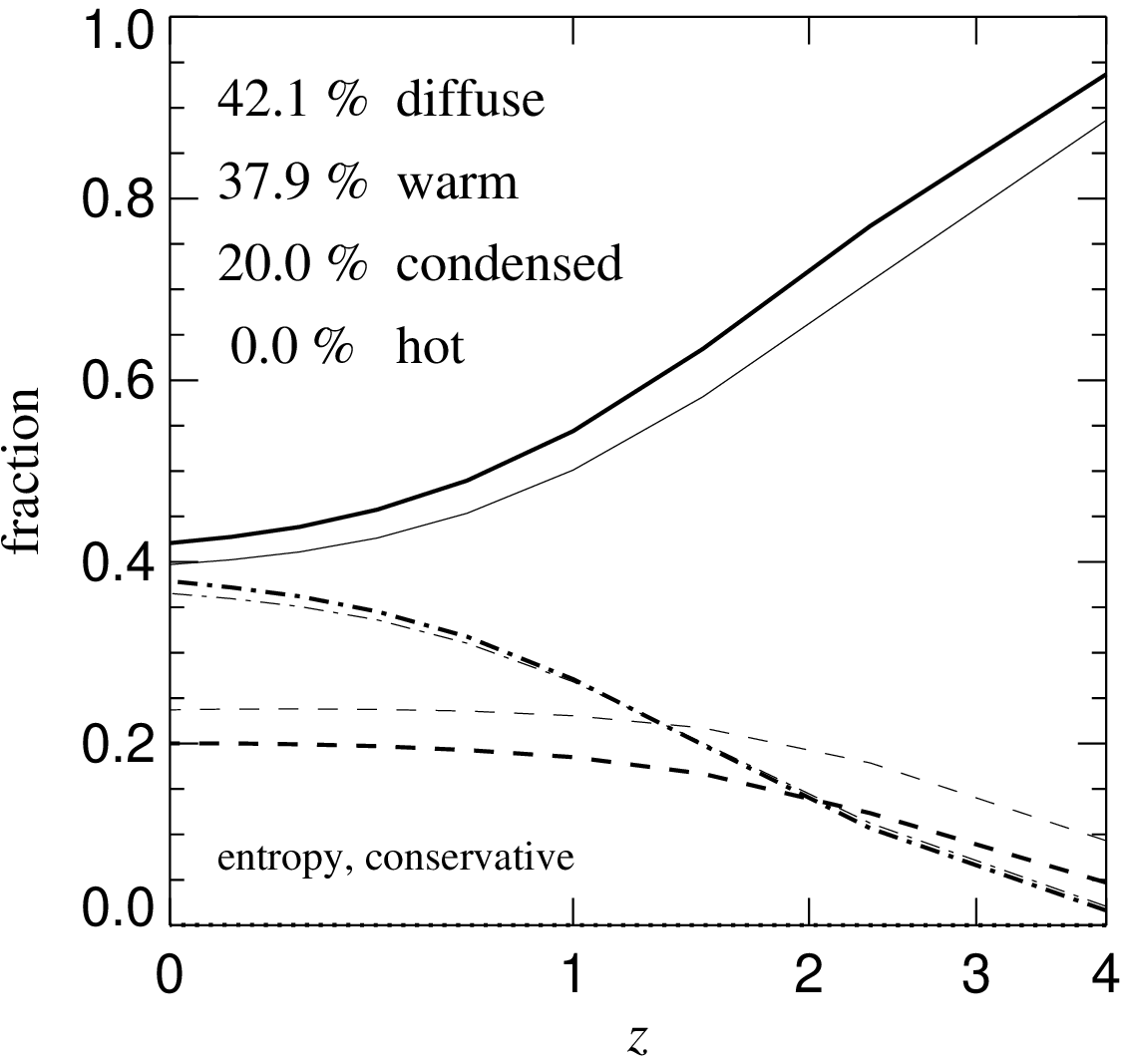}}\vspace*{-.4cm}\\
\caption{Mass fractions of gas in different phases as a function of
redshift.  Gas with temperatures below $10^5\,{\rm K}$ is considered
to be either `diffuse' (solid), if it is at low density
($\rho<1000\,\overline\rho\,$), or `condensed' (dashed) otherwise.
Gas with intermediate temperatures $10^5 - 10^7\,{\rm K}$ is labelled
`warm' (dot-dashed), and for $T>10^7\,{\rm K}$ `hot' (dotted).  The
left panel shows simulations where the thermal energy equation with
geometric symmetrisation was integrated, while the right panel gives
results for our new conservative entropy method. In both panels, thick
lines refer to a resolution of $2\times 50^3$, thin lines to $2\times
100^3$. Note that the entropy method cools substantially less gas, which
is also reflected in the differing fractions of condensed and warm gas.
\label{figPhaseFrac}}
\ec
\end{figure*}

\begin{figure*}
\bc
\vspace*{-0.4cm}%
\resizebox{8.6cm}{!}{\includegraphics{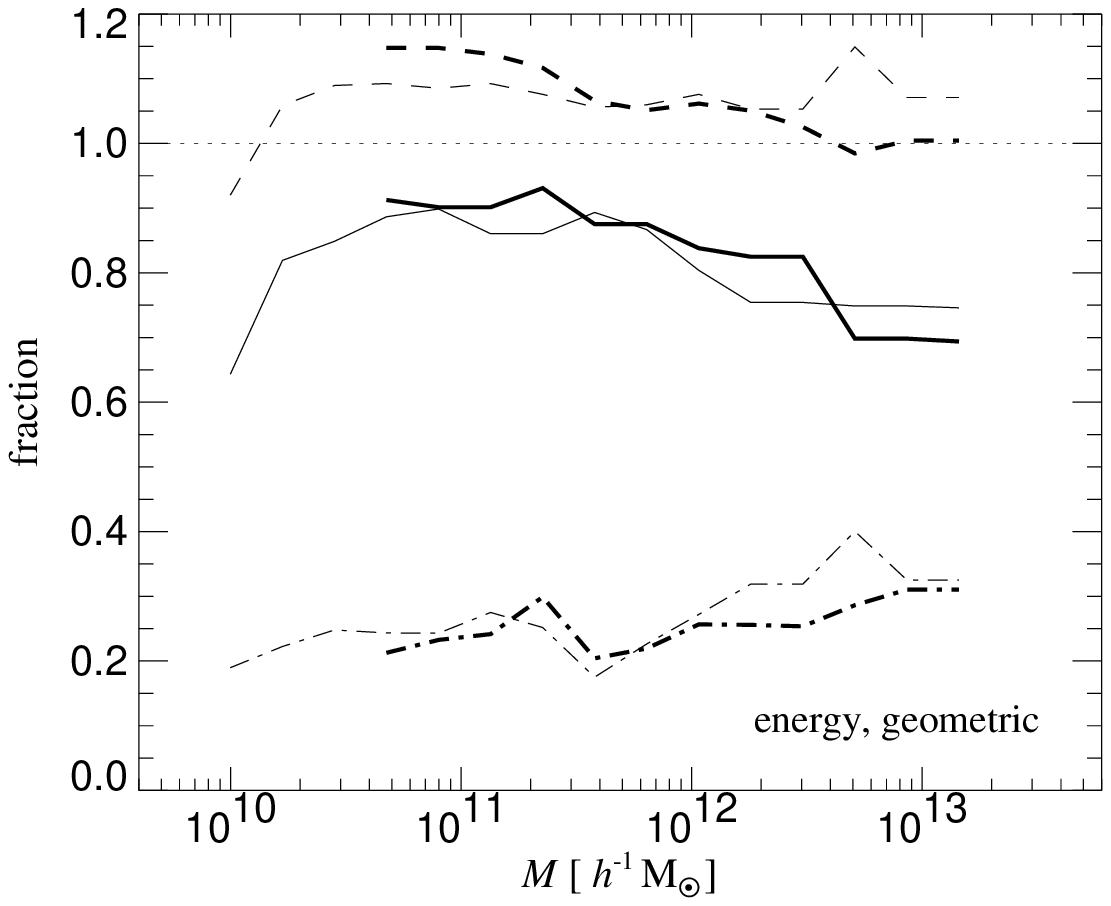}}%
\resizebox{8.6cm}{!}{\includegraphics{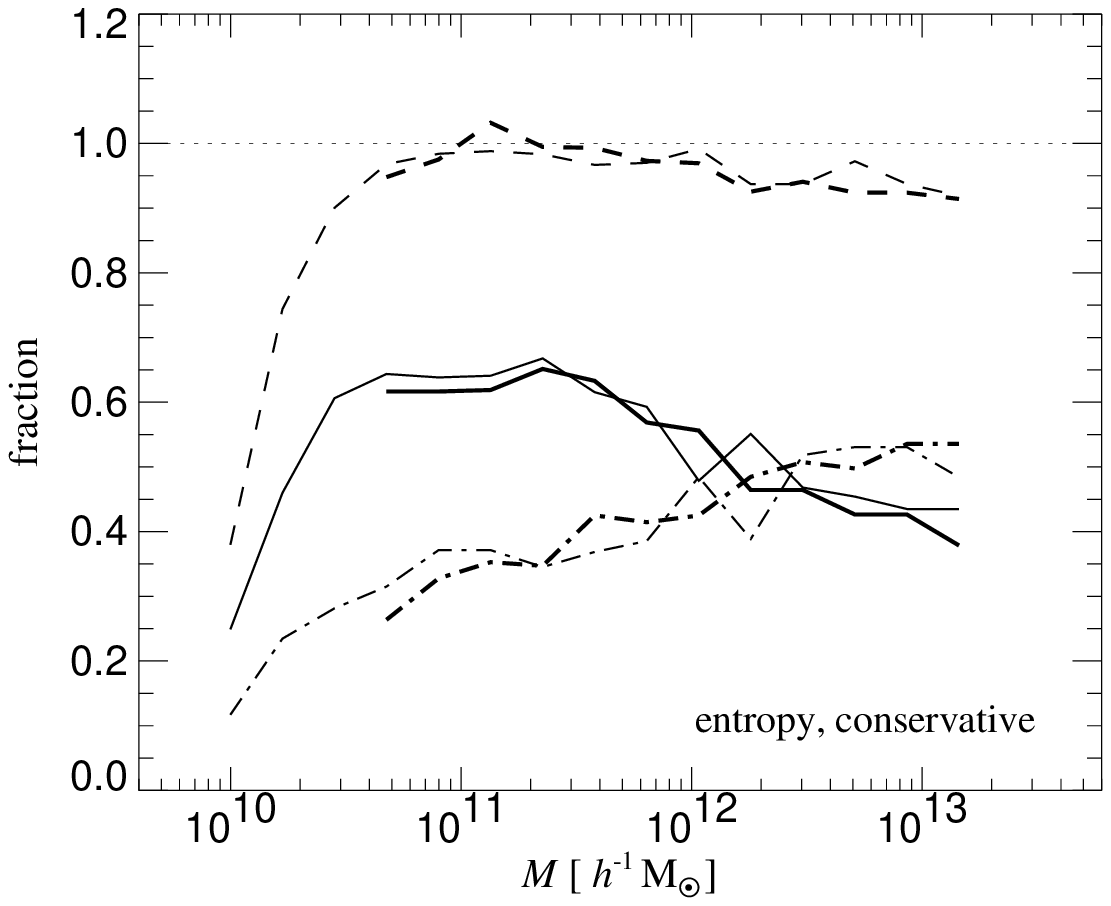}}\vspace*{-.3cm}\\
\resizebox{5.6cm}{!}{\includegraphics{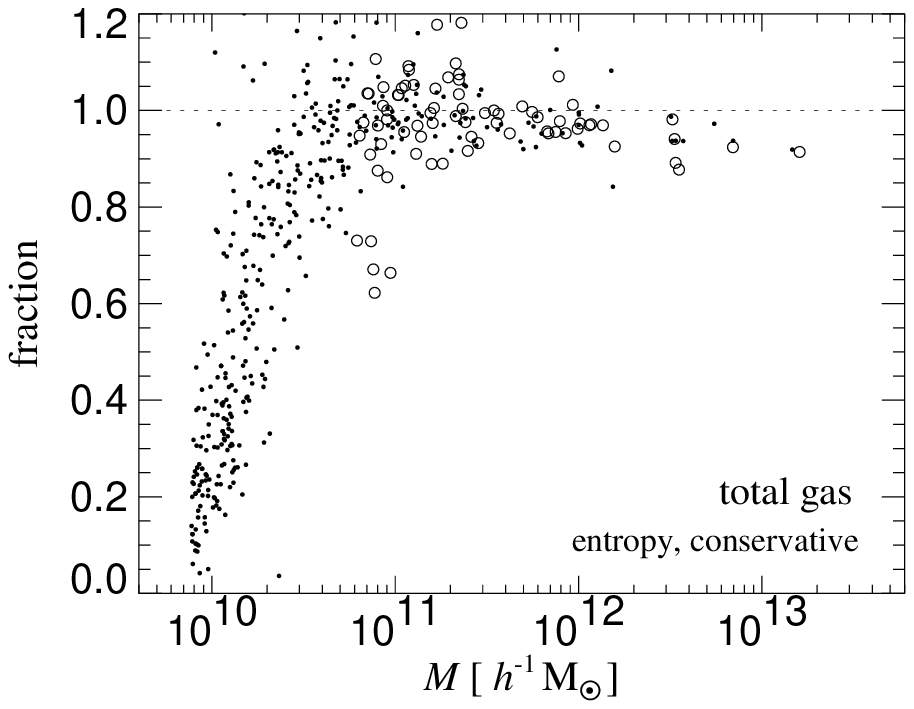}}%
\resizebox{5.6cm}{!}{\includegraphics{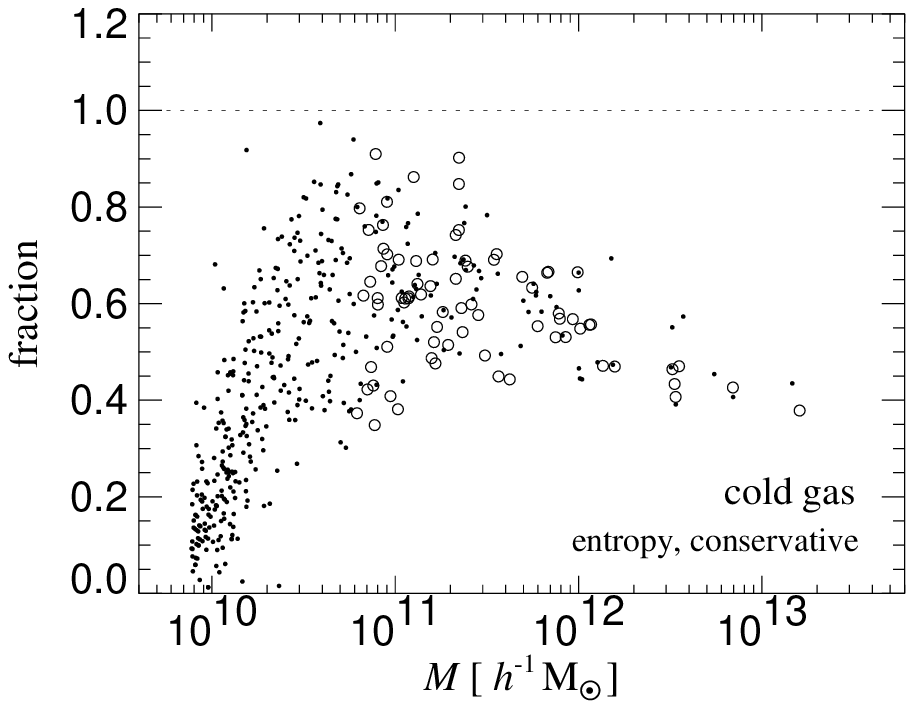}}%
\resizebox{5.6cm}{!}{\includegraphics{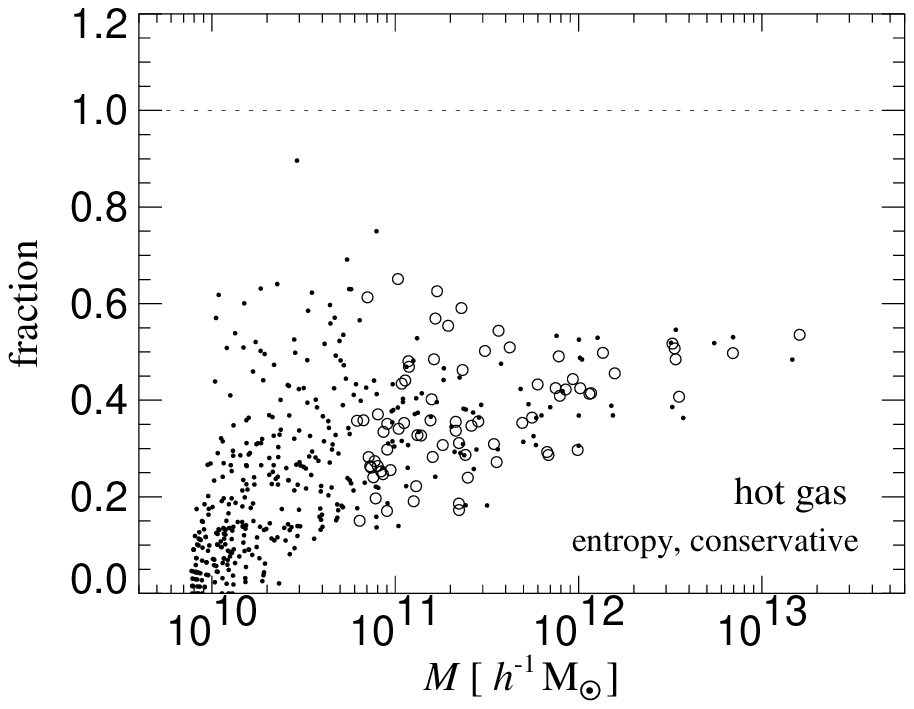}}\vspace*{-0.2cm}\\
\caption{Fraction of gas mass in halos as a function of virial mass,
measured at $z=0$. The panels on top compare results at obtained for
two different formulations of SPH at resolutions of $2\times 50^3$
(thick lines) and $2\times 100^3$ (thin lines). In each case, we show
fractions of total (dashed lines), cold (solid lines), and hot
(dot-dashed lines) gas within the virial radius, normalized to the
universal baryon fraction.  The plotted values are averages obtained
for groups of halos in logarithmic bins of size $\Delta \log M =
0.226$ in mass. The bottom three panels directly show the gas
fractions of individual halos in the entropy simulations. Hollow and
filled circles give results for the $2\times 50^3$ and $2\times 100^3$
resolutions, respectively.
\label{figHaloFraction}}
\ec
\end{figure*}

\begin{figure*}
\bc
\resizebox{8cm}{!}{\includegraphics{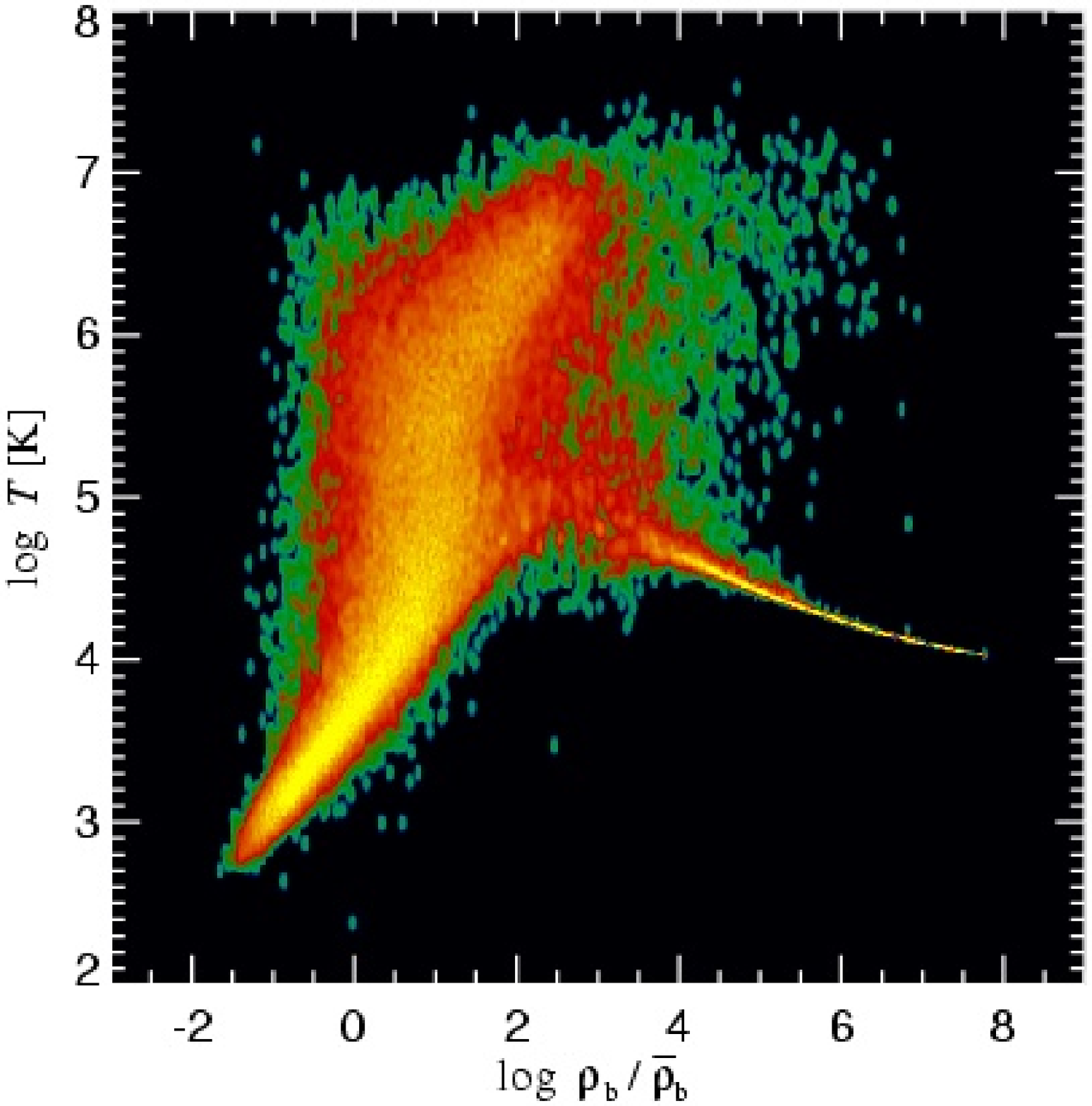}}%
\resizebox{8cm}{!}{\includegraphics{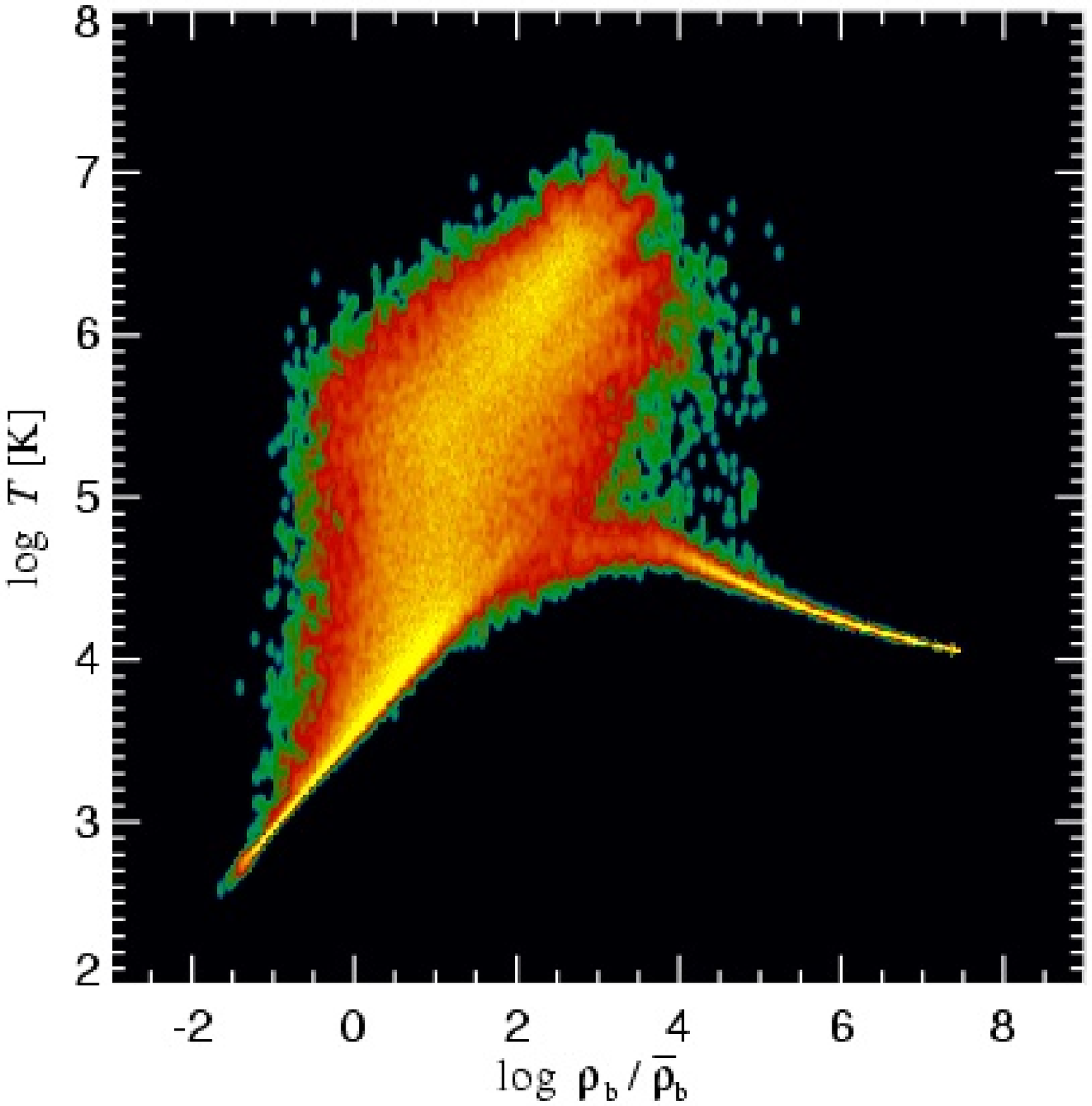}}\\
\caption{Phase-space diagrams of the density-temperature plane at
$z=0$ for two $2\times 50^3$ cosmological simulations with cooling.  The number
density of SPH particles is shown colour-coded. The left panel gives
the result for an integration of the thermal energy, using geometric
symmetrisation, while the right panel shows the result for an
integration of the entropy equation with our new conservative
approach.
\label{figPhaseSpace}}
\ec
\end{figure*}

\section{Cosmological results}

\begin{figure*}
\bc
\resizebox{8cm}{!}{\includegraphics{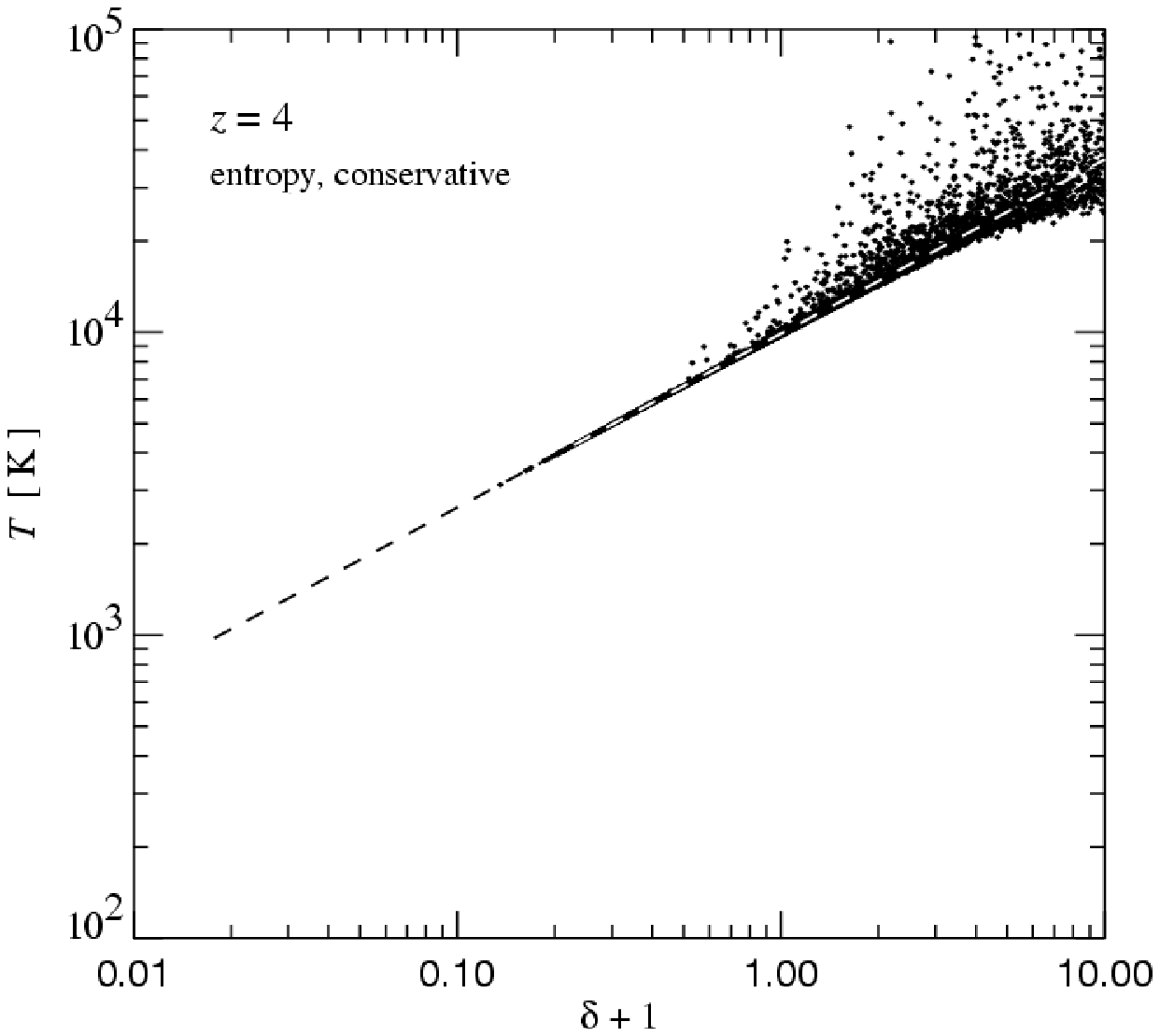}}%
\resizebox{8cm}{!}{\includegraphics{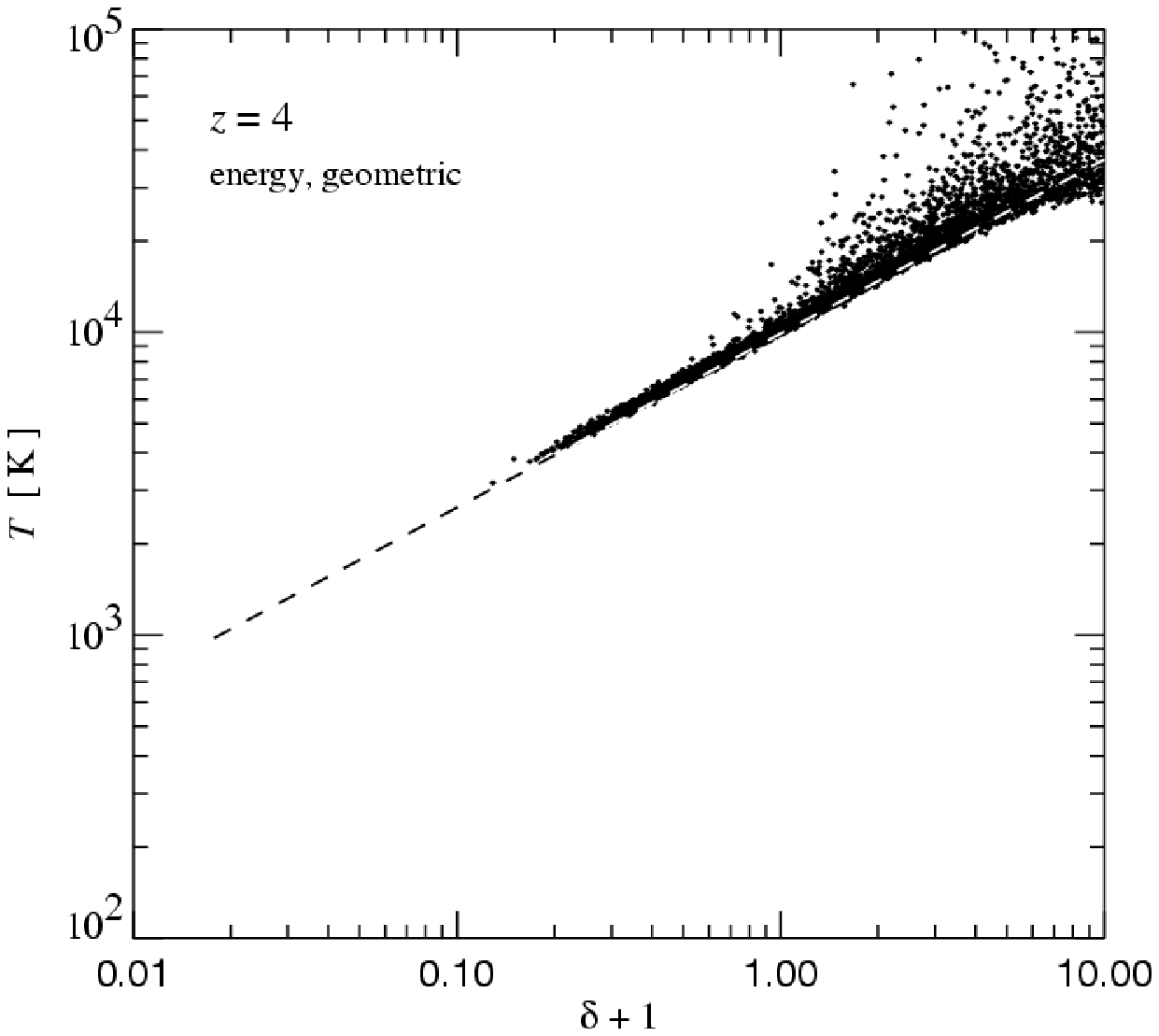}}\\
\resizebox{8cm}{!}{\includegraphics{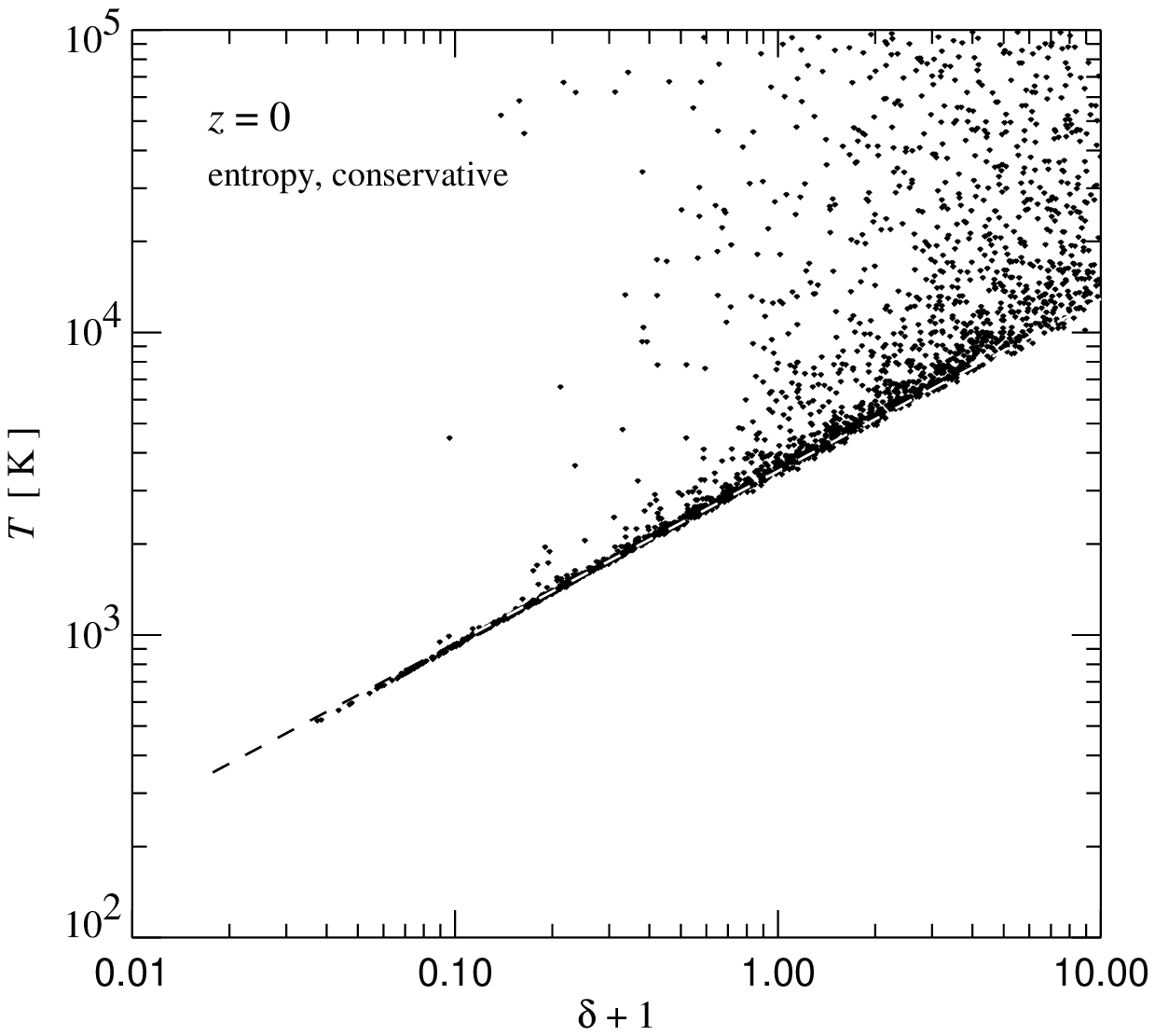}}%
\resizebox{8cm}{!}{\includegraphics{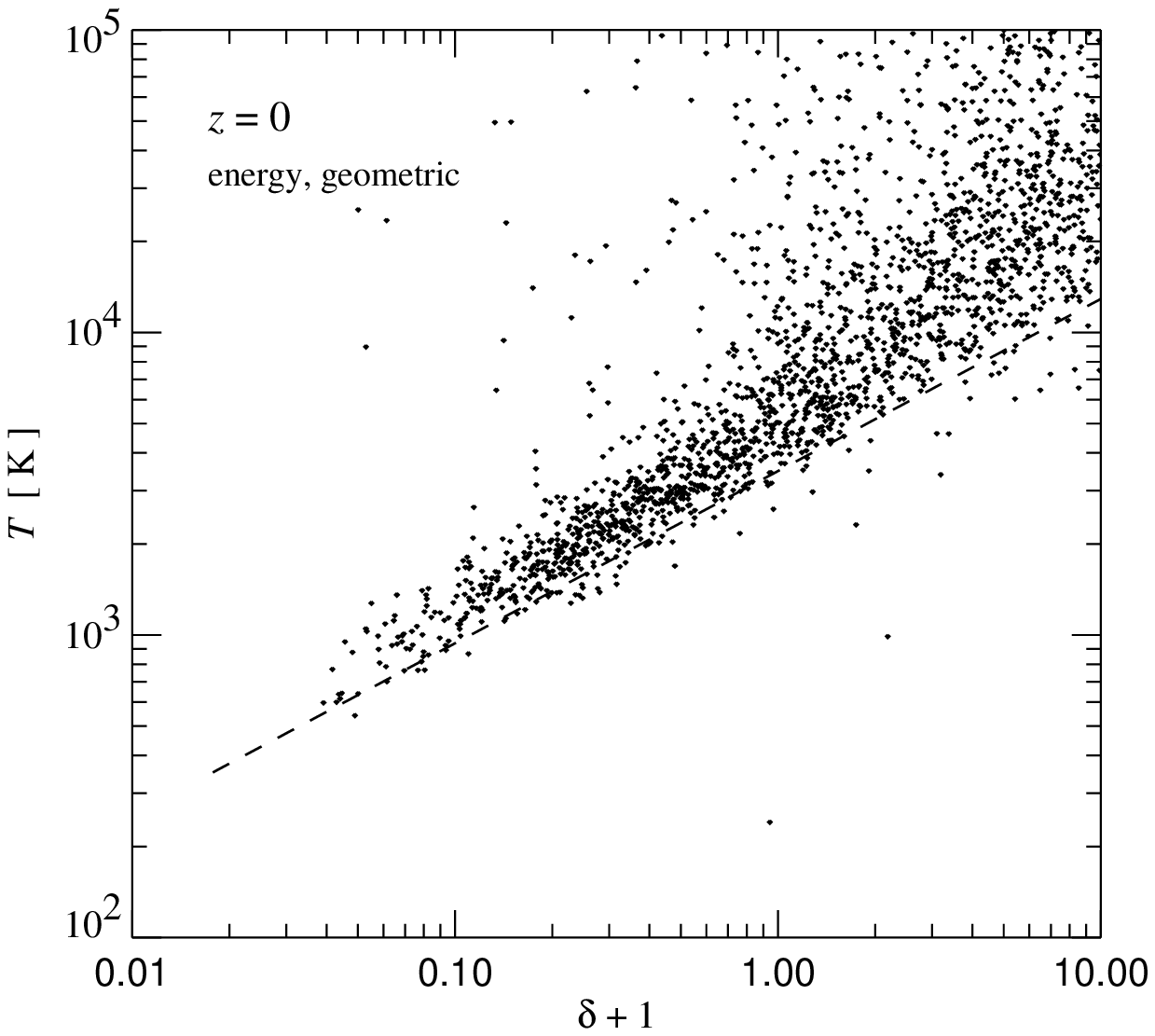}}\\
\caption{Temperature-density relations for the photoionized IGM at two
different redshifts, and for two different methods of solving the SPH
equations.  For graphical clarity, only a random subset of 5\% of
the particles is shown in each case. In the top panels at $z=4$, a
line has been fitted to the particles in the entropy run (conservative
formulation), marking the equation of state of unshocked IGM gas.  It
has amplitude $T_0=9925\,{\rm K}$ and slope $\alpha=0.57$. The same
line is also plotted in the corresponding thermal energy run
(geometric symmetrisation employed).  In the bottom panels, the fit to
the entropy run is described by $T_0= 3550\,{\rm K}$, $\alpha=0.58$,
and again, the same line is drawn for the $z=0$ thermal energy run.
\label{figEqs}}
\ec
\end{figure*}

Based on the results presented above, we expect a significant
difference in the fraction of gas that can cool in halos of a given
size in full cosmological simulations when SPH is formulated either in
terms of the thermal energy equation or the entropy equation.  We
examine this further using moderate-sized test simulations at two
different mass resolutions, containing a total of $2\times 50^3$ and
$2\times 100^3$ dark matter and SPH particles, respectively.  We
consider a canonical $\Lambda$CDM cosmology with parameters
$\Omega_0=0.3$, $\Omega_\Lambda=0.7$, $\Omega_b=0.04$, $h=0.67$, and
$\sigma_8=0.9$ within a periodic box of length $11.3\,h^{-1}{\rm Mpc}$
per side.  We include radiative heating and cooling processes for a
primordial mix of helium and hydrogen in the way described by
\citet{Ka96}.  An external UV background field similar to that
advocated by \citet{Ha96} photoionizes the gas and reionization takes
place at redshift $z\simeq 6$ \citep[for details, see][]{Da99}.  For
the purposes of the present study, we have not renormalized the
ionizing background to match e.g.~the redshift evolution of the mean
opacity of the Lyman-$\alpha$ forest, as is commonly done \citep[see
e.g.][]{He96}.  While boxes of this size are too small to be
representative of the universe at low redshifts, they suffice to
investigate systematic differences arising from different formulations
of SPH.

In Figure~\ref{figPhaseFrac}, we show the fraction of baryons residing
in different phases as a function of redshift.  These phases are
defined as in \citet{Dave2001}: Gas with temperatures above
$10^7\,{\rm K}$ is referred to as `hot', gas with temperatures between
$10^5\,{\rm K}$ and $10^7\,{\rm K}$ is `warm', and at lower
temperatures gas is either labelled `diffuse' if it is at
overdensities $\rho < 1000\,\overline\rho\,$, or `condensed' (cold)
for $\rho>1000\,\overline\rho\,$.  The left panel in
Figure~\ref{figPhaseFrac} shows the result when the thermal energy
equation is integrated with geometric mean symmetrisation, while the
right panel is for our new conservative entropy method. In each panel,
we show the results obtained for the $2\times 50^3$ resolution as thick
lines, while thin lines give the results for the $2\times 100^3$
simulations.

When the two SPH methods are compared, it is obvious that there is a
very large difference in the amount of gas that becomes cold and
dense. The entropy method predicts almost a factor of 2 less cold gas
for these simulations, which is reflected in a corresponding
difference in the amounts of gas in the warm phase.  The fraction of
gas in the diffuse phase is nearly unaffected, however. There is
virtually no hot gas in these simulations because the box size is too
small to contain massive galaxy clusters. Note that the fraction of
cold gas keeps growing for the geometric method by a large amount even
at low redshift, while it stays nearly constant for the entropy method
for redshifts below $z \simeq 2$. This is likely related to the
instability of density discontinuities in the geometric symmetrisation
technique, as discussed at the end of Section~\ref{sectioncool}. At
low redshift, this allows large halos to keep cooling efficiently
because their gas easily `overlaps' with the phase of cold dense gas
that has accumulated at the centres of these halos.

As far as convergence is concerned, the total amount of cold gas
becomes slightly larger in both SPH variants when better mass
resolution is employed.  For example, the fraction of cold gas goes up
from 20.0\% to 23.7\% in the entropy method when moving from the
$2\times 50^3$ to the $2\times 100^3$ resolution. Note however that
the higher resolution runs are able to probe halos of smaller mass
than were previously resolved. The gas that cools in these newly seen
objects can possibly account to a large extent for the difference in
the total amount of cold gas. A better assessment of convergence is
therefore obtained if we compare the content of cold gas in individual
halos as a function of their mass. To this end, we find halos in the
simulations as follows: We first construct a catalogue of group
candidates using a friends-of-friends algorithm with linking length
0.2, applied to the dark matter only. For each dark matter group that
contains at least 64 particles, we define a group centre as the
position of the particle with the minimum gravitational potential.  We
then determine a spherical `virial radius' around this centre such
that the enclosed `virial mass' of gas and dark matter has an
overdensity of 180 with respect to the background. For each of the
resulting halos, we also compute its fractional content of gas,
$(M_{\rm gas}/M_{\rm vir})\times (\Omega_0/\Omega_b)$, normalized to
the universal baryon fraction. Similarly, we define a fraction of cold
gas, where cold is here taken to be gas colder than $5\times
10^4\,{\rm K}$, and we define a `hot' gas content simply as the
difference of the total and cold gas fractions.

In Figure~\ref{figHaloFraction}, we show these gas fractions as a
function of virial mass, comparing results for the $2\times 50^3$ and
$2\times 100^3$ simulations when the geometric symmetrisation is used,
or the new entropy formulation is employed. In the two panels on top,
we have used logarithmic bins in mass to compute mean gas fractions
for halos in each mass bin. This suppresses the scatter, which we
separately illustrate for the entropy runs in the lower three panels
by plotting indiviudal halos as symbols.

Overall, the agreement between the low- and high-resolution runs is
gratifyingly good for the entropy method, as seen in the top right
panel of Figure~\ref{figHaloFraction}.  Even at the resolution limit
of $6.15\times 10^{10}\, h^{-1}{\rm M}_\odot$ of our $2\times 50^3$
halo catalogue, the fraction of cold gas is well reproduced, apart
perhaps from a small bias towards a lower amount of cold gas, which is
however expected based on our earlier results for the cooling in
isolated gas spheres. Interestingly, at the low mass end, the
high-resolution catalogue shows a drop of the total gas fraction below
the universal value, while a similar effect is not observed for the
low-resolution run.  This depletion of the total gas content is
presumably caused by the UV background that we included
\citep{Gnedin00}. At redshift $z=0$, this effect appears to be strong
enough to start reducing the gas content in objects with virial
temperatures below $T\sim 8\times 10^4\, {\rm K}$.

When the cold gas fractions for the geometric symmetrisation technique
are considered, it becomes again clear that this method cools gas too
efficiently. Here the low resolution run actually tends to produce
{\em more} cold gas at a given mass scale than the high resolution
run, again in agreement with the trends seen in the collapse
simulations of isolated gas spheres discussed earlier. Note that the
difference in the cold and total gas fractions between the two
different resolutions is also markedly larger than seen for the
entropy runs.  Interestingly, the total baryon fraction is also pushed
beyond the universal value in this method, indicating that the strong
cooling also affects the accretion flow into the halos.

The difference in the phase space distributions of the gas between the
different SPH techniques may also be shown directly in a
density-temperature diagram.  In Figure~\ref{figPhaseSpace}, we show
where the SPH particles lie in the overdensity-temperature plane.
Compared to the simulation that integrates the entropy, the plume of
shock heated gas is clearly less populated in the run that evolved the
energy equation.  This gas has cooled and moved to the locus of dense
gas at $T\simeq 10^4\,{\rm K}$.  The amount of diffuse, low-density
gas is the same in the two simulations.  However, the entropy run
clearly exhibits much less scatter in the density-temperature relation
of the photoionized IGM.  \citet{Hui97} have studied the expected
density-temperature relation of the photoionized intergalactic medium.
At low overdensities ($\delta < 5$), the gas is expected to follow a
power-law `equation of state' \be T= T_0 (1+\delta_{\rm b})^{\alpha},
\ee where $\delta_b$ is the baryonic overdensity.  The amplitude $T_0$
and the slope $\alpha$ at a given epoch depend on the details of the
reionization history, and on cosmology.  Because gas with density
around or below the mean is not expected to have been shocked, very
little scatter is expected around this relation, as long as the
ionising radiation field is homogeneous.

We examine the scatter in the equation of state of the IGM in
Figure~\ref{figEqs}.  In the top panels we compare the
temperature-density relation for the two runs at redshift $z=4$, and
in the bottom panels at redshift $z=0$.  While at redshift $z=4$ the
scatter in the thermal energy equation run is only marginally larger
than that in the entropy run, it has grown substantially by redshift
$z=0$.  This happens because the adiabatic cooling of the particles
due to the expansion of the universe is perturbed by noise in the SPH
estimates.  The particles see only a finite number of neighbours that
move away from them and generate the adiabatic cooling.  While they
cool at {\em nearly} the correct rate on average, some of them will do
so more rapidly or more slowly than average and, as a result,
particles tend to diffuse away from the adiabat they are expected to
follow.  In addition, for a small number of smoothing neighbours and a
nearly homogeneous particle distribution, there can be a small bias in
the mean of the estimates of the local velocity divergences, resulting
in a mean temperature that does not decline exactly in proportion to
$a^{-2}$, where $a$ is the scale factor.  In fact, this happens for
the thermal energy run, which arrives at an IGM equation of state with
a slightly higher value of $T_0$.  This is a result of a small
systematic underestimate of the cooling rate due to the adiabatic
expansion of the universe.  Ultimately, this is again a manifestation
of the violation of entropy conservation in this approach.  Note that
our code computes the rate of change of thermal energy for each SPH
particle in physical coordinates, thus the situation can be compared
to a homogeneously expanding polytrope where the smoothing lengths
increase with the expansion.  As \citet{He93} has shown, the entropy
will be violated under these circumstances.  On the other hand, if the
entropy is integrated, the adiabatic cooling is automatically exact.
Particles will only move to higher temperatures if they are shocked.
The entropy computation thus proceeds essentially as in the
`semi-analytic' method outlined by \citet{Hui97}, and should represent
the more accurate result.

\section{Discussion}

We have derived a new formulation of SPH that, when appropriate,
manifestly conserves both entropy and energy if the smoothing lengths
are adjusted to a constant local mass resolution.  Unlike previous
attempts to include $\nabla h$-terms, the equations of motion retain a
remarkably simple form.  In our approach, $\nabla h$-terms do not have
to be evaluated explicitly, mitigating additional sources of noise.
This improves on previous schemes where these terms depend on the
single particle with the maximum distance among the set of neighbours.

We have investigated the robustness of several formulations of SPH
when applied to the problem of cooling in halos under conditions of
poor resolution, and when dealing with strong explosions triggered by
local energy injection, as it occurs in certain feedback scenarios.
As such, these test problems are highly relevant for current attempts
to model galaxy formation in cosmological SPH simulations.

We find that the standard formulations of SPH in terms of the thermal
energy can lead to substantial overcooling in poorly resolved halos
when cooling flows are not adequately resolved.  This can lead to
insufficient compressional heating in the accretion flow, or expressed
differently, to a violation of entropy conservation in the adiabatic
region of the flow.  In the cooling problem, this leads to an
accelerated accretion, an effect that becomes more substantial for
halos resolved with few particles, and for temperatures in a range
where the cooling function declines with temperature.  The effect is
particularly severe for a geometric symmetrisation of the pressure
terms, which can also cause unstable behaviour of strong density
discontinuities.

Point-like energy injection can lead to unphysical negative
temperatures in the standard formulation of SPH.  When the
temperatures of these particles are prevented from becoming negative,
a reasonable explosion still takes place once the initial energy has
spread over a smoothing volume, but a substantial violation of total
energy conservation results.  When the pressure terms are symmetrised
with a geometric mean, the integration of the thermal energy equation
is more stable, but then unphysical behaviour may result for very
strong gradients in energy density.  A better alternative is to use
the asymmetric form of the thermal energy equation, which avoids such
artifacts.

On the other hand, when the entropy equation is integrated, point
explosions are also treated very well.  Even when the initial
explosion energy is deposited in a smooth way, the blast waves exhibit
less scatter than for an integration of the thermal energy equation.
However, the total energy is not conserved well in this case, and
shows fluctuations of order several percent, although there are no
secular trends of energy violation.  This is not surprising.
\citet{He93} has shown that for standard formulations of SPH,
simultaneous conservation of energy {\em and} entropy is not manifest.
Depending on whether SPH is solved in terms of energy or entropy, one
of the conservation laws is only strictly fulfilled in the continuum
limit.  However, in this paper we have derived a new formulation of
SPH that solves this problem and conserves both energy and entropy.
In addition, this new method is nearly as easy to implement
numerically as the standard approach.

We believe that the results presented in this study clearly favour the
entropy formulation of SPH for cosmological simulations of galaxy
formation.  It provides the best technique among the ones we have
tested for modeling the process of point-like energy injection, which
is relevant for certain feedback algorithms.  It also avoids
artificial overcooling in poorly resolved halos and reduces the
scatter in the density-temperature relation of the gas in the low
density Ly-$\alpha$ forest.

\section*{Acknowledgements}

We thank Simon White and Simone Marri for instructive discussions and
critical comments that were helpful for the work on this paper.  We
are indebted to David Weinberg for useful suggestions, and to
the anonymous referee for providing a highly useful report that helped
to improve the paper.  This work was supported in part by NSF grants
ACI96-19019, AST-9803137, and PHY 9507695.

\bibliographystyle{mnras}
\bibliography{paper_entropy}

\end{document}